\documentclass[fleqn,usenatbib]{mnras}

\usepackage{newtxtext,newtxmath}

\usepackage[T1]{fontenc}

\DeclareRobustCommand{\VAN}[3]{#2}
\let\VANthebibliography\thebibliography
\def\thebibliography{\DeclareRobustCommand{\VAN}[3]{##3}\VANthebibliography}

\usepackage{graphicx}	
\usepackage{amsmath}	
\usepackage{xcolor}
\usepackage{tabularx}
\usepackage{mathrsfs}

\newcolumntype{Y}{>{\centering\arraybackslash}X}

\title[The Disk in I Zw 1]{A Phenomenological Study of the Accretion Disk in the Super-Eddington AGN I Zw 1}

\author[F. Drewes et al.]{Farin Drewes,$^{1}$\thanks{E-mail: \href{mailto:n.c.drewes@soton.ac.uk}{n.c.drewes@soton.ac.uk}} Roberta Vieliute,$^{2}$ Juan V. Hern\'{a}ndez Santisteban,$^{2}$ Keith Horne,$^{2}$ Aaron J. Barth,$^{3}$ 
\newauthor Edward M. Cackett,$^{4}$ Encarni Romero Colmenero,$^{5,6}$ Michael R. Goad,$^{7}$ Shai Kaspi,$^{8}$ Hermine Landt,$^{9}$ 
\newauthor Paulina Lira,$^{10}$ Hagai Netzer,$^{8}$ Marianne Vestergaard,$^{11}$ Hartmut Winkler$^{12}$ 
\\\\
$^{1}$School of Physics \& Astronomy, University of Southampton, University Road, Southampton SO17 1BJ, UK\\
$^{2}$SUPA Physics and Astronomy, University of St. Andrews, Fife KY16 9SS, UK\\
$^3$Department of Physics and Astronomy, 4129 Frederick Reines Hall, University of California, Irvine, CA 92697-4575, USA\\
$^4$Wayne State University, Department of Physics \& Astronomy, 666 W Hancock St, Detroit, MI 48201, USA\\
$^5$South African Astronomical Observatory, P.O Box 9, Observatory 7935, Cape Town, South Africa\\
$^6$Southern African Large Telescope Foundation, P.O Box 9, Observatory 7935, Cape Town, South Africa\\
$^7$School of Physics and Astronomy, University of Leicester, University Road, Leicester, LE1 7RH, UK\\
$^{8}$School of Physics and Astronomy and Wise Observatory, Tel-Aviv University, Tel-Aviv 6997801, Israel\\
$^9$Centre for Extragalactic Astronomy, Department of Physics, Durham University, South Road, Durham DH1 3LE, UK \\
$^{10}$Departamento de Astronomía, Universidad de Chile, Camino el Observatorio 1515, Santiago, Chile\\
$^{11}$DARK, The Niels Bohr Institute, University of Copenhagen, Jagtvej 155, DK-2200 Copenhagen, Denmark\\
$^{12}$Department of Physics, University of Johannesburg, PO Box 524, 2006 Auckland Park, South Africa
}

\date{Accepted XXX. Received YYY; in original form ZZZ}

\pubyear{\the\year{}}

\begin{document}
\label{firstpage}
\pagerange{\pageref{firstpage}--\pageref{lastpage}}
\maketitle

\begin{abstract}
The structure of the accretion disk in AGN is still an unsolved question, especially how it may change with Eddington ratio. Here we examine the accretion disk in the super-Eddington AGN I~Zw~1 using reverberation mapping of the optical continuum. We use three years of optical monitoring with Las Cumbres Observatory at sub-day cadence in $uBgVriz_s$. The lag-wavelength spectrum, calculated using the cross correlation method and PyROA, shows a $u$-band excess. PyROA lags are equally well fitted with a thin and slim disk profile. The UV/optical AGN spectral energy distribution is consistent with a thin disk. The disk size at $4495\:\textnormal{\AA}$ for a thin disk model is $4.23\pm0.24\:\mathrm{ld}$ and for a slim disk model is $1.71\pm0.09\:\mathrm{ld}$, larger by a factor of $2-4$ than the fiducial disk size of $1.07\pm0.15\:\mathrm{ld}$ as determined using the Eddington ratio. We find evidence of different size scales probed with different variability timescales. Lags evaluated at longer variability timescales increase as do frequency-resolved lags at low frequencies, which we interpret as an additional secondary reprocessor at large radii consistent with the broad-line region (BLR) in I~Zw~1. The high frequency lags, predicted well with just a disk, are fit with a thin disk profile and a size of $0.61\pm0.37\:\mathrm{ld}$. This indicates that the actual disk size may be on the order of the fiducial size. We also collate the most extensive set of directly measured internal sizes of an AGN, from optical to mid-infrared with reverberation mapping and optical interferometry. Assuming that the disk is indeed the fiducial size, these show little evidence that the accretion disk extends into the BLR significantly, tentatively disfavouring the failed radiatively accelerated dust driven outflow BLR formation model.
\end{abstract}




\section{Introduction}

Active galactic nuclei (AGN) are some of the most luminous objects in the Universe, capable of outshining the entire stellar population of their host galaxy. The source of this power is the most efficient process of mass-energy conversion in the Universe: accretion onto a black hole and the associated release of gravitational potential energy through radiation and outflows \citep{salpeter1964,lyndenbell1969,Laha2021}. In AGN, the accreting material orbits in a disk around the black hole. The fundamental analytic approximation of this structure is that of a thin disk: geometrically thin ($H/R \ll 1$) and optically thick, with a radial temperature profile of $T \propto R^{-3/4}$ \citep{novikov1973,shakura1973}. However, the exact nature of this accretion disk is still under discussion, particularly when applied to AGN \citep{Antonucci2015,Lawrence2018}.

Even though AGN disks are too small to be resolved in the IR/optical/UV wavelengths, currently and presumably in the near future, their stochastic variability enables us to probe this temperature profile through reverberation (echo) mapping \citep[RM,][]{Collier1999,cackett2007}. In the typical assumption of the internal AGN structure, a central variable X-ray source illuminates the accretion disk from above, which then reprocesses and re-emits this light according to the disk temperature profile, called the lamppost model \citep{Haardt1991}. Accordingly, light curves measured at different wavelengths are dominated by the light reprocessed at different radii. These light curves are similar in shape but smoothed, lower in amplitude, and shifted in time, with the lag between them being the light travel time between the different emission radii. 
Measuring the lag between light curves at different wavelengths, $\tau(\lambda)$, enables us to measure distances across the disk.

Disk reverberation studies have all shown correlated continuum light curves, with short wavelength light curves temporally leading those at longer wavelengths on the order of days \citep[e.g.,][]{cackett2007,cackett2018,fausnaugh2016,fausnaugh2018,edelson2017,hernandezsantisteban2020}. This implies the presence of a structure with a radial temperature profile reprocessing light from a central, variable source. In the vast majority of objects, the lag-wavelength spectrum is well fit with a thin disk $\tau(\lambda) \propto \lambda^{4/3}$ relation \citep[e.g.,][]{Collier1999,cackett2007,fausnaugh2016,edelson2019}. However, several frequently found discrepancies suggest that the accretion disk and observations are significantly more complicated. First, there is the `$u/U$-band excess'. This is an excess lag observed in the $u/U$-band, at times accompanied by a similarly increased lag in the $r/R$-band/$i/I$-band. In addition, there is the `too-large disk' problem. The normalisation of the lag-wavelength relation lets us infer a size, or alternatively an accretion rate, of the disk \citep[under the assumption of the classical thin disk-lamppost model,][]{cackett2007,fausnaugh2016}. The majority of studies have found a disk several times larger than predicted by their bolometric luminosity, usually around a factor of $2-3$ \citep[also observed via microlensing e.g.,][]{Morgan2010}. Another feature often noticed is continuum variability on different timescales \citep{hernandezsantisteban2020,vincentelli2021,cackett2023,donnan2023,miller2023,lewin2024}. While variability on a daily scale appears to be related to disk reprocessing, additional variability on longer timescales (tens – hundreds of days) has also been found and contains important information about the internal structure of the AGN. Recently, frequency\footnote{Frequency in this paper is only used to refer to Fourier frequencies, i.e. the variability frequency, and not the frequency of radiation. Radiation energy is only referred to using its corresponding wavelength.}-resolved Fourier analysis of continuum light curves has shown great power to unravel different sources of reverberation signals \citep{uttley2014,cackett2022,lewin2023,lewin2024,panagiotou2025}.

It has been frequently noted and demonstrated that these discrepancies can be created by a contamination of the disk continuum by diffuse continuum (DC) emission from a second reprocessing region at larger distances, usually interpreted as the broad line region \citep[BLR,][]{korista2001,korista2019,lawther2018,Netzer2022}. The BLR is a region of high velocity and density clouds at larger radii than the accretion disk. This BLR diffuse continuum is particularly strong at the Balmer and Paschen jumps, i.e. the $u/U$- and $i/I$-bands. Further, the BLR continuum likely ‘artificially’ increases lags through a second reprocessing at larger distances at all wavelengths and may therefore be partially responsible for the too-large disk problem. 

Of course, one of the main reasons for the complexity of these studies is that the underlying disk may not be a simple thin disk, but rather more complex, such as a slim disk \citep{abramowicz1988}. It is generally presumed that AGN accretion mechanisms and disk structures change at the Eddington limit. In the basic picture, sub-Eddington AGN as described above are presumed to host thin disks, and super-Eddington AGN slim disks. This transition is probably rather gradual as advective cooling becomes more and more important, even at moderate to high sub-Eddington rates. 

There have been fewer super-Eddington objects studied, but they mostly present with the same features as sub-Eddington objects. The lag-wavelength spectra are well fit with the thin disk profile of $\tau(\lambda) \propto \lambda^{4/3}$ and the slim disk profile of $\tau(\lambda) \propto \lambda^{2}$ corresponding to the slim disk temperature profile $T\propto R^{-1/2}$ \citep[inside the photon trapping radius; discussed in further detail in Section \ref{disk_struc},][]{wang1999,cackett2020,donnan2023,thorne2025}. In addition, an excess in the $u/U$-band lag has been observed, and there is evidence of the too-large disk problem. The SED is used to analyse the temperature profile of the underlying disk, where a thin disk has a long-wavelength tail with $F_{\nu} \propto \nu^{1/3}$, while a slim disk has $ F_{\nu} \propto \nu^{-1}$ \citep{wang1999b}. SED constraints generally favour the thin disk \citep{cackett2020,donnan2023}.

AGN disks and the study thereof is complex. Therefore, we aim to foremost phenomenologically describe the disk structure of the prototype narrow line Seyfert 1 galaxy I Zwicky 1 (I~Zw~1). I~Zw~1 is super-Eddington with a central black hole mass of $\log(M_{\rm BH}/{\rm M}_\odot)\sim 6.97$, accreting at an Eddington ratio of $\dot{m}_E \sim 2$ and at a redshift of $z\sim0.061$ \citep[$255\:\mathrm{Mpc}$,][]{huang2019,drewes2025}. This enables us to observe an accretion disk in an extreme state, which can reveal more information about the underlying physics. The structure of this AGN has also been extensively studied across the electromagnetic spectrum, including the X-ray, UV/optical observations and reverberation mapping of the BLR, and radial mapping of the dusty `torus' \citep{silva2018,rogantini2022,huang2019,juranova2024,burtscher2013,gravity2024,drewes2025}. This enables us to also put our results in the wider context of multi-wavelength AGN structure. 

We have obtained photometric monitoring of I~Zw~1 using Las Cumbres Observatory \citep{Brown:2013}, over three years in 7 optical bands with an average cadence of $\simeq0.75$ days. We set out three aims for this paper. Firstly, we will phenomenologically describe the accretion disk in I~Zw~1. Next, we will discuss our observations in the context of other super-Eddington and sub-Eddington disks to compare and contrast their underlying disk structure. Lastly, we will examine the multi-wavelength structure of I~Zw~1, for the first time assembling physical sizes from the optical accretion disk to the mid-IR dust. In Section 2, we describe our data collection and reduction processes. In Section 3, we conduct the time series analysis of the light curves, using the cross correlation and PyROA methods, as well as Fourier analysis. In Section 4, we analyse the SED. In Section 5, we discuss our results in relation to a phenomenological description of the disk structure, in the context of other AGN disk studies, and the multi-wavelength structure of I~Zw~1.

\section{Data Collection and Reduction}\label{sec_data}

In this section we describe our data collection and reduction procedures. Our main data set is four years of optical photometric observations of I~Zw~1 with a sub-day cadence from the Las Cumbres Observatory (LCO). We supplement this with UV/optical observations from \textit{XMM-Newton} OM and \textit{Swift} UVOT. Even though only one of these observations was taken concurrently with the LCO monitoring, we include these data to characterise the UV emission generally. The observation logs are presented in Table~\ref{tab:obs}.

\begin{table*}
    \centering
    \begin{tabularx}{\textwidth}{Y|YYY}
    Facility & Observation Date & Filters & Cadence (days) \vspace{1ex}\\ \hline \hline \\[-1ex]
    LCO & Year 1: 02.07.2020 -- 09.02.2021 & $uBgVriz_s$ & 1.14 \\
    & Year 2: 23.05.2021 -- 19.02.2022 & & 0.86 \\
    & Year 3: 09.06.2022 -- 15.02.2023 & & 0.56 \\
    & Year 4: 23.05.2023 -- 08.02.2024 & & 0.91\vspace{1ex}\\
    \hline \\[-1ex]
    \textit{XMM-Newton} OM & 19.01.2015& \textit{UVW2, UVM2, UVW1, U, B, V}&\\
    &21.01.2015& \textit{UVW2, UVM2, UVW1, U, B, V}&\\
    &12.01.2021& \textit{UVW2, UVW1, U, B, V}\vspace{1ex}&\\
    \hline \\[-1ex]
    \textit{Swift} UVOT & 13.07.2023 & \textit{UVW2, UVM2, UVW1, U, B, V} &\\
    \end{tabularx}
    \caption{Observation log for the data used in this paper. Details about the observations and reductions are in Section~\ref{sec_data}.}
    \label{tab:obs}
\end{table*}

\begin{figure*}
    \centering
    \includegraphics[trim=0cm 0.25cm 0cm 0cm,clip,width=\textwidth]{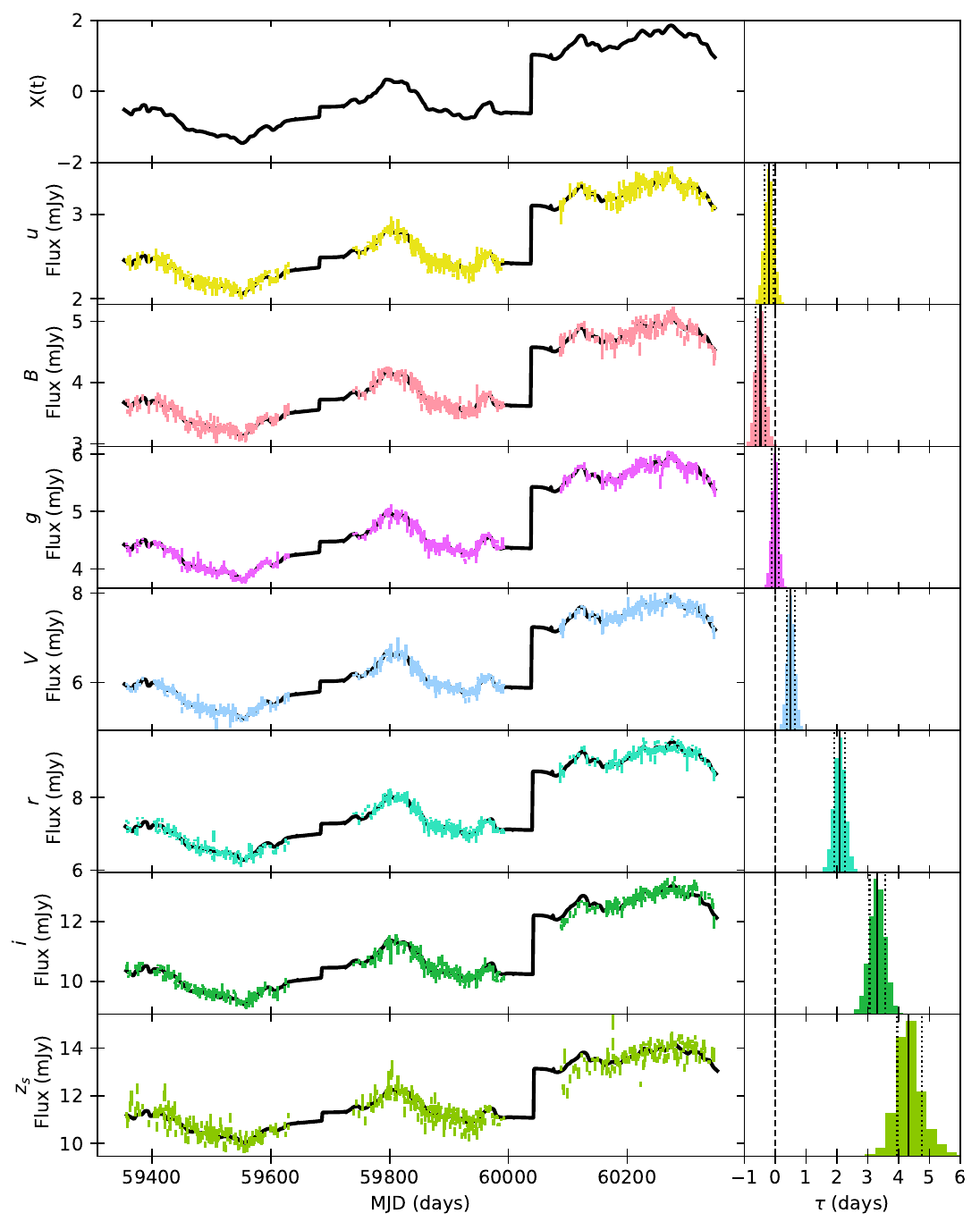}
    \caption{The LCO light curves in all bands for Years 2--4, the PyROA model (solid line), and its $68$\% confidence interval shown in grey, including the reference light curve $X(t)$ (top panel). The right panel shows the marginalised posterior distributions for the inter-band lags as calculated by PyROA, with its mean (solid line) and $68$\% confidence interval denoted by the dotted lines. Lags are measured with respect to the $g$-band.}
    \label{fig:lc}
\end{figure*}

\subsection{Las Cumbres Observatory}
The multiband observations of the light of curve of I~Zw~1 were acquired with the Las Cumbres Observatory robotic telescope network \citep{Brown:2013}. The observations were made under LCO Key Projects KEY2020B-006 and KEY2023B-001 (PI: J. V. Hern\'{a}ndez Sanstisteban). An overview of the observations can be found on the AGN Variability Archive (AVA)\footnote{\href{https://alymantara.com/ava/}{alymantara.com/ava/}}. The telescopes are 1\:m telescopes and host Sinistro CCD cameras with a field of view of $26.5'\times 26.5'$ and a resolution of $0\farcs{3}{8}{9}\:\mathrm{pix}^{-1}$. We obtained high cadence photometry in seven filters: Bessell $BV$, SDSS $ugri$, and Pan-STARSS $z_s$. At each observation, two exposures were made with exposure times $(BVugriz_s) = 2\times(30,30,120,30,30,30,60)\:\mathrm{s}$. Fig.~\ref{fig:spec} shows the average observed spectrum of I~Zw~1 across the LCO campaign overlaid with the LCO filter transmission curves. While the campaign on I~Zw~1 is still ongoing, here we look at the first four years of data. In these four years, we had four observing seasons, each between 250 – 270 days long. Apart from Year 1 (1.14 days), all other years have observations at a sub-day cadence. The average cadence of the latter three years is 0.75 days. A summary of the observations is presented in Table~\ref{tab:obs}. We do not use Year 1 due to its particularly low cadence and lower variability amplitude, and generally worse light curve quality and instead only use Years 2, 3, and 4 in our analysis. We downloaded the data from the LCO archive, flat-fielded and bias corrected internally using {\sc banzai} \citep{mccully2018}. Sources are extracted using {\sc sextractor} with an extraction aperture radius of $5''$ \citep{bertin1996}. This is to ensure that the effect of the variable PSF over the observing season due to atmospheric changes is minimal. The background is calculated using a global background model made by smoothing the image in a 200 point pixel mesh. We use the field stars to calculate the zero-point, using the AAVSO Photometric All-Sky Survey (APASS) DR10 for all filters except $u$ \citep{henden2018}. The $u$-band is calibrated using the Sloan Digital Sky Surveys (SDSS) DR16 \citep{ahumada2020}. For additional details on the data reduction process see \citet{hernandezsantisteban2020,donnan2023}.

To calibrate and adjust the light curves between the different telescopes we use a new intercalibration method, PyTICS, based on comparison field stars described in \citet{vieliute2025}\footnote{\href{https://github.com/Astroberta/PyTICS/}{github.com/Astroberta/PyTICS/}}. This method corrects for effects from the small systematic differences of the telescopes in the LCO network, such as filter transmissions and camera sensitivities, and the large range of observing conditions, including variations in weather, airmass, seeing, and sky transparency. PyTICS also estimates the additional uncertainty that arises from these effects, for example increasing noise for telescope faults or bad nights. This allows for more reliable identification and down-weighting of bad data points in contrast to simply using large offsets from the average AGN light curve. The correction factors and their uncertainties are calculated incrementally through an iterative process. They are then blindly applied to the AGN light curve, conserving the AGN’s inherent variability structure. This is done independently in each filter. For further details on the algorithm, see \citet{vieliute2025}.

\subsection{\textit{XMM-Newton} OM} \label{sec:xmm}
I~Zw~1 was observed with the optical monitor (OM) onboard \textit{XMM-Newton} in 2015 and 2021 \citep{jansen2001, mason2001}. Here we use the two observations from 2015 (obsIDs 0743050301 and 0743050801) and one observation from 2021 (obsID 0851990101). In 2015 the object was observed with the \textit{UVW2, UVM2, UVW1, U, B, V} filters while in 2021 only the \textit{UVW2, UVW1, U, B, V} filters were used (see Table~\ref{tab:obs}). We downloaded the processed data from the XMM-Newton Science Archive and extracted the relevant information. The data were processed with the Science Analysis System (SAS) version 18.0.0. Fluxes were extracted from a source region calculated by the \verb|omdetect| task, which detects sources and then performs aperture photometry with an extraction aperture the size of the source. The extraction apertures for I~Zw~1 vary between 2\farcs{0} and 2\farcs{5}.

\subsection{\textit{Swift} UVOT} \label{sec:swift}
We use archival ultraviolet/optical data from the Neil Gehrels {\it Swift} Observatory \citep{Gehrels:2004} serendipitously taken throughout the optical LCO monitoring. One epoch from 2023 was processed through the standard UVOT \citep{Roming:2005} pipeline (v.4.5) to obtain photometric measurements with a $5^{\prime\prime}$ radius aperture centred at the coordinates of the AGN, to be consistent with the ground-based photometry and avoid different host-galaxy contributions. The background estimation was selected from a $30^{\prime\prime}$ radius aperture located in a nearby blank part of the image.

\section{Time Series Analysis}\label{sec:tsa}

The light curves of I~Zw~1 over all three years show both short and longer term variability. They are also correlated – distinctive features are repeated across the seven bands (see Fig.~\ref{fig:lc}). The reference band is usually taken to be the shortest wavelength. However, in our case this is the $u$-band for which we expect significant BLR contamination, which will then dilute the disk signal in the lags from all other bands. We chose to measure these lags using as reference the $g$-band as this is the shortest wavelength that will have the least BLR contamination, with the highest data quality. 

Multiple methods can be used to calculate time lags, but here we focus on two: the cross correlation function (CCF) and PyROA \citep{donnan2021,donnan2023}. Cross correlation is the most direct mathematical method, shifting light curves with respect to a reference light curve and evaluating how well they correlate as a function of lag. PyROA instead constructs a reference light curve $X(t)$ from the running optimal average of all the data and directly models each light curve band to retrieve a time delay. In the following sub-sections we further describe these methods and evaluate the time lags. All lags discussed in this work are observed lags.

\begin{table*}
    \centering
    \begin{tabularx}{\textwidth}{ccYYYYYYY}
    &&\multicolumn{2}{c}{Year 2}&\multicolumn{2}{c}{Year 3}&\multicolumn{2}{c}{Year 4}&PyROA\\
    Band & $\lambda_{\mathrm{eff}}$ ($\textnormal{\AA}$) & $\tau_\mathrm{peak}$ & $\tau_\mathrm{cent}$&$\tau_\mathrm{peak}$&$\tau_\mathrm{cent}$&$\tau_\mathrm{peak}$&$\tau_\mathrm{cent}$&$\tau$\vspace{1ex}\\ \hline \hline \\[-1ex]
    $u$&$3540$&$0.20^{+0.20}_{-0.55}$&$-1.90_{-1.07}^{+1.26}$&$0.15_{-0.35}^{+0.10}$&$-1.82_{-0.80}^{+0.82}$&$-0.25^{+0.10}_{-0.10}$&$-1.96_{-0.45}^{+0.64}$&$-0.19^{+0.15}_{-0.15}$\vspace{1ex}\\
    $B$&$4361$&$-0.20_{-0.15}^{+0.45}$&$-2.38_{-1.16}^{+1.09}$&$0.10_{-0.25}^{+0.10}$&$-1.04_{-0.92}^{+0.79}$&$-0.05_{-0.10}^{+0.10}$&$-0.46_{-0.69}^{+0.56}$&$-0.48_{-0.16}^{+0.15}$\vspace{1ex}\\
    $g$&$4770$&$0.00_{-0.00}^{+0.00}$&$-0.02_{-0.92}^{+0.96}$&$0.00_{-0.00}^{+0.00}$&$0.01_{-0.64}^{+0.64}$&$0.00_{-0.00}^{+0.00}$&$0.00_{-0.48}^{+0.46}$&$0.00^{+0.11}_{-0.11}$\vspace{1ex}\\
    $V$&$5448$&$-0.25_{-0.10}^{+0.55}$&$1.38_{-1.49}^{+1.20}$&$0.15_{-0.05}^{+0.10}$&$1.37_{-0.77}^{+0.72}$&$-0.05_{-0.05}^{+0.40}$&$0.63_{-0.59}^{+0.65}$&$0.50_{-0.13}^{+0.13}$\vspace{1ex}\\
    $r$&$6215$&$5.35_{-2.80}^{+1.25}$&$7.92_{-1.84}^{+1.41}$&$0.95_{-0.65}^{+1.40}$&$1.81_{-0.76}^{+0.70}$&$1.30_{-0.75}^{+2.30}$&$2.83_{-0.62}^{+0.56}$&$2.09_{-0.18}^{+0.18}$\vspace{1ex}\\
    $i$&$7545$&$2.60_{-3.10}^{+2.50}$&$7.87_{-2.06}^{+1.67}$&$2.25_{-0.95}^{+2.10}$&$6.15_{-0.83}^{+0.67}$&$1.40_{-0.90}^{+1.35}$&$7.14_{-2.02}^{+3.67}$&$3.31_{-0.25}^{+0.24}$\vspace{1ex}\\
    $z_s$&$8700$&$9.55_{-2.08}^{+0.85}$&$10.8_{-2.2}^{+1.4}$&$0.20_{-0.55}^{+2.65}$&$4.77_{-1.75}^{+1.87}$&$6.85_{-1.10}^{+5.45}$&$8.05_{-2.10}^{+4.55}$&$4.32_{-0.36}^{+0.44}$\vspace{1ex}\\
    \hline
    \end{tabularx}
    \caption{Lags in days (observed frame) measured between each light curve and the reference light curve in the $g$-band. The interpolated cross correlation function was applied to each year of data separately. From this, the lag at the peak of the CCF $\tau_\mathrm{peak}$, and the centroid  above $0.8r_\mathrm{peak}$, $\tau_\mathrm{cent}$, are calculated (see Section~\ref{ccf}). PyROA was applied to Years 2--4 simultaneously (see Section~\ref{pyroa}). Uncertainties are the $68$\% confidence intervals.}
    \label{tab:lags}
\end{table*}

\subsection{Cross Correlation} \label{ccf}

\begin{figure}
    \centering
    \includegraphics[width=0.48\textwidth]{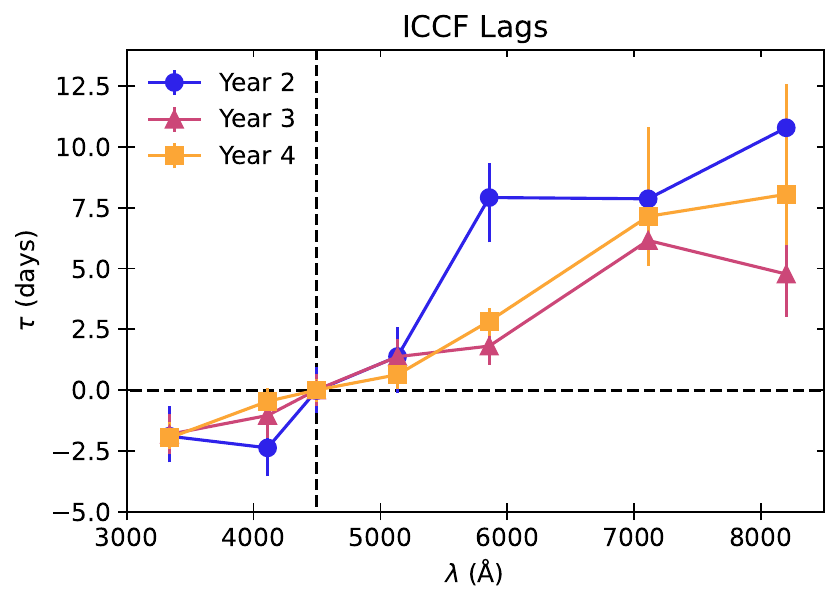}
    \caption{Lag spectrum for Years 2--4 as calculated using ICCF, using the centroid lag $\tau_\mathrm{cent}$ and with reference to the $g$-band in the AGN rest frame. Year 2 is denoted by the circles, Year 3 by the triangles, and Year 4 by the squares. Lags plotted here are presented in Table~\ref{tab:lags}. Lags increase with wavelength for all years.}
    \label{fig:iccf_lags}
\end{figure}

Here we use the interpolated cross correlation function (ICCF) method with flux randomisation/random subset sampling (FR/RSS) as demonstrated in \citet{peterson2004}, using the PyCCF code \citep{sun2018} to retrieve the inter-band lags. This method accounts for uneven sampling of the light curves and uncertainties in the data, and estimates relatively reliable errors on the lags. Notably, however, it has been shown that ICCF overestimates errors \citep{cackett2018,yu2020}.

We use RSS to create a large number ($N_\mathrm{sim}$) of realisations of the light curves, which is sampling with substitution. The errors on the data points are scaled according to the number of times that point has been picked. Then, applying FR, we add Gaussian noise where the mean is the flux of the point and sigma is the error on the flux. The ICCF is evaluated by linearly interpolating one light curve first, and then cross correlating both curves. Next, the second light curve is interpolated and both curves are again cross correlated. The final ICCF is the average CCF of both instances. To determine the lag, we calculate both the lag at the peak $\tau_\mathrm{peak}$ and the centroid lag $\tau_\mathrm{cent}$ (above $0.8r_{\mathrm{peak}}$) of the final CCF. We repeat this process $N_\mathrm{sim} = 10^4$ times. We use a lag search interval of $-50$ to $50$ days, with a sampling step of 0.05 days. The final lags are determined as the median of the peak and centroid distributions, with uncertainties from the 16\% and 84\% quantiles. We analysed each year individually, as the considerable gap between each observing season ($\sim100$ days) is not approximated well by linear interpolation when considering variability on the timescale of days. All lags (peak and centroid) for Years 2--4 are presented in Table~\ref{tab:lags}. The peak correlation coefficients $r_\mathrm{peak}$ are further tabulated in Table~\ref{tab:rmax}. We also use the centroid lag, $\tau_{\mathrm{cent}}$, as the main ICCF-derived lag from here on. The lag-wavelength spectrum for Years 2--4 as calculated using ICCF is plotted in Fig.~\ref{fig:iccf_lags}. 

\begin{figure}
    \centering
    \includegraphics[width=0.48\textwidth]{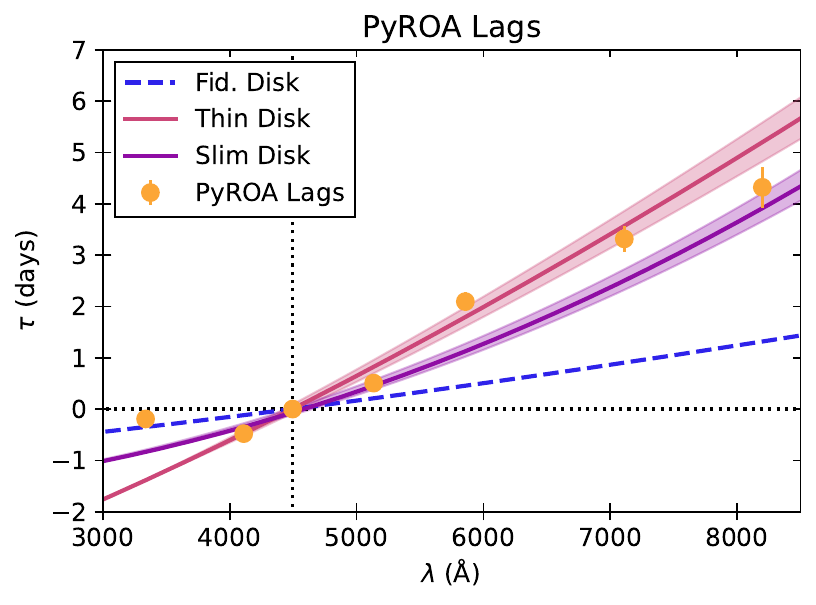}
    \caption{Lag spectrum as calculated using PyROA simultaneously for all three years in the AGN rest frame, with most of the errorbars too small to be visible. As in Fig.~\ref{fig:iccf_lags}, lags increase with wavelength. A thin disk profile with $\tau \propto \lambda^{4/3}$ and a slim disk profile with $\tau \propto\lambda^{2}$ is fitted to these data (Table~\ref{tab:fits}). These are shown with the solid lines and their error regions are shaded. As a comparison, the fiducial thin disk profile for I~Zw~1 with its mass and bolometric luminosity is illustrated with the dashed line, with $\tau_0 = 1.07\:\textnormal{days}$ as calculated in Section~\ref{disc:rm_signals_source}.}
    \label{fig:pyroa_lags}
\end{figure}

\begin{figure}
    \centering
    \includegraphics[width=0.48\textwidth]{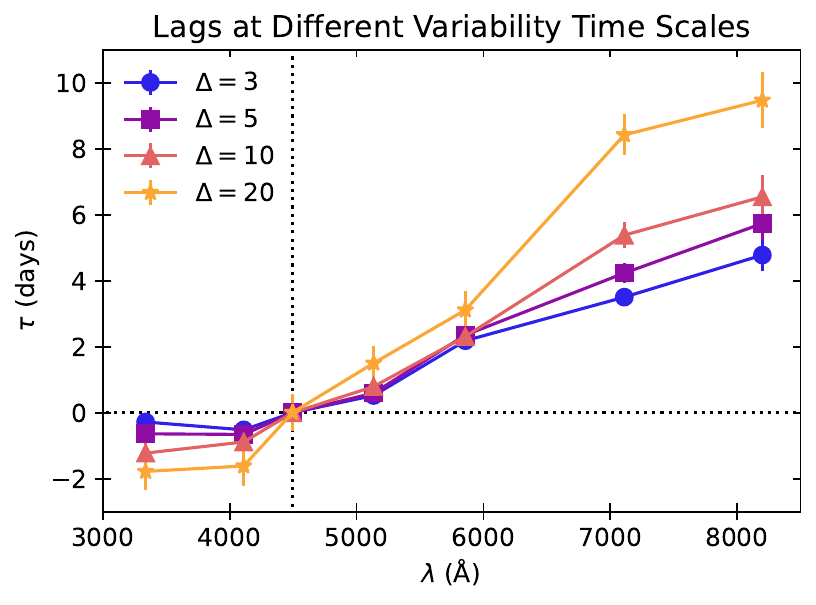}
    \caption{Lags calculated using PyROA while varying the variability stiffness parameter $\Delta$ with $\Delta = 3, 5, 10, 20\:\mathrm{days}$ in the AGN rest frame. $\Delta = 3$ is denoted by the circles, $\Delta = 5$ by the squares, $\Delta=10$ by the triangles, and $\Delta = 20$ by the stars. As $\Delta$ increases, the reference light curve stiffens and longer variability timescales are probed. The plot shows that as that variability timescale increases, the magnitudes of the lags increase.}
    \label{fig:var_ts_lags}
\end{figure}

\begin{table}
    \centering
    \begin{tabularx}{0.48\textwidth}{YYYY}
    &$\beta$&$\tau_0$ (days)&$y_0$\vspace{1ex}\\ \hline \hline \\[-1ex]
    thin disk & $4/3$ & $4.23\pm0.24$ & $1.00\pm0.02$\vspace{1ex}\\
    slim disk & 2 & $1.71\pm0.09$&$1.03\pm0.04$\vspace{1ex}\\
    free $\beta$ & $2.51\pm0.34$ & $1.19_{-0.26}^{+0.33}$ & $1.04\pm0.06$\vspace{1ex}\\
    \hline
    \end{tabularx}
    \caption{The fit results for three different disk profiles to the lag-wavelength spectrum calculated by using PyROA, based on Eq. \ref{eq:disk}. For the thin and slim disk profiles $\beta$ was fixed while the free $\beta$ fit varied $\beta$. Uncertainties are the 68\% confidence intervals.}
    \label{tab:fits}
\end{table}

\subsection{PyROA} \label{pyroa}
For a more detailed analysis of the light curves, we use PyROA \citep{donnan2021,donnan2023}. PyROA uses a running optimal average to model the light curves, and MCMC methods to estimate the uncertainties of the light curve parameters. To take advantage of all available information, all light curves are fitted simultaneously. From this, PyROA calculates a dimensionless reference light curve $X(t)$, with a mean of zero, and a variance of one. To translate the reference light curve to the observed ones, we use the basic light curve model in PyROA. In this, $X(t)$ is scaled and re-normalised to the mean flux of the observed flux, and shifted in time by the lag $\tau$ as given by a delta function response,
\begin{equation}
    F_i(t) = A_iX(t-\tau_i) + B_i
\end{equation}
after Eq. 9 in \citet{donnan2021}. $F_i(t)$ is the model flux of the light curve in band $i$, $A_i$ is the RMS flux, $B_i$ is the mean flux, and $\tau_i$ is the lag. The flexibility of the reference light curve is controlled by a Gaussian memory function with a width $\Delta$. As $\Delta$ increases, $X(t)$ stiffens and cannot respond well to rapid variability in the light curves. Accordingly, we can analyse different variability timescales by controlling $\Delta$ and the timescales over which the light curves are fit. The localised delta function response does not reflect the spatially extended response of an accretion disk. We further investigate the spatially extended response later in this section but do not consider it in our basic fit. \citet{donnan2023} showed that including a disk-like spatially extended response gave results within errors of a delta function response. In addition, \citet{thorne2025} shows that removing long-scale variations before analysis with PyROA does not change the measured lags significantly. As the fundamental model, we simultaneously fitted the data from Years 2, 3, and 4 with PyROA. The lag of $X(t)$ is set to zero in the $g$-band such that lags are measured with respect to the $g$-band. This fit optimizes $A_i$, $B_i$, and $\tau_i$ for each band and a common $\Delta$. The light curves are modelled with an evenly spaced grid of 1000 time intervals. The results from this fit and the entire set of light curves are presented in Fig.~\ref{fig:lc} and the lag-wavelength spectrum in Fig.~\ref{fig:pyroa_lags}. The lags are also included in Table~\ref{tab:lags}. This fit gives $\Delta = 2.55\pm0.07\:\mathrm{days}$ when left as a free parameter. To examine behaviour of the lag spectrum at different variability timescales, we repeated this process but setting $\Delta$ to $\Delta = 3, 5, 10, 20\:\mathrm{days}$, as this parameter controls how sensitive the fit is to certain features in the light curves. The lag-wavelength spectra from this are shown in Fig.~\ref{fig:var_ts_lags} and the lags are tabulated in Table~\ref{tab:var_lags}. As the best fit $\Delta$ of our initial model is also $\simeq 3\:\mathrm{days}$, the first of these variability timescale fits is essentially equivalent to our initial fit. 

Figs.~\ref{fig:iccf_lags} and \ref{fig:pyroa_lags} show that independent of the method for lag-calculation, there is a positive relationship between the lag and the wavelength. When calculating the PyROA lags separately for each year as shown in Fig. \ref{fig:iccf_pyroa}, both methods show the same trends (e.g. there is a $r$-band excess relative ot the general power law trend in Year 2 present in both methods). The PyROA lags are smaller than the ICCF lags likely due to the fact that PyROA tends to concentrate on the smallest reasonable variability timescale, i.e., $\Delta$ tends to the smallest reasonable value. PyROA ignores long-term trends in the light curve and the long tail of the transfer function. On the other hand, the ICCF lags are measured using the centroid, which is generally drawn from an asymmetric CCF, with a significant contribution at larger lag values. This means that the ICCF lags are skewed towards larger delays. ICCF errors are also larger than the PyROA ones. This is to be expected as the FR/RSS implemented in PyCCF overestimates the uncertainties \citep[][see also discussion in \citealp{donnan2021}]{cackett2018}. 

From here on we will concentrate our analysis and discussion on the PyROA lags as with this method we can fit a larger data set and therefore more information simultaneously. With ICCF, only individual years are accessible and the results of these show significant scatter from year to year. This indicates that these results are more susceptible to variations in quality of the light curves between and in observing season. In addition, we are interested in the disk at small sizes and lags, therefore PyROA is more likely to extract more relevant lags than ICCF. In general, the ICCF results appear to be less robust than PyROA, in the context of measuring the disk.

To characterize the lag-wavelength spectrum we fit $\tau(\lambda)$ according to the standard disk profile power law parametrised as
\begin{equation}
\tau(\lambda) = \tau_0 \left[ \left(\frac{\lambda}{\lambda_0}\right)^{\beta} - y_0 \right], \label{eq:disk}
\end{equation}
where $\lambda_0 = 4770/(1+z)\:\textnormal{\AA}$, as lags are calculated in reference to the $g$-band. The factor $y_0$ normalises the lag at $\lambda=\lambda_0$ to zero and is therefore expected to be 1. For a thin disk $\beta = 4/3$ and for a slim disk $\beta = 2$ \citep{shakura1973,wang1999}. We also perform this fit leaving $\beta$ as a free parameter. We use a MCMC routine with 16 walkers and $10^5$ steps, discarding the first 5000 steps and thinning by 50\% to reduce autocorrelation. The fitting results with 68\% confidence intervals are shown in Table~\ref{tab:fits}. The thin and slim disk fits are also overplotted in Fig.~\ref{fig:pyroa_lags}. The other disk profile has an extremely similar track in the wavelength space explored here. We have also plotted the track for the fiducial disk in I~Zw~1 with $\tau_0 = 1.07\:\textnormal{days}$, based on the mass and bolometric luminosity (see Section~\ref{disc:rm_signals_source}).

\subsubsection{The lag-frequency spectrum}\label{frl}

\begin{figure*}
    \centering
    \includegraphics[width=\textwidth]{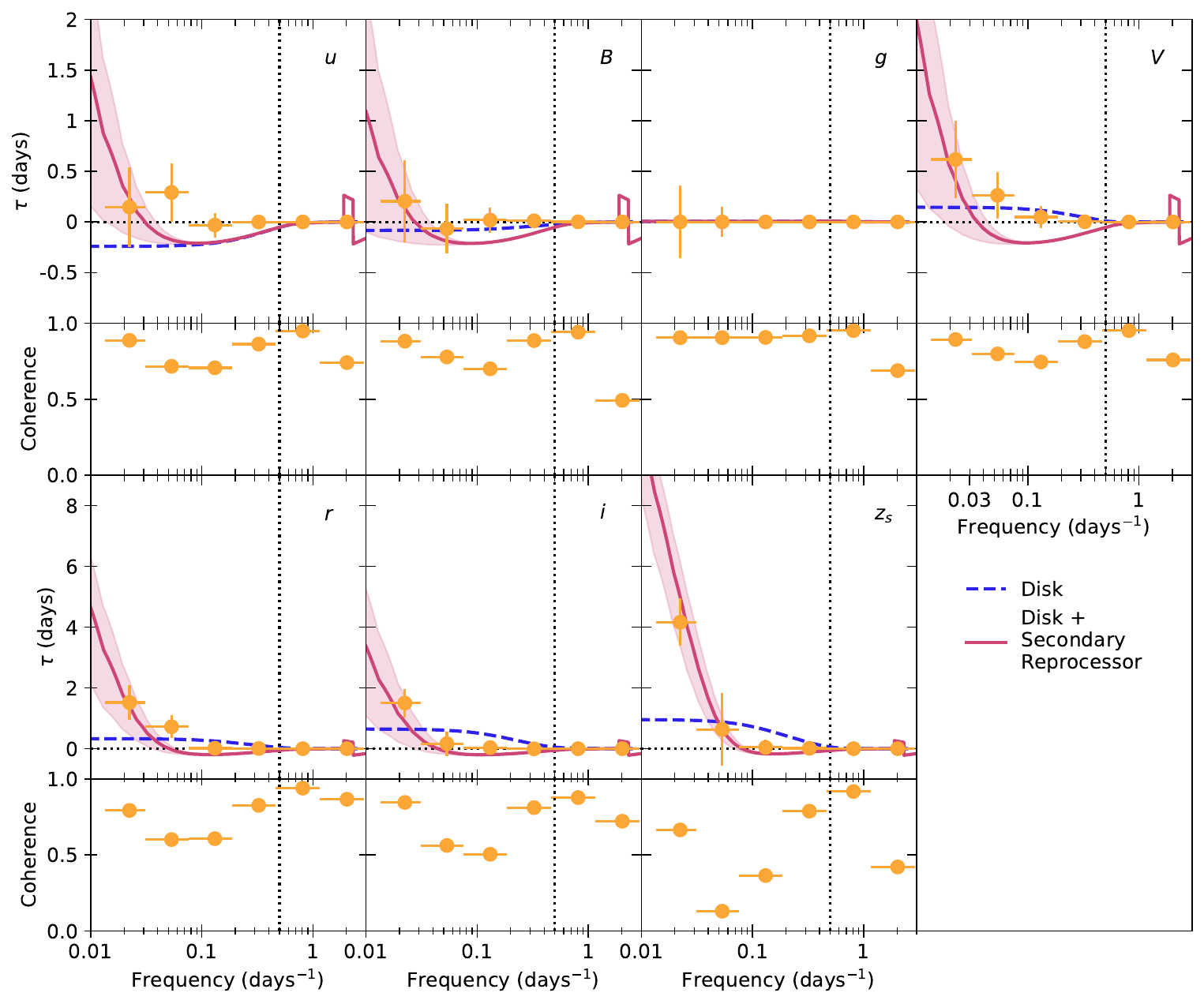}
    \caption{The lag-frequency and coherence spectra for all bands derived from Fourier analysis. The studied frequency range is $0.013-2.9\:\mathrm{days}^{-1}$ ($1.5\times10^{-7}-3\times10^{-5}\:\mathrm{Hz}$), however data at frequencies above $0.5\:\mathrm{days}^{-1}$ (vertical dotted lines) is uninformative as the ROA washes out the variations (and correlates adjacents points, also increasing its coherence). The data points are plotted with circles, the simple thin disk is denoted with the dashed line, and the disk+secondary reprocessor model is denoted by the solid line. The model's uncertainty envelope is shown by the shaded region. The median delay of the secondary reprocessor in this model is $\tau_M = 20\:\mathrm{days}$.}
    \label{fig:frl}
\end{figure*}

\begin{figure}
    \centering
    \includegraphics[width=0.48\textwidth]{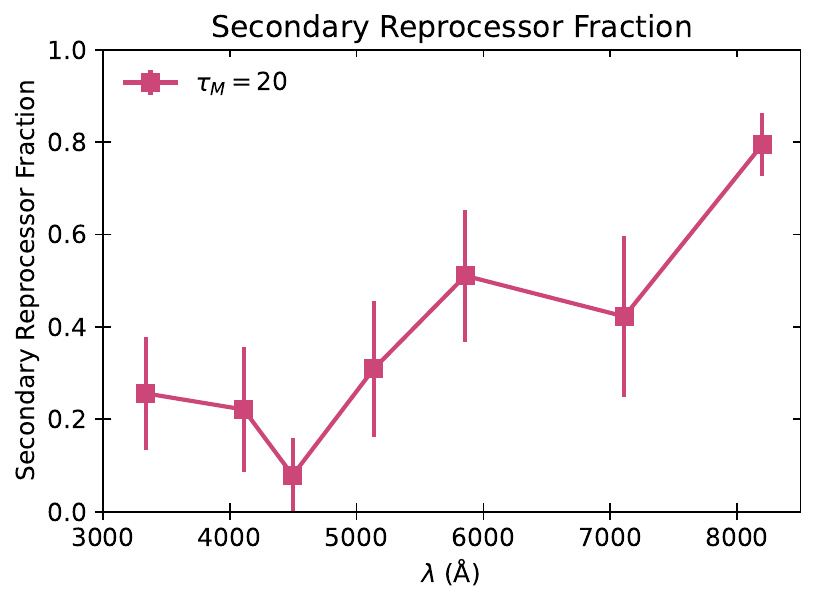}
    \caption{The fractional contribution of the secondary reprocessor ($f(\lambda)$ in Eq. \ref{eq:tot_tf}) as a function of wavelength as fitted with the disk+secondary reprocessor model in Fig.~\ref{fig:frl}, with the median delay of the secondary reprocessor at $\tau_M = 20\:\mathrm{days}$.}
    \label{fig:sp_frac}
\end{figure}

\begin{figure}
    \centering
    \includegraphics[width=0.48\textwidth]{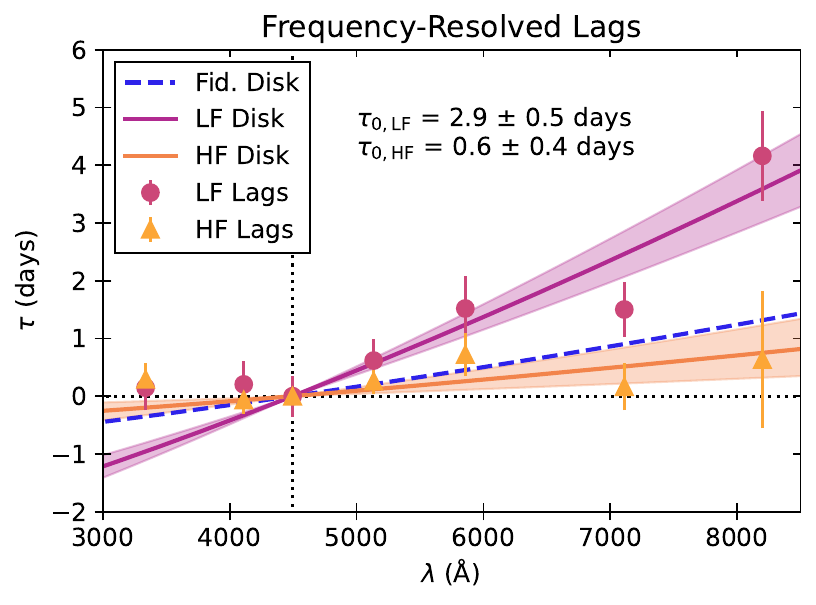}
    \caption{The high frequency (HF, $0.031-0.076\:\mathrm{days}^{-1}$) and low frequency (LF, $0.013-0.031\:\mathrm{days}^{-1}$) lag-wavelength spectra in the AGN rest frame. The high frequency lags are plotted with the triangles and the orange line and shaded region denote the thin disk fit and associated uncertainty region with $\tau_{0,\mathrm{HF}}=0.61\pm0.37\:\mathrm{days}$. The low frequency lags are represented by the circles and the magenta line and shaded region denote the thin disk fit and associated uncertainty region with $\tau_{0,\mathrm{LF}}=2.92\pm0.47\:\mathrm{days}$. The fiducial thin disk profile discussed in Section~\ref{disc:rm_signals_source} is plotted with the dashed line.}
    \label{fig:frl_fit}
\end{figure}

A powerful tool for the analysis of times series are Fourier techniques. Fig.~\ref{fig:var_ts_lags} shows that the lags in I~Zw~1 change with the variability timescales, with longer lags at longer variability timescales. To further probe this behaviour, we look at the frequency-resolved lags, similar to \citet{cackett2022,lewin2023,lewin2024}. We create evenly sampled light curves by using PyROA to model the light curves independently, with $\Delta$ set to 2~days to retain as much short term variability as possible, without overfitting the white noise level. We then calculate the cross-spectrum from the Fourier transforms of the reference light curve and light curve of interest. The lag is evaluated from the phase of the cross-spectrum, per frequency bin \citep[for further details see][]{uttley2014}. In addition, we calculate the coherence, which indicates the fraction of variability in the reference and the light curve of interest that can be described by a linear transformation.

We use three segments in our analysis, each segment covering one year of data. These segments have a length of $155\:\mathrm{days}$ each. We use seven frequency bins, covering the frequency range between $0.006-2.9\:\mathrm{days}^{-1}$ ($7\times10^{-8}-3\times10^{-5}\:\mathrm{Hz}$). The lag-frequency and coherence spectra for all bands are plotted in Fig.~\ref{fig:frl}. In this plot, we have already removed the lowest frequency bin, $\sim0.01\:\mathrm{days}^{-1}$, as it is sensitive to small variations in the binning parameters. This is likely due to bias effects that have been shown to exist for the lowest frequency bin \citep[e.g.,][]{cackett2022}. The frequency-resolved lags in the $g$-band are flat at zero because it is used as the reference band. Further, all frequencies above $0.5\:\mathrm{days}^{-1}$ are uninformative since the highest frequency we modelled our light curves with was on a variability timescale of $\Delta = 2\:\mathrm{days}$, which we denote with the vertical dotted lines in Fig.~\ref{fig:frl}. The PyROA modelling washes out the variability at higher frequencies and correlates adjacent points. This effect is illustrated in the coherence spectra: above $0.5\:\mathrm{days}^{-1}$, the coherence is $\sim1$ and then drops towards lower frequencies. This frequency region is where real signal is probed in the light curves and it rises again towards the lowest frequencies, as expected. There is a clear presence of a variability-dependant lag behaviour, with lags significantly increasing in the lowest frequency bin, especially at the longest wavelengths. To analyse this behaviour, we evaluate the predicted frequency-resolved lags for a simple thin disk and fit a model of a thin disk with an additional secondary reprocessor.

To produce the frequency-resolved lag expectation of this model, we require the corresponding transfer functions for the accretion disk and the secondary reprocessor models for each band observed. For the accretion disk, we used the thin disk transfer function, $\psi_{\rm{disk}}$ \citep[see details in][]{Collier1999,cackett2007,Starkey2016}, which is a function of black hole mass, mass accretion rate and inclination angle. We parametrise the transfer function of the secondary reprocessor $\psi_{\mathrm{SR}}$ as a log-normal distribution as implemented in \citet{cackett2022} and \citet{lewin2023,lewin2024}:
\begin{equation}
    \psi_{\rm SR}(t) = \frac{1}{S \sqrt{2\pi}t} \exp\left[-\frac{\ln{(t/\tau_M)^2}}{2S^2}\right]\,,\label{eq:blr_tf}
\end{equation}
where $\tau_M$ is the median delay of the reprocessor; the distance at which the median reprocessor response occurs. The standard deviation $S$ is that of the $\ln{t}$ function, and does not indicate the physical size of the reprocessor. As $S$ increases at a fixed median delay, the response at short delays increases. Therefore, $S$ is better understood as a measure of the skewness of the transfer function towards short delays; that is how well $\tau_M$ characterise the peak of the response. For small $S$, the median delay is approximately at the peak of the response function and the response clusters around the median delay. For large $S$, the peak rapidly shifts towards short delays, the function spreads out, and the response is increasingly dominated by short delays. This also implies that for larger $S$, the median delay needs to be increased greatly to account for the increase of response at short delays. The total transfer mixture model, $\psi_{\rm{tot}}$, of both reprocessors is given by:
\begin{equation}
    \psi_{\rm{tot}}(t) = (1-f(\lambda))\psi_{\rm{disk}}(t) + f(\lambda)\psi_{\rm SR}(t)\,,\label{eq:tot_tf}
\end{equation}
where $f(\lambda)$ is the fractional contribution of the secondary reprocessor. The responding light curve and the is a convolution of the driving light curve and the transfer function. The cross-spectrum of the responding and driving light curves can then be expressed in terms of the Fourier transforms of the driving light curve and the transfer function. Finally, the frequency-resolved lags are evaluated from the phase of the cross-spectrum \citep[for a detailed explanation see][]{cackett2022}.

The thin disk is the predicted disk based on the Eddington ratio of I~Zw~1 of 1.95, with an approximate face-on inclination, as outlined in Section~\ref{disc:rm_signals_source} and plotted with the dashed blue line in Fig.~\ref{fig:frl}. This disk largely reproduces lags in the second lowest frequency bin ($0.031-0.076\:\mathrm{days}^{-1}$) but fails to predict the lag behaviour at even lower frequencies, which consistently lie above it. To model the lags at the lowest frequencies, we fit the thin disk and an additional BLR-like secondary reprocessor according to Eq. \ref{eq:blr_tf} and \ref{eq:tot_tf}, optimising $f(\lambda)$. We also derived uncertainties envelopes of this model. The median delay of the secondary reprocessor is varied between $\tau_M = 10-50\:\mathrm{days}$, which makes little difference to the goodness-of-fit, so we choose to show the fits for a size of $\tau_M = 20$ days. At shorter and longer distances the goodness of fit drops significantly. The standard deviation $S$ is fixed at 1, as results from previous studies of other AGN show $S=1-2$ with the majority at $\sim1$ \citep{cackett2022,lewin2023,lewin2024}. Notably, setting the standard deviation to 2 also reproduces the data well, with an increase in $f(\lambda)$. Since the response amplitude at longer lags has decreased relative to shorter lags, the fractional contribution needs to increase to compensate for this. The disk+secondary reprocessor model is denoted with the solid magenta line in Fig.~\ref{fig:frl} and clearly accounts for the higher lags at the lowest frequencies. This fit also estimates the fractional contribution $f(\lambda)$ from the secondary reprocessor, which is plotted in Fig.~\ref{fig:sp_frac} as function of wavelength (for $\tau_M = 20\:\mathrm{days}$). For different values of $\tau_M$, the shape of the contribution fraction spectrum is similar, but the magnitude of the contribution varies. 
At larger median delays, the secondary reprocessor fraction increases (within errors); this has little physical meaning and 
is probably because we do not sample low enough frequencies to constrain these models. The fits in Fig.~\ref{fig:frl} indicate that the reverberation signals in the second lowest frequency bin (`high frequency', $0.031-0.076\:\mathrm{days}^{-1}$) are dominated by the disk and in the lowest frequency bin (`low frequency', $0.013-0.031\:\mathrm{days}^{-1}$) by the secondary reprocessor. We extract the lag-wavelength spectra of these two bins and fit them with a thin disk according to the method described in Section~\ref{pyroa}, with $y_0$ fixed to 1 to constrain the fits. For the high frequency fit, $\tau_{0,\mathrm{HF}} = 0.61\pm 0.37\:\mathrm{days}$ and for the low frequency fit $\tau_{0,\mathrm{LF}}=2.92\pm0.47\:\mathrm{days}$. The lag-wavelength spectra with their fits and the fiducial thin disk profile according to Section~\ref{disc:rm_signals_source} are plotted in Fig.~\ref{fig:frl_fit}.

\section{SED Analysis} \label{sed}

\begin{figure}
    \centering
    \includegraphics[width=0.48\textwidth]{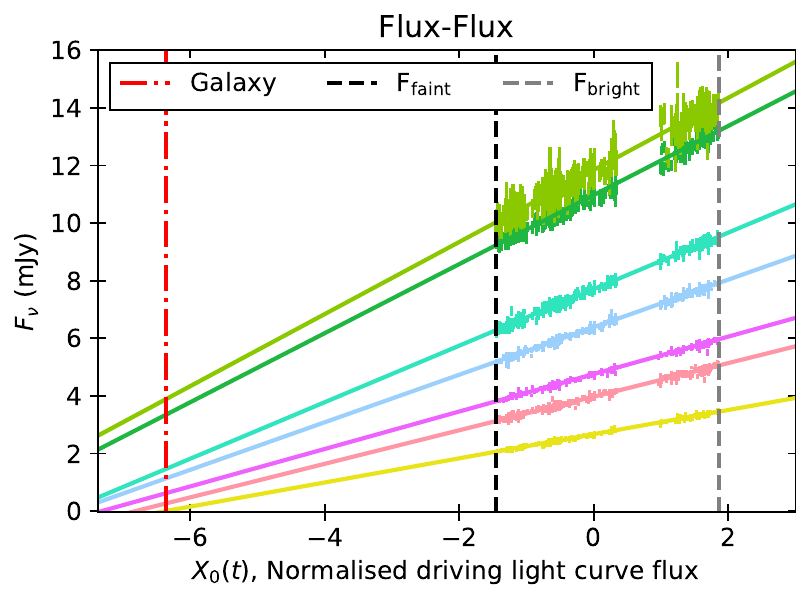}
    \caption{The flux-flux ($F_\nu - X(t)$) plot, used to determine the host galaxy contribution. The $X(t)$ at which the first filter intersects $F_\nu=0$ (here $u$) is taken as $X_\mathrm{gal}$ (red dash-dotted line). The flux in all other bands at $X_\mathrm{gal}$ is the host galaxy contribution in those bands. The colours correspond to the individual filter light curves in Fig.~\ref{fig:lc}, from bottom to top $uBgVriz_s$.}
    \label{fig:fluxflux}
\end{figure}

To further explore the structure of the disk, we analyse the optical SED. This covers the long-wavelength accretion disk emission, predicted to be $F_\nu \propto \nu^{1/3}$ for a thin disk, and $F_\nu \propto \nu^{-1}$ for a slim disk \citep{wang1999b}. We use a flux-flux analysis to isolate the constant host galaxy contribution from the variable AGN flux \citep{Winkler1992,Winkler1997}. Accordingly, we examine the $F_\nu - X(t)$ (flux – reference light curve) space, $X(t)$ taken from PyROA \citep{donnan2023}. The light curves are corrected for Galactic extinction using $E(B-V)=0.057$ and then de-redshifted to the AGN rest frame \citep{fitzpatrick1999,schlafly2011}. 

The data in each filter are well-described by a straight line, indicating that the AGN SED does not change shape with changes in flux within our observation window (Fig.~\ref{fig:fluxflux}). We fit a line to the data in each band, and extrapolate these to $F_\nu = 0$, essentially `winding down' the AGN. The $X(t)$ at which the first filter intersects at $1\sigma$ above $F_\nu = 0$ is $X_{\mathrm{gal}}$. Subsequently, we evaluate the flux of all other bands at $X_\mathrm{gal}$. This is the lower limit host galaxy contribution. The band which reaches $F_\nu=0$ first is the $u$-band, the shortest wavelength we have available. Contrary to our assumptions, we do still expect significant host galaxy contribution in the $u$-band due to circumnuclear star formation and are therefore underestimating the host galaxy flux \citep{fei2023}. The SED fluxes can be found in Table~\ref{tab:sed} in the Appendix.

The resulting SED (not pictured) is flat, even slightly decreasing towards shorter wavelengths. This was similarly noticed by \citet{juranova2024} and is a sign of intrinsic absorption. 
\citet{juranova2024} used a custom extinction law fit based on the SMC bar average model by \citet{gordon2003}. Their extinction law is flattened below $1550\:\textnormal{\AA}$. This curve, lacking the Galactic $2175\:\textnormal{\AA}$ bump and flattened in the UV, matches well with average extinction curve found by \citet{gaskell2004} for quasars. 

\begin{figure*}
    \centering
    \includegraphics[width=0.9\textwidth]{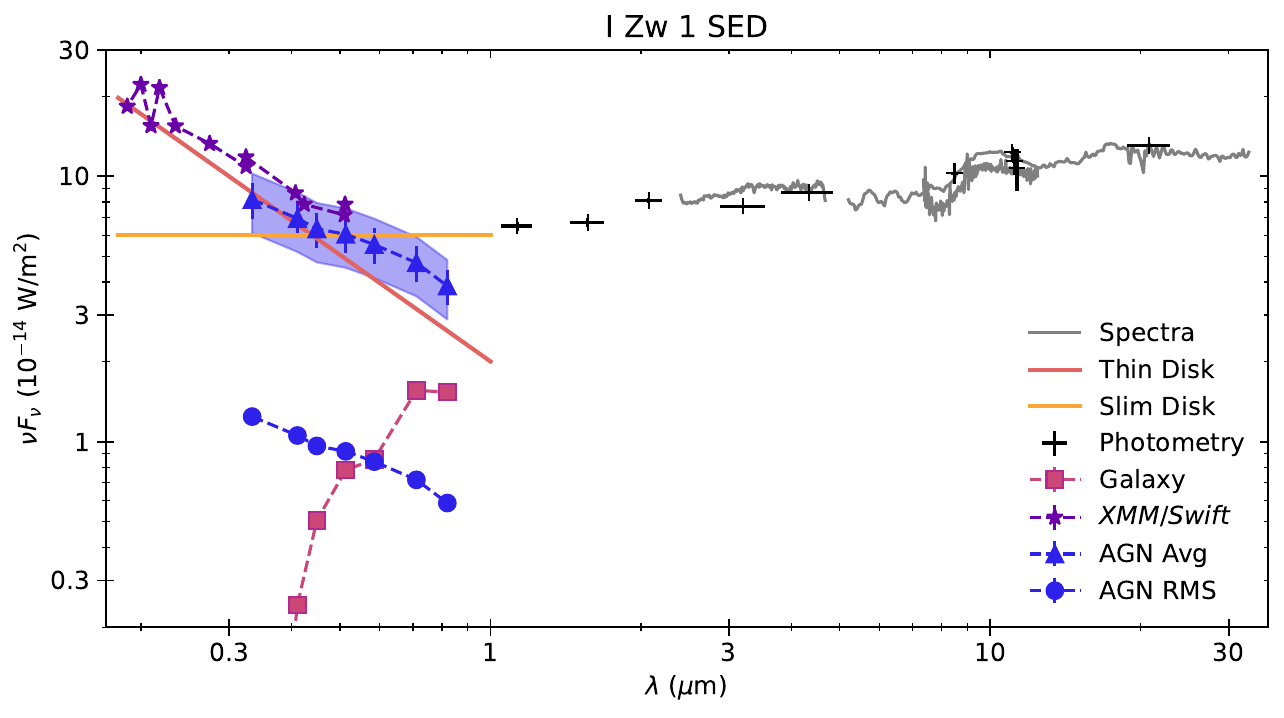}
    \caption{The near-UV to mid-infrared SED of I~Zw~1, in the rest frame of the AGN, and corrected for Galactic (IR) and intrinsic extinction (UV/optical). The LCO AGN flux during Years 2, 3, and 4 spans the shaded region with the average flux denoted by the triangles with errorbars given by the RMS flux. The RMS of the AGN is denoted by the circles, and has the same slope as the average AGN. The host galaxy flux determined using the flux-flux technique is shown by the squares, and is only corrected for Galactic extinction. The stars show the \textit{XMM-Newton} OM and \textit{Swift} UVOT measurements and the IR photometry points and spectra are taken from \citet{drewes2025}. The SED slopes of the thin and slim disks are also overplotted using thick lines.}
    \label{fig:sed_all}
\end{figure*}
\begin{figure}
    \centering
    \includegraphics[width=0.48\textwidth]{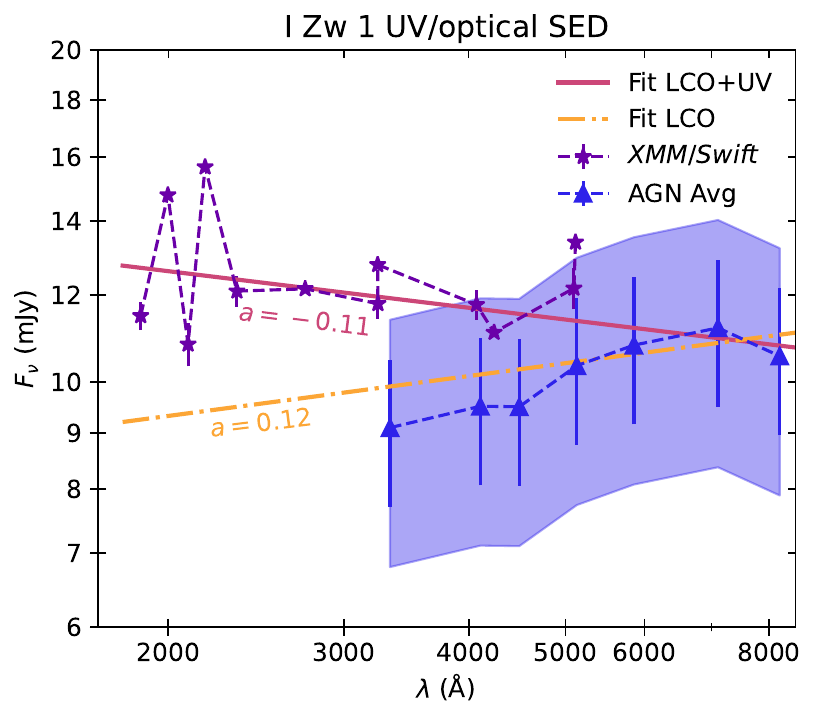}
    \caption{The UV/optical SED of I~Zw~1 including the average LCO AGN flux (triangles), with errorbars representing the RMS flux, and the \textit{XMM-Newton} and \textit{Swift} flux (stars). The fits for the slope are performed according to $F_\nu \propto \lambda^{a}$, where a thin disk has $a = -1/3$ and a slim disk $a = 1$. The fit to only the LCO average AGN data is denoted by the dash-dotted line with $a = 0.12\pm0.13$ and the fit to all of the data is the thick line with $a = -0.11\pm0.12$.}
    \label{fig:sed_fits}
\end{figure}

We correct our LCO data for intrinsic extinction according to \citet{juranova2024}. The resulting mean and RMS AGN fluxes and host galaxy flux are plotted in Fig.~\ref{fig:sed_all}. We use the same model to correct the \textit{XMM-Newton} OM and \textit{Swift} UVOT data described in Section~\ref{sec:xmm} and \ref{sec:swift} for intrinsic extinction. We are focusing on the average flux, so we average the data between the \textit{XMM-Newton} and \textit{Swift} in the corresponding filters, with uncertainties taken as the standard deviation, and we use these data in fitting. The UV data plotted in Figs.~\ref{fig:sed_all} and \ref{fig:sed_fits} are the weighted mean of the three \textit{XMM-Newton} epochs as well as the \textit{Swift} data separately. Further, we subtracted host galaxy contribution from the bands for which have this information from the LCO flux-flux analysis ($UBV$). To compare the UV/optical slope to the thin and slim disks we overlaid their profiles (with arbitrary normalisation). To provide a more expansive picture of the AGN emission, we added the IR spectrum from \citet{drewes2025}. This was only corrected for Galactic extinction using the ISM profile from \citet{chiar2006} and the Galactic $A_K / E(B-V) = 0.36$ ratio from \citet{fitzpatrick1999}. 

The complete SED is plotted in Fig.~\ref{fig:sed_all}. Both the average AGN and RMS AGN flux follow the same slope. In comparison with the example slopes of a thin and a slim disk, the slope of the AGN emission is not easily classifiable. Accordingly, we fit the optical and UV slope. Similarly to the lags, we use a MCMC routine with 16 walkers and $10^5$ steps, discarding the first 5000 steps and thinning by 50\% to avoid autocorrelation. These fits are parametrised as $F_\nu \propto \lambda^a $ where a thin disk has $a = -1/3$ and a slim disk $a = 1$. First, we only fit the average AGN LCO data which gives a slope of $a = 0.12 \pm 0.13$. Including the average \textit{XMM-Newton} and \textit{Swift} measurements steepens the slope to $a = -0.11\pm0.12$. The relevant data and fits are plotted in Fig.~\ref{fig:sed_fits}.

\section{Discussion}

In this section we discuss our results in the context of our aims to phenomenologically describe the accretion disk in I~Zw~1, examine the results in the context of other super- and sub-Eddington AGN, and analyse the multi-wavelength internal structure of I~Zw~1. Accordingly, we first examine the actual sources of the reverberation signals we measure. In the context of those results we then analyse the underlying accretion disk structure based on the lag-wavelength spectra and the UV/optical SED profile. Finally, we collate the optical to mid-IR size-wavelength relation of I~Zw~1 and discuss the resulting implications on BLR formation models. 

\subsection{The Source(s) of Reverberation Signals} \label{disc:rm_signals_source}

\subsubsection{The Accretion Disk}
The lag-wavelength spectrum as shown in Figs.~\ref{fig:iccf_lags} and \ref{fig:pyroa_lags} clearly displays larger lags with increasing wavelengths, implying an emission region with a radially stratified temperature profile. Moreover, it necessitates the existence of non-local large-scale communication in the AGN. Independent of the source of variability, it must be in some way communicated over large distances, e.g. in a reprocessing scenario or large-scale fluctuations throughout the entire disk \citep{cai2018,hagen2024}. If we assume the classical lamppost-disk reprocessing model we can estimate the size of the disk from the AGN properties, further assuming an underlying thin disk. The following assumptions and relations are probably not valid for super-Eddington objects, particularly those concerning bolometric luminosities. However, we include these calculations for completeness and to compare with other studies on super-Eddington objects that also use standard bolometric conversions \citep[e.g.,][]{cackett2020,thorne2025}. We use Eq. 12 from \citet{fausnaugh2016} to evaluate a fiducial value for $\tau_0$. We derive the bolometric luminosity $L_\mathrm{bol}$ from the $5100\: \textnormal{\AA}$ luminosity $\lambda L_{5100\:\textnormal{\AA}}$ according to \citet{trakhtenbrot2017} \citep[for a further discussion see Section 4.1 of][]{drewes2025}. The intrinsic luminosity of I~Zw~1, $\lambda L_{5100\:\textnormal{\AA}} = 10^{44.50}\:\mathrm{erg\,s}^{-1}$, gives a bolometric luminosity of $L_\mathrm{bol} = 10^{45.36}\:\mathrm{erg\,s}^{-1}$ \citep{trakhtenbrot2017,huang2019}. As bolometric luminosity corrections often have large scatter and uncertainty partly due to unknown underlying SED shapes, we include an uncertainty of 20\% on the bolometric luminosity \citep[e.g.,][]{gordon2006}. We derive an Eddington ratio of $\dot{m}_E = 1.95 \pm 0.48$, propagating errors on the bolometric luminosity and measured black hole mass. Next, we calculate the accretion efficiency $\eta$ from $\dot{m}_E = \eta \dot{\mathscr{M}}$ where $\dot{\mathscr{M}}$ is the dimensionless accretion rate according to Eq. 2 from \citet{du2015}. With $\dot{\mathscr{M}} \simeq 130$ we calculate $\eta = 0.015 \pm 0.004$, for a face-on disk. This is similar to the accretion efficiency calculated for another super-Eddington AGN, Mrk~142 and what is generally expected in slim disk flow \citep{cackett2020}. As is common practice, we take the ratio of external to internal heating $\kappa$ to be 1. The correction factor $X$ for the conversion between wavelength and temperature for a certain radius is often taken as $X=2.49$, which describes the flux-weighted mean radius \citep{fausnaugh2016}. Using this value, and considering all relevant uncertainties, we calculate a fiducial $\tau_0 = 1.07 \pm 0.15\:\mathrm{days}$ as the predicted disk size at $4495\:\textnormal{\AA}$. Alternatively, considering the emission-weighted mean radius gives $X=4.97$ and accordingly $\tau_0 = 2.70\pm0.38\:\mathrm{days}$.

Our fitted $\tau_0$ for a thin disk profile is $4.23\pm0.24\:\mathrm{days}$, larger by a factor of $4.0\pm0.6$ to $1.6\pm0.2$, depending on the value of $X$ used. Using our fitted $\tau_0$, we get a large dimensionless mass accretion rate of $\dot{\mathscr{M}} = 9970$. Assuming $\eta = 0.015$, this implies $\dot{m}_E = 150$ and hyper super-Eddington accretion. A similar offset was also found in the other super-Eddington AGN, Mrk~142, PG~1119+120, and 3C 273 \citep{cackett2020,donnan2023,thorne2025}. We must note, however that the expression for $\tau_0$ is only valid for sub-Eddington objects. Nevertheless, even in sub-Eddington objects studies consistently find the disk measured through RM to be larger by a factor of $2-3$, similar to our case \citep[e.g.,][]{edelson2015,fausnaugh2016,cackett2018,hernandezsantisteban2020,miller2023}. From this we instead infer that the source(s) of the RM signals we measure are rather similar in super- and sub-Eddington AGN.

It has been shown that an inclined disk can reproduce some of the results of continuum reverberation mapping; namely overall longer lags and skewed lag distributions. The lag distribution of an inclined disk peaks at smaller lags that correspond to fast variability and has a long tail at larger lags which responds to slower variability \citep{Starkey2016}. This emulates the skewed CCF distributions that are commonly found and have a long tail at large lags, such that $|\tau_\mathrm{peak}| < |\tau_\mathrm{cent}|$ \citep[e.g.,][]{edelson2017,edelson2019,hernandezsantisteban2020}. Table~\ref{tab:lags} shows that this behaviour is also present in I~Zw~1. To further investigate the possibility of an inclined disk, we compare the frequency-resolved lags with the prediction for a thin disk at an inclination of 60° and with $\dot{m}_E = 50$, plotted in Fig.~\ref{fig:frl_mdot50}. This disk does reproduce the majority of the data well. However, the lowest frequencies in the $z$ and especially $u$-band are not modelled well by this inclined disk. In addition, the skewed lag distributions can result from other factors, most relevantly a secondary reprocessor located at larger distances.

\subsubsection{A Continuum Secondary Reprocessor}

In fact, there is ample evidence that there is more than one source of continuum reverberation signals in I~Zw~1. The second source besides the fiducial accretion disk is a secondary reprocessor emitting continuum flux. This secondary reprocessor is located at larger radii than the disk and responds to the driving light curve with larger lags, increasing the overall observed lags. The continuum emitted is usually assumed to be diffuse continuum, from hydrogen bound-free recombination and free-free emission processes \citep{korista2001,korista2019,Netzer2022}. Accordingly, this diffuse continuum is particularly strong around the Balmer ($3650\:\textnormal{\AA}$) and Paschen ($8210\:\textnormal{\AA}$) jumps, which map to our $u$- and $i$-band filters. The continuum strength in these features then maps to the lag-wavelength spectrum and we expect to see an increase in the $u$- and possibly $i$-band lags, the $u/i$-band excess \citep[e.g.,][]{cackett2018}. In fact, the lag-wavelength spectrum in Fig.~\ref{fig:pyroa_lags} does appear to show a $u$-band excess. The $u$-band lag is noticeably larger than predicted by both fits. 

Further, the secondary reprocessor is expected to increase lags at all wavelengths. The addition of a secondary reprocessor at larger delays is predicted to skew the lag distributions and add a long tail at larger lags, shifting the centroid to larger lags such that $|\tau_\mathrm{peak}| < |\tau_\mathrm{cent}|$ \citep{lawther2018}. As aforementioned, Table~\ref{tab:lags} shows that in I~Zw~1 the peak lag is consistently smaller than the centroid lag. These longer lags should also be particularly noticeable at longer variability timescales. As the secondary reprocessor sits at larger radii it also covers a larger area than the disk and therefore responds to lags over a longer timescale. Accordingly, variability on longer timescales is expected to be more connected to the secondary reprocessor. In Fig.~\ref{fig:var_ts_lags}, we show the lag-wavelength spectrum as a function of the variability timescale, with an increasing range $\Delta$ of 3, 5, 10, and 20 days. As the variability timescale increases, the lags steadily increase. At a variability timescale of 20 days, at longer wavelengths, the lags make a marked jump, increasing to $\sim 10$ days and doubling compared to the short timescale values. From this we infer that as we increase our studied variability timescale, we are increasingly looking at material reverberating at larger distances, the secondary reprocessor. This effect also appears in the frequency-resolved lags in Fig.~\ref{fig:frl}: at a frequency of $0.03\:\mathrm{days}^{-1}$ (30 days) lags make a jump to larger values, especially at long wavelengths and in the $u$-band. A thin disk with an Eddington ratio of 1.95 derived from the bolometric luminosity consistently under-predicts these lowest frequency lags. Notably, even the highly accreting and inclined disk cannot reproduce the positive low frequency $u$-band lag (Fig.~\ref{fig:frl_mdot50}). It appears that at low frequencies, lags systematically come from a secondary reprocessor at larger radii. Modelling these lags, with a disk and a secondary reprocessor, gives the median delay of the secondary reprocessor at $10-50$ days. The resulting fractional contribution spectrum of this secondary reprocessor is shown in Fig.~\ref{fig:sp_frac}. Here, the contribution fraction rises towards the $u$-band, which suggests the imprint of the Balmer jump of the diffuse continuum emission on the lags at low frequencies, and is present independent of the median delay of the secondary reprocessor. The low frequency ($0.013-0.031\:\mathrm{days}^{-1}$) lag-wavelength spectrumm, which is likely associated with the secondary reprocessor and shown in Fig.~\ref{fig:frl_fit}, when fitted with a thin disk profile gives $\tau_{0,\mathrm{LF}}=2.92\pm0.47\:\mathrm{days}$. This is larger than the fiducial disk size of 1.07 days and is unlikely to arise primarily from the disk. Rather, we conclude that a secondary reprocessor is needed to reproduce these low frequency lags, which are on the order of the overall best-fit lags, indicating that there is significant contribution from a secondary reprocessor in the overall best-fit lags.

There are two general theories for the nature of this secondary reprocessor: the classical, virially bound BLR or a wind \citep{korista2001,korista2019,lawther2018,hagen2024}. Of course, depending on interpretation the difference between these might solely be semantic as the BLR itself has been shown to be mainly in Keplerian motion but includes inflows and/or outflows \citep[e.g.,][]{gravity2018,bentz2021}. Here we differentiate between the BLR winds detected in absorption ($v_\mathrm{out} = 1950\:\mathrm{km\,s}^{-1}$) and the BLR as described by the reverberation mapping of the H$\beta$ emission line \citep{juranova2024,huang2019}. Assuming that the minimum launching radius of the wind is set by the escape velocity, approximated by the virial velocity, the wind is launched from at least 10--25\:ld (light days, $0.01-0.02\:\mathrm{pc}$). In contrast, the size of the BLR estimated using H$\beta$ is $37.2_{-4.9}^{+4.5}\:\mathrm{ld}\sim 0.031\:\mathrm{pc}$ (I~Zw~1 lies on the $R_\mathrm{BLR}\propto L^{1/2}$ relation). The modelled size of the secondary reprocessor based on the frequency-resolved lags is $10-50\:\mathrm{ld}$ ($0.008-0.04\:\mathrm{pc}$). The low frequency lags and the secondary reprocessor model from our Fourier analysis are consistent with the region inhabited by the BLR and associated winds, likely indicating that indeed the BLR is responsible. However, we cannot distinguish the location of the secondary reprocessing material between the BLR winds and the BLR radius traced by the H$\beta$ line. It is probable that this material is not actually located at one specific distance but rather spread throughout the entire $0.01-0.04\:\mathrm{pc}$ region.

A similar picture of contribution from a secondary processor can be seen in other super-Eddington as wells as sub-Eddington AGN. Super-Eddington AGN Mrk~142 and PG~1119+120 have $u/U$-band excesses, and it is a well-documented feature in sub-Eddington AGN \citep{cackett2020,donnan2023,fausnaugh2016,hernandezsantisteban2020}. In fact, in NGC 4593 the Balmer jump is clearly resolved in the lag-wavelength spectrum \citep{cackett2018}. Further, PG~1119+120 shows much larger lags, especially in the $u$-band, on a variability timescale of 100 days \citep{donnan2023}. Again, this is also seen in a number of sub-Eddington AGN \citep{pahari2020,vincentelli2021}. The frequency-resolved lags of (sub-Eddington) Mrk~335 and Mrk~817 show a marked increase in lags at low frequencies, below $0.01\:\mathrm{days}^{-1}$ and $0.05\:\mathrm{days}^{-1}$ respectively \citep{lewin2023,lewin2024}. The secondary reprocessors modelled for these objects have sizes that concur with BLR sizes derived from reverberation mapping of H$\beta$ in both cases. 

There is consistent evidence that a secondary reprocessing diffuse continuum is mixed into general continuum lags and lengthens these in a multitude of objects. Notably, this secondary reprocessor significantly contributes to the continuum lags across the entire optical spectrum, not just where it is especially strong in the $u/U$-band. This can be seen when studies remove the $u/U$-band lag to remove the impact of the diffuse continuum, but fits still produce disk sizes several times too large \citep{fausnaugh2018,cackett2020}. On the other hand, using solely the high frequency lags which are expected to be significantly less contaminated by the secondary reprocessor results in disk sizes which are in line with expectations \citep{lewin2023,lewin2024}. Substantial contribution across the optical spectrum is also evidenced by our results: low frequency lags which require a secondary reprocessor to replicate are on the order of the overall best-fit PyROA lags. Further, the modelled fractional contribution of the secondary reprocessor in Fig.~\ref{fig:sp_frac} is significant and above $\sim20\%$ in the majority of bands. In addition, the high frequency lag-wavelength spectrum in Fig.~\ref{fig:frl_fit} is fit with a notably smaller size, more similar to the fiducial thin disk. It is evident that the presence of this secondary reprocessor is a consequence of the basic Seyfert AGN structure. In general, there appears to be a secondary reprocessor at larger radii co-spatial with the BLR emitting diffuse continuum that considerably lengthens observed continuum lags at all wavelengths, independent of Eddington ratio. 

\subsubsection{\ion{Fe}{ii}}

I~Zw~1 has strong \ion{Fe}{ii} emission, which can be seen in the average spectrum in Fig.~\ref{fig:spec} \citep[e.g.,][]{boroson1992,veron-cetty2004}. For comparison, we also plot the \ion{Fe}{ii} template constructed by \citet{veron-cetty2004} based on I~Zw~1, arbitrarily scaled. Finally, we overlay the LCO filter transmission curves. With this combination, it is apparent that the \ion{Fe}{ii} emission is particularly strong in the $BgV$-bands. \ion{Fe}{ii} emission arises from the BLR and reverberates in response to the continuum \citep{gaskell2022}. Studies have shown the \ion{Fe}{ii} lag to be around twice the size of the H$\beta$ lag, placing it towards the outer BLR \citep[e.g.,][]{vestergaard2005,barth2013,hu2015,zhang2019}. However, the variability amplitude of \ion{Fe}{ii} is dampened relative to H$\beta$, mainly because it sits at larger distances and reverberates over a larger area. As such, \ion{Fe}{ii} can introduce additional reverberation signals in I~Zw~1 at lags approximately twice the size of H$\beta$, $\sim80\:\mathrm{days}$, in primarily the $BgV$-bands. This then indicates that our choice of the $g$-band as an `uncontaminated’ reference band may be problematic. Using an UV reference band might then provide a more `uncontaminated’ option, which we unfortunately do not have. A UV reference light curve might also directly uncover the signature of \ion{Fe}{ii} reverberation in the lag spectrum through increased lags in the $BgV$-bands, similar to the $u$-band excess as a sign of the Balmer jump from the diffuse continuum. However, I~Zw~1 also shows significant \ion{Fe}{ii} and \ion{Fe}{iii} emission in the UV -- whether this further affects reverberation signals is unknown \citep{vestergaard2001}. Further, detecting evidence of \ion{Fe}{ii} reverberation in frequency-resolved lags like we have shown with the diffuse continuum is probably not feasible with our data. This is because if the fiducial \ion{Fe}{ii} is located at a delay of $\sim80\:\mathrm{days}$, our light curve segment size is less than double that (155 days), making it extremely difficult to detect without a longer season of intensive reverberation mapping.

\subsection{The Underlying Accretion Disk Structure}\label{disk_struc}

To examine our results in the correct context with regards to the underlying accretion disk structure, we first discuss whether we can actually probe a slim disk profile in I~Zw~1 using optical/UV reverberation mapping and SED profile analysis. A slim disk, which heats up as radiative cooling becomes ineffective and therefore puffs up, is only expected to dominate inside the photon trapping radius \citep{wang1999}. The photon trapping radius $R_\mathrm{tr}$ can be estimated using the dimensionless accretion rate $\dot{\mathscr{M}}$ where we here take $\dot{\mathscr{M}} = 130$ as calculated above. Using Eq. 17 from \citet{donnan2023}, we find $R_\mathrm{tr} \sim 187 R_{S}$ (Schwarzschild radii). Emission from this accretion disk regime will peak at $817\:\textnormal{\AA}$ \citep[Eq. 7 in][]{cackett2020}. In this study we consider $\lambda > 3000\:\textnormal{\AA}$ for reverberation mapping, far away from the putative photon trapping radius, and therefore should probably not expect the lag to follow $\lambda^2$. Even in the SED we only consider $\lambda > 1700\:\textnormal{\AA}$, while the slim disk with $F_\nu \propto \nu^{-1}$ should only appear at shorter wavelengths. Indeed, \citet{kubota2019} show that the SED profile of an accretion disk partly in the slim disk state is almost identical to that of a thin disk at the wavelengths considered here. It is therefore possible that we are not in fact probing the slim disk in I~Zw~1, even with it being super-Eddington. Similar situations have been found in the super-Eddington AGN Mrk~142 and PG~1119+120 \citep{cackett2020,donnan2023}. Instead, we might be looking at an outer disk region.

Looking at the underlying disk structure of this outer disk region in I~Zw~1, we can look at the lag-wavelength spectrum, which reflects the radial temperature profile of the disk, and the SED slope in the UV/optical, which contains the long-wavelength tail of the disk emission. Parameter results for the power law fits to the lag-wavelength spectrum of the PyROA lags are shown in Table~\ref{tab:fits}. 
Both fits with a fixed $\beta$ -- the thin and slim disk -- shown in Table~\ref{tab:fits} perform similarly, with comparable fit parameter errors ($\sim5\%$) and $y_0 \sim 1$ as expected. As the lag-wavelength spectrum in Fig.~\ref{fig:pyroa_lags} illustrates, the data can also not sufficiently differentiate between these different profiles. The free $\beta$ fit has larger fit parameter uncertainties due to the larger number of fit parameters. This agnostic fit has a shallow slope, i.e. temperature falls of more slowly with radius in the disk \citep[similar to][]{cackett2020}. It also has the smallest disk size, within errors of the fiducial size. In addition, it shows that the disk size and $\beta$ are degenerate to an extent, as $\tau_0$ decreases as $\beta$ increases. There is additional information about the disk structure in the frequency-resolved lags at high frequencies ($0.031-0.076\:\mathrm{days}^{-1}$) which we expect to be dominated by the disk rather than the secondary reprocessor. These lags are reproduced by the fiducial thin disk generally well in Fig.~\ref{fig:frl}. The extracted high frequency lag-wavelength spectrum is shown in Fig.~\ref{fig:frl_fit} compared with the fiducial thin disk profile and size. While the uncertainties and scatter are larger, there is still a trend of increasing lag with wavelength, probing a radially stratified temperature profile. Assuming a thin disk, the disk size of these high frequency lags is $\tau_{0,\mathrm{HF}}=0.61\pm0.37\:\mathrm{days}$. This is smaller than but consistent within errors with the fiducial disk size of 1.07 days and shows that at higher variability frequencies we probe increasingly smaller structures in the AGN. In addition, it indicates the disk size based on the overall best-fit PyROA lags is likely to be too large, and the actual disk sizes tend towards the fiducial disk size or even smaller, possibly resembling a thin disk.

Considering the SED, both the RMS and mean AGN profiles have the same slope (Fig.~\ref{fig:sed_all}). This implies that both the variable components and the dominant constant component in the AGN have the same SED shape, with dominant emission from the accretion disk. The variable SED also shows no bluer-when-brighter behaviour to within measurement uncertainties. The exact shape of the SED profile is difficult to determine, depending on the amount of internal extinction applied and host galaxy contribution. First, if we only consider the LCO derived SED we find that for $F_\nu \propto \lambda^a$, $a = 0.12\pm0.13$. This is closer to a thin disk than a slim disk. When we then add UV/optical data from archival \textit{XMM-Newton} and \textit{Swift} observations the resulting slope steepens to $a = -0.11\pm0.12$, shifting even closer to a thin disk slope with $a=-1/3$. Notably, the \textit{XMM-Newton} data was collected before our campaign, and only one \textit{Swift} data set is simultaneous with our observations. Long term monitoring with ASAS-SN shows the optical variability in I~Zw~1 to be $\sim3-6\%$, so even earlier \textit{XMM-Newton} data should be within the flux range covered by our campaign \citep{huang2019}. There appears to be an offset between the \textit{XMM-Newton} and \textit{Swift} $UBV$ and the LCO data, which may be due to intrinsic variability. Shifting up the LCO SED to match the optical \textit{XMM}/\textit{Swift} data would result in a redder SED slope ($a\sim0$). Further, if we consider only the \textit{Swift} points (stars at lower flux in Fig.~\ref{fig:sed_fits}), the result would be similar, with an essentially flat SED. We are likely underestimating the host galaxy contribution, especially in the bluer bands, as we anchor the flux-flux decomposition in the $u$-band. While host galaxy spectra are generally red, there can be significant contribution in the blue optical and UV when there is active star formation. In fact, I~Zw~1 hosts a nuclear starburst within 1\:kpc, which is within our extraction aperture \citep{fei2023}. However, fitting a galactic template, \citet{juranova2024} found a red spectrum and negligible contribution at shorter wavelengths, especially in the UV. The red slope of the LCO SED can also be considered in the context of contribution from diffuse continuum. The diffuse continuum spectrum increases from the local minimum at the Balmer jump to the local maximum at the Paschen jump, which corresponds to our $BgVriz_s$-bands \citep{korista2019}. Therefore, in a case of significant DC contribution (as we are likely to be observing RM signals), this could account for a reddened SED slope at these wavelengths. However, this would also mean an even larger contribution in the $u$-band, which covers the Balmer jump. As the $u$-band has the lowest flux (Fig.~\ref{fig:sed_fits}), it is difficult to argue for this case. While the inclusion/exclusion of certain data does change the SED slope, all results are closer to a thin disk than a slim disk. This agrees with the proposition that we are looking at an outer disk region behaving like a thin disk, similar to the geometry proposed by \citet{kubota2019} for super-Eddington AGN which includes an inner slim disk and an outer thin disk.

The exact SED slope is highly sensitive to the amount of intrinsic extinction applied. The estimates of the internal extinction in I~Zw~1 vary considerably: from $0.1\:\mathrm{mag}$ in \citet{laor1997}, \citet{rudy2000}’s compromise value of $0.13\:\mathrm{mag}$, their calculated value of $0.19\:\mathrm{mag}$, to $0.206\:\mathrm{mag}$ as fitted for by \citet{juranova2024} (assuming a thin disk SED). A higher extinction appears more likely in I~Zw~1. \cite{rudy2000} calculated a $E(B-V)=0.19\:\mathrm{mag}$ based on \ion{O}{i} emission line ratios, even though they then adopted a lower value based on \citet{laor1997}. This is consistent with the value of $0.206\:\mathrm{mag}$ found by \citet{juranova2024} using an independent technique and data taken more than 15 years later. In addition, the extinction law we use \citep[and fitted by][]{juranova2024} is characteristic for quasars, eschewing the Galactic $2175\:\textnormal{\AA}$ bump and flattening out below $1550\:\textnormal{\AA}$ \citep{gaskell2004}. In its uncorrected state the SED of I~Zw~1 resembles the red slopes often reported for quasars which are sometimes cited as evidence for an inherently different SED shape in quasars \citep[e.g.,][]{davis2007,trammell2007}. However, upon correcting for internal extinction in I~Zw~1, the slope becomes markedly bluer and approaches that of a thin disk. It is reasonable that internal extinction is also present in other quasars, leading to artificially redder slopes, but nonetheless harbour a standard disk.

Clearly, we cannot make any conclusions as to the exact underlying structure of the accretion disk. We can however say that we are probably observing an optically thick, thermally emitting disk with a radial temperature profile, in which the temperature decreases with radius. This is of course under the assumption of the lamppost model (irrespective of the exact X-ray-UV/optical relationship), where lags between light curves are directly related to the physical distances across the disk and in the AGN \citep{cackett2007,fausnaugh2016,Kammoun2021}. If there is another method of communication over large scales in the disk, such as large-scale temperature fluctuations the lags are likely to have a different more complex relationship with physical distance \citep[e.g.,][]{cai2018}. The lag-wavelength spectrum in I~Zw~1 holds little differentiating power between different radial temperature profiles. For one, this is due to the comparatively small wavelength range probed in this study even though even across a larger wavelength range, another study on a super-Eddington object has similarly failed to find conclusive evidence for either a thin or slim disk profile \citep{cackett2020}. The other reason, of course, is the significant contribution in the lag spectrum from the secondary reprocessor. On the other hand, the SED quite plainly tends towards a thin disk slope, similar to the super-Eddington AGN Mrk~142 \citep{cackett2020}. This can be well-explained by a slim disk only existing within the photon trapping radius, which we do not probe here, and an outer disk region. This outer disk region might be like a standard thin disk, implying that we are indeed probing similar disk structures in sub- and super-Eddington AGN with UV/optical reverberation mapping. Therefore, we should not expect a turnover in results when crossing the Eddington limit. Indeed, a gradual change in disk structure occurs as $\dot{m}_E < 1$ increases, with super-Eddington AGN presenting the extremes of this transition.

\subsection{The Inner AGN Structure in I~Zw~1}

\begin{figure}
    \centering
    \includegraphics[width=0.48\textwidth]{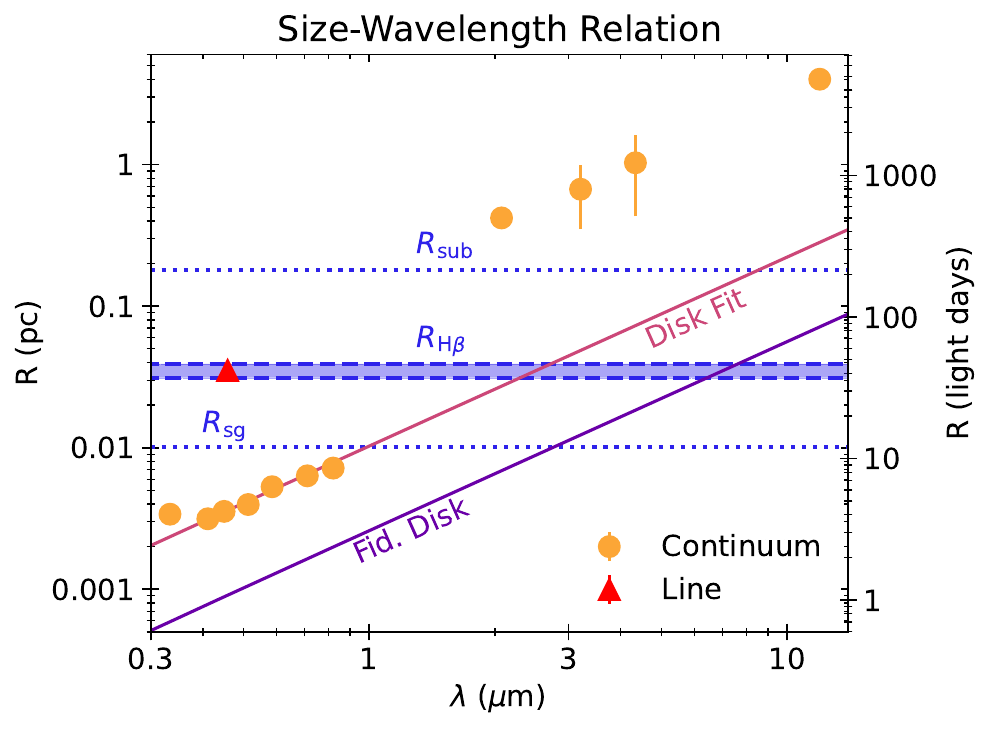}
    \caption{The directly measured size-wavelength relation in I~Zw~1, from the optical to the mid-IR. Circles represent sizes measured using continuum emission, using reverberation mapping for the optical and optical interferometry for the infrared, which are taken from \citet{drewes2025}. The triangle and the shaded region represents the radius $R_{\mathrm{H}\beta}$ measured using the H$\beta$ emission line with reverberation mapping by \citet{huang2019}. As the optical continuum lags were measured with respect to the $g$-band, we add the disk size in the $g$-band, $\tau_0$, to get the absolute size. Here we assume thin disk so we use $\tau_0 = 4.23\:\mathrm{days}$. We also add this size and the measured $V$-band lag to $R_{\mathrm{H}\beta}$ as the H$\beta$ lag was evaluated with reference to the $V$-band. This fitted thin disk profile is also plotted, and the fiducial thin disk profile with $\tau_0 = 1.07\:\mathrm{days}$. Finally, we indicate the sublimation radius $R_\mathrm{sub} = 0.18\:\mathrm{pc}$ from \citet{drewes2025} and the self-gravitating radius of the disk $R_\mathrm{sg} = 12\:\mathrm{ld}$ using the dotted lines, which are assumed to be the outer and inner boundaries of the BLR respectively \citep{lobban2022}.}
    \label{fig:size_wav}
\end{figure}

We have here collected the most extensive set of directly measured sizes of internal AGN components from the optical disk through to the mid-IR dusty `torus' \citep[this work,][]{huang2019,drewes2025}. This is plotted in Fig.~\ref{fig:size_wav}, which also includes the thin disk fit from Table~\ref{tab:fits}, a thin disk with the fiducial size of $\tau_0 = 1.07\:\mathrm{ld}$, the self-gravitating radius $R_\mathrm{sg}$ of the disk, and the sublimation radius $R_\mathrm{sub}$. As the lag we measure and that is described by Eq. \ref{eq:disk} is the lag relative to the $g$-band, we correct the optical continuum lags by adding the absolute $g$-band size as fitted for a thin disk, $\tau_0 = 4.23\:\mathrm{ld}$. The H$\beta$ lag was determined relative to the $V$-band so we correct it by adding our resulting absolute $V$-band lag \citep{huang2019}. Represented by the orange circles below $1\:\mu\mathrm{m}$ in Fig.~\ref{fig:size_wav}, we detect optical emission of a fiducial accretion disk structure out to $\sim0.01\:\mathrm{pc}$. This is also where the self-gravitating radius of the accretion disk is predicted to be, at $12\:\mathrm{ld}\simeq0.01\:\mathrm{pc}$, as denoted by the lower dotted line \citep{lobban2022}. The self-gravitating radius of the disk is where self-gravity in the disk starts to dominate over radiation pressure and the disk is expected to fragment. Therefore, we are indeed likely probing the outer disk region and possibly even the outer edges of the accretion disk. Beyond these radii, we enter the region of the BLR. A measure of the region occupied by BLR gas is the radius derived from reverberation mapping of the H$\beta$ emission line at $\sim 37\:\mathrm{ld}$ ($0.031\:\mathrm{pc}$) \citep{huang2019}. This is indicated in Fig.~\ref{fig:size_wav} with the red triangle, and the shaded region covers the uncertainty. However, the BLR likely extends to both smaller and larger radii than this as the BLR has a radial ionization structure, with higher ionization lines found at smaller radii than lower ionization ones \citep[e.g.,][]{clavel1991}. Classically, the inner boundary of the BLR is approximately given by the outer edge of the accretion disk and the outer boundary by the sublimation radius. The secondary reprocessor as fitted to the frequency-resolved lags covers the H$\beta$ radius but also extends significantly inwards ($0.01-0.04\:\mathrm{pc}$), likely probing a larger region of the BLR. Notably, we see little indication of continuum reprocessing at a distance of $\sim100\:\mathrm{ld}$, as that fit is strongly disfavoured. There is also evidence of winds launched from the BLR, a warm X-ray absorber with $v_\mathrm{out} \simeq 1750\:\mathrm{km\,s}^{-1}$ and an outflowing wind detected through absorption lines in the BLR spectrum with $v_\mathrm{out} \simeq 1950\:\mathrm{km\,s}^{-1}$ \citep{silva2018,rogantini2022,juranova2024}. Assuming that the outflow velocity must be at least the escape velocity as given by the virial velocity, the minimum launching radii of these outflowing components are $0.01-0.03\:\mathrm{pc}$. This places them towards the inner part of the fiducial BLR. 

The outer edge of the BLR is delimited by the sublimation radius at $0.18\:\mathrm{pc}$, denoted by the upper dotted line in Fig.~\ref{fig:size_wav} \citep{gravity2020,drewes2025}. This is consistent with the inner dust radius of $224\:\mathrm{ld}$ ($0.19\:\mathrm{pc}$) determined by \citet{landt2023} using spectral fitting. This also describes the inner rim of the dusty `torus'. The infrared continuum measurements using optical interferometry of the hot dust ($2.2-4.6\:\mu\mathrm{m}$) indicates the presence of a wind launching region: a ‘puffed-up’ inner region of the dust disk where a dusty wind is launched through infrared radiation pressure \citep{honig2019,drewes2025}. Reverberation mapping of the dust at $3.4-4.5\:\mu\mathrm{m}$ yields lags smaller than the optical interferometric measured sizes by a factor of $2.5-3.5$ \citep{lyu2019}. This is similar to the behaviour observed between $K$-band reverberation and interferometric measurements, which are usually offset by a factor of $2 - 2.5$ \citep[e.g.,][]{gravity2024}. An integral reason for this is the fact that these two measurements trace different sizes: reverberation mapping measures the response-weighted radius and optical interferometry the flux-weighted radius. Geometries that produce such offsets such as a bowl-shaped/concave inner dust structure have also been explored \citep[e.g.,][]{gravity2024}. However, the reverberation mapping results must be treated carefully as the lags are consistent with time intervals between observations of the IR light curve, indicating that the results may be spurious. Finally, we then trace the cooler parts of the dust in the mid-IR out to $\sim4\:\mathrm{pc}$ \citep{burtscher2013,drewes2025}. 

Using this size-wavelength relation, we can examine the failed radiatively accelerated dust driven outflow BLR formation theory \citep{czerny2011,baskin2018}. In this model, the accretion disk extends into the BLR and out to the dusty ‘torus’. Temperatures of the accretion disk in this region are $\sim 1000\:\mathrm{K}$, which is less than the sublimation temperature ($\sim1500\:\mathrm{K}$). Dust forms in the accretion disk atmosphere and is launched due to radiation pressure. At a certain height above the disk, the dust will sublime again due to strong irradiation and fall back down into the disk. The BLR is then constituted out of this failed wind. A central component of this theory is the presence of the accretion disk in the BLR. Extrapolating our fitted thin disk size to lower temperatures around $1000\:\mathrm{K}$ ($\sim1-2\:\mu\mathrm{m}$, $JHK$-bands), as shown with the solid lines in Fig.~\ref{fig:size_wav}, does extend it into the BLR. This is just the inner parts of the BLR. However, this is for a disk fit based on lags that are expected to include a significant contribution from a secondary reprocessor at larger radii, with a fitted $\tau_0$ four times larger than the fiducial $\tau_0$. If we consider a thin disk with the fiducial size, which is likely closer to the true size, also shown in Fig.~\ref{fig:size_wav}, the accretion disk emitting in these wavelengths is outside of the BLR as denoted by the self-gravitating radius. Going down to the disk size as fitted to high frequency lags, smaller again by half, locates the relevant accretion disk sections even further inwards. 

Similarly, \citet{thorne2025} found that in 3C 273 the accretion disk is likely to overlap in its $K$-band region with the BLR, based on extending their reverberation mapping disk fit to longer wavelengths, up to $130-170\:\mathrm{days}$ in the $K$-band. The BLR radius is $145\pm35\:\mathrm{days}$ as determined using optical interferometry \citep{gravity2018}.
The expected size in the $K$-band was estimated using the lag-wavelength spectrum fit that was several times larger than predicted. As discussed in Section~\ref{disc:rm_signals_source}, this likely includes diffuse continuum reprocessing at larger scales. Frequency-resolved lags show results approaching the fiducial disk size \citep{lewin2023,lewin2024}. The larger fiducial $\tau_0$ in 3C 273 of $8.95\:\mathrm{days}$ gives an accretion disk $K$-band size of $88\:\mathrm{ld}$, significantly smaller than the BLR size from optical interferometry \citep{thorne2025,gravity2018}. As in I~Zw~1, the extension of the accretion disk into the BLR beyond its self-gravitating radius to temperatures $\sim1000\:\mathrm{K}$ relies on the assumption that the measured continuum lags overwhelmingly arise from the disk. However, based on our results and literature, this is likely to be false. Instead, assuming the predicted disk size, places this region of the disk at approximately smaller radii, inconsistent with the measured BLR. Nonetheless due to the stratified nature of the BLR and its large spatial range, we are generally unsure of the location of the inner extent of the BLR, i.e. it can possibly exist well within the self-gravitating radius. This would change the interpretation of the extent of the disk within the BLR as presented here. Further, uncertainties in lag measurements introduce additional uncertainties in the positions of the components that we calculated.

Furthermore, should we expect to be able to detect direct $K$-band emission from a disk in the BLR? The dominant $K$-band emission region in an AGN is the hot dust at the sublimation region. If we consider the ratio of the luminosity of the disk in the BLR to that of the torus in the $K$-band, we can approximate this as the ratio of the disk to torus area. This is approximately $r^2_\mathrm{disk}/r^2_\mathrm{torus}$, where $r_\mathrm{disk}$ is the distance of the disk to the centre and $r_\mathrm{torus}$ is the distance of the torus. In I~Zw~1, if the $K$-band emission region of the disk is in the BLR, we can take $r_\mathrm{disk} \sim R_{\mathrm{H}\beta} = 0.031\:\mathrm{pc}$ \citep{huang2019}. The measured size of the torus $K$-band emission region is 0.42 pc, which gives $ r^2_\mathrm{disk}/r^2_\mathrm{torus } \simeq 1\%$ \citep{gravity2024,drewes2025}. In 3C 273, $ r_\mathrm{disk} \sim R_\mathrm{BLR}= 0.12\:\mathrm{pc}$ and $r_\mathrm{torus} = 0.57\:\mathrm{pc}$, giving $ r^2_\mathrm{disk}/r^2_\mathrm{torus } \simeq 4\%$ \citep{gravity2018,gravity2020}. These results indicate that the putative disk $K$-band emission is only a small fraction of the dominant torus emission, and that it is incredibly difficult to detect the presence of a disk in the BLR even in spatially resolved studies such as reverberation mapping and optical interferometry.

\section{Conclusions}
In this paper, we present the reverberation mapping results of the optical continuum in the super-Eddington AGN I~Zw~1 and collate the most extensive set of directly measured internal sizes of an AGN. We use the cross correlation method and PyROA to evaluate the lags of three years of $uBgVriz_s$ light curves. These lag-wavelength spectra are fitted with a thin and a slim disk, and with a free power law index profile to probe the underlying accretion disk structure. We also calculate the lag as a function of variability timescale and the frequency-resolved lags. Finally, we isolate the AGN SED and fit the UV/optical SED profile. These results show that:
\begin{enumerate}
	\item There is a continuum emitting secondary reprocessor at large radii consistent with the BLR, increasing lags significantly across all wavelengths, and likely resulting in an artificially inflated fitted accretion disk size.
	\item The evidence for this secondary reprocessor consists of longer lags at longer variability timescales, and the need for an additional, secondary reprocessing component to reproduce the lags at low frequencies.
    \item There are also indications that the source of this secondary reprocessor is diffuse continuum emission from hydrogen, such as the characteristic $u$-band excess in the lag-wavelength spectrum and an increase in the secondary reprocessor fraction in the $u$-band based on the modelling of the frequency-resolved lags, which both suggest the presence of the Balmer jump.
	\item We cannot determine the underlying accretion disk profile, that is, we cannot distinguish between a standard thin disk and a slim disk that is expected for super-Eddington AGN based on the lag-wavelength spectrum. The UV/optical SED profile trends towards a thin disk.
	\item The actual disk size is likely to be on the order of the fiducial disk size ($\sim1.07\:\mathrm{ld}$ at $4495\:\textnormal{\AA}$) or even smaller ($\sim0.6\pm0.4\:\mathrm{ld}$ at $4495\:\textnormal{\AA}$), based on the high frequency lag-wavelength spectrum.
	\item There is little difference between the results of the disk reverberation mapping for this super-Eddington object and other super- and sub-Eddington objects from literature. This indicates that there are very similar structures and processes responsible for the variable UV/optical continuum emission independent of accretion rate and/or that our observations and analysis techniques are not able to access the parameter spaces in which differences will manifest. 
    \item Considering the evidence that continuum reverberation mapping measures a combination of the disk and diffuse continuum at larger distances, and adjusting the fiducial disk size for this, the size-wavelength relation from the optical to the mid-infrared shows limited evidence that the accretion disk extends into the BLR significantly. This might tentatively disfavour the failed radiatively accelerated dust driven outflow BLR formation model.
\end{enumerate}
Fourier analysis and frequency-resolved lags are shown to be capable of disentangling different signals that contribute to the continuum lags. Presumably, there is a combination of structures we measure when we perform disk reverberation mapping experiments. The key to isolating the underlying accretion disk structure is in the further development of these Fourier techniques and their optimisation for AGN light curves. Further, there lies significant power in combining directly measured sizes of the internal AGN structures from reverberation mapping and optical interferometry to study the interplay and relationship between internal structural components. The sample of objects with BLR and hot dust size measurements from optical interferometry is about to exponentially increase with the full commissioning of GRAVITY+ \citep{gravity2022}. Mid-infrared size measurements of the warm dust, especially the never before accessed $3-5\:\mu\mathrm{m}$ range, will also increase significantly as MATISSE also benefits from GRAVITY+. In the further future, the establishment of a km- or tens of km-baseline optical interferometer might enable direct observations of the accretion disk.

\section*{Acknowledgements}
We thank the anonymous referee for their helpful and construc-
tive comments that improved the manuscript. FD thanks Sebastian F. H{\"o}nig for useful discussions and Jan for proofreading the manuscript. FD acknowledges support from Science and Technology Facilities Council (STFC) studentship ST/W507805/1. RV acknowledges support from STFC studentship ST/Y509589/1. JVHS acknowledges support from STFC grant ST/V000861/1. ERC acknowledges support from the National Research Foundation of South Africa. MV gratefully acknowledges financial support from the Independent Research Fund Denmark via grant number DFF 3103-00146 and from the Carlsberg Foundation (grants CF21-0649 and CF23-0417).

This work makes use of observations from the Las Cumbres Observatory global telescope network. We acknowledge the use of public data from the Swift data archive and the XMM-Newton Science Archive.
This research made extensive use of {\sc astropy}, a community-developed core Python package for Astronomy \citep{Astropy-Collaboration:2013aa} and {\sc matplotlib} \citep{Hunter:2007aa}.

\section*{Data Availability}
The raw datasets were derived from sources in the public domain: LCO archive \url{https://archive.lco.global}, XMM-Newton Science Archive \url{https://nxsa.esac.esa.int/nxsa-web/}, and {\it Swift} archive \url{https://www.swift.ac.uk/swift_live}. The inter-calibrated light curves are available on Zenodo at \url{https://dx.doi.org/10.5281/zenodo.16599407}.



\bibliographystyle{mnras}
\bibliography{references} 

@ARTICLE{salpeter1964,
       author = {{Salpeter}, E.~E.},
        title = "{Accretion of Interstellar Matter by Massive Objects.}",
      journal = {\apj},
         year = 1964,
        month = aug,
       volume = {140},
        pages = {796-800},
          doi = {10.1086/147973},
       adsurl = {https://ui.adsabs.harvard.edu/abs/1964ApJ...140..796S}
}

@ARTICLE{shakura1973,
       author = {{Shakura}, N.~I. and {Sunyaev}, R.~A.},
        title = "{Black holes in binary systems. Observational appearance.}",
      journal = {\aap},
         year = 1973,
        month = jan,
       volume = {24},
        pages = {337-355},
       adsurl = {https://ui.adsabs.harvard.edu/abs/1973A&A....24..337S}
}

@INPROCEEDINGS{novikov1973,
       author = {{Novikov}, I.~D. and {Thorne}, K.~S.},
        title = "{Astrophysics of black holes.}",
    booktitle = {Black Holes (Les Astres Occlus)},
         year = 1973,
       editor = {{Dewitt}, C. and {Dewitt}, B.~S.},
        month = jan,
        pages = {343-450},
       adsurl = {https://ui.adsabs.harvard.edu/abs/1973blho.conf..343N}
}

@ARTICLE{fausnaugh2016,
       author = {{Fausnaugh}, M.~M. and {Denney}, K.~D. and {Barth}, A.~J. and {Bentz}, M.~C. and {Bottorff}, M.~C. and {Carini}, M.~T. and {Croxall}, K.~V. and {De Rosa}, G. and {Goad}, M.~R. and {Horne}, Keith and {Joner}, M.~D. and {Kaspi}, S. and {Kim}, M. and {Klimanov}, S.~A. and {Kochanek}, C.~S. and {Leonard}, D.~C. and {Netzer}, H. and {Peterson}, B.~M. and {Schn{\"u}lle}, K. and {Sergeev}, S.~G. and {Vestergaard}, M. and {Zheng}, W. -K. and {Zu}, Y. and {Anderson}, M.~D. and {Ar{\'e}valo}, P. and {Bazhaw}, C. and {Borman}, G.~A. and {Boroson}, T.~A. and {Brandt}, W.~N. and {Breeveld}, A.~A. and {Brewer}, B.~J. and {Cackett}, E.~M. and {Crenshaw}, D.~M. and {Dalla Bont{\`a}}, E. and {De Lorenzo-C{\'a}ceres}, A. and {Dietrich}, M. and {Edelson}, R. and {Efimova}, N.~V. and {Ely}, J. and {Evans}, P.~A. and {Filippenko}, A.~V. and {Flatland}, K. and {Gehrels}, N. and {Geier}, S. and {Gelbord}, J.~M. and {Gonzalez}, L. and {Gorjian}, V. and {Grier}, C.~J. and {Grupe}, D. and {Hall}, P.~B. and {Hicks}, S. and {Horenstein}, D. and {Hutchison}, T. and {Im}, M. and {Jensen}, J.~J. and {Jones}, J. and {Kaastra}, J. and {Kelly}, B.~C. and {Kennea}, J.~A. and {Kim}, S.~C. and {Korista}, K.~T. and {Kriss}, G.~A. and {Lee}, J.~C. and {Lira}, P. and {MacInnis}, F. and {Manne-Nicholas}, E.~R. and {Mathur}, S. and {McHardy}, I.~M. and {Montouri}, C. and {Musso}, R. and {Nazarov}, S.~V. and {Norris}, R.~P. and {Nousek}, J.~A. and {Okhmat}, D.~N. and {Pancoast}, A. and {Papadakis}, I. and {Parks}, J.~R. and {Pei}, L. and {Pogge}, R.~W. and {Pott}, J. -U. and {Rafter}, S.~E. and {Rix}, H. -W. and {Saylor}, D.~A. and {Schimoia}, J.~S. and {Siegel}, M. and {Spencer}, M. and {Starkey}, D. and {Sung}, H. -I. and {Teems}, K.~G. and {Treu}, T. and {Turner}, C.~S. and {Uttley}, P. and {Villforth}, C. and {Weiss}, Y. and {Woo}, J. -H. and {Yan}, H. and {Young}, S.},
        title = "{Space Telescope and Optical Reverberation Mapping Project. III. Optical Continuum Emission and Broadband Time Delays in NGC 5548}",
      journal = {\apj},
         year = 2016,
        month = apr,
       volume = {821},
       number = {1},
          eid = {56},
        pages = {56},
          doi = {10.3847/0004-637X/821/1/56},
archivePrefix = {arXiv},
       eprint = {1510.05648},
 primaryClass = {astro-ph.GA},
       adsurl = {https://ui.adsabs.harvard.edu/abs/2016ApJ...821...56F}
}

@ARTICLE{cackett2018,
       author = {{Cackett}, Edward M. and {Chiang}, Chia-Ying and {McHardy}, Ian and {Edelson}, Rick and {Goad}, Michael R. and {Horne}, Keith and {Korista}, Kirk T.},
        title = "{Accretion Disk Reverberation with Hubble Space Telescope Observations of NGC 4593: Evidence for Diffuse Continuum Lags}",
      journal = {\apj},
         year = 2018,
        month = apr,
       volume = {857},
       number = {1},
          eid = {53},
        pages = {53},
          doi = {10.3847/1538-4357/aab4f7},
archivePrefix = {arXiv},
       eprint = {1712.04025},
 primaryClass = {astro-ph.HE},
       adsurl = {https://ui.adsabs.harvard.edu/abs/2018ApJ...857...53C}
}

@ARTICLE{fausnaugh2018,
       author = {{Fausnaugh}, M.~M. and {Starkey}, D.~A. and {Horne}, Keith and {Kochanek}, C.~S. and {Peterson}, B.~M. and {Bentz}, M.~C. and {Denney}, K.~D. and {Grier}, C.~J. and {Grupe}, D. and {Pogge}, R.~W. and {De Rosa}, G. and {Adams}, S.~M. and {Barth}, A.~J. and {Beatty}, Thomas G. and {Bhattacharjee}, A. and {Borman}, G.~A. and {Boroson}, T.~A. and {Bottorff}, M.~C. and {Brown}, Jacob E. and {Brown}, Jonathan S. and {Brotherton}, M.~S. and {Coker}, C.~T. and {Crawford}, S.~M. and {Croxall}, K.~V. and {Eftekharzadeh}, Sarah and {Eracleous}, Michael and {Joner}, M.~D. and {Henderson}, C.~B. and {Holoien}, T.~W. -S. and {Hutchison}, T. and {Kaspi}, Shai and {Kim}, S. and {King}, Anthea L. and {Li}, Miao and {Lochhaas}, Cassandra and {Ma}, Zhiyuan and {MacInnis}, F. and {Manne-Nicholas}, E.~R. and {Mason}, M. and {Montuori}, Carmen and {Mosquera}, Ana and {Mudd}, Dale and {Musso}, R. and {Nazarov}, S.~V. and {Nguyen}, M.~L. and {Okhmat}, D.~N. and {Onken}, Christopher A. and {Ou-Yang}, B. and {Pancoast}, A. and {Pei}, L. and {Penny}, Matthew T. and {Poleski}, Rados{\l}aw and {Rafter}, Stephen and {Romero-Colmenero}, E. and {Runnoe}, Jessie and {Sand}, David J. and {Schimoia}, Jaderson S. and {Sergeev}, S.~G. and {Shappee}, B.~J. and {Simonian}, Gregory V. and {Somers}, Garrett and {Spencer}, M. and {Stevens}, Daniel J. and {Tayar}, Jamie and {Treu}, T. and {Valenti}, Stefano and {Van Saders}, J. and {Villanueva}, S., Jr. and {Villforth}, C. and {Weiss}, Yaniv and {Winkler}, H. and {Zhu}, W.},
        title = "{Continuum Reverberation Mapping of the Accretion Disks in Two Seyfert 1 Galaxies}",
      journal = {\apj},
         year = 2018,
        month = feb,
       volume = {854},
       number = {2},
          eid = {107},
        pages = {107},
          doi = {10.3847/1538-4357/aaaa2b},
archivePrefix = {arXiv},
       eprint = {1801.09692},
 primaryClass = {astro-ph.GA},
       adsurl = {https://ui.adsabs.harvard.edu/abs/2018ApJ...854..107F}
}

@ARTICLE{edelson2017,
       author = {{Edelson}, R. and {Gelbord}, J. and {Cackett}, E. and {Connolly}, S. and {Done}, C. and {Fausnaugh}, M. and {Gardner}, E. and {Gehrels}, N. and {Goad}, M. and {Horne}, K. and {McHardy}, I. and {Peterson}, B.~M. and {Vaughan}, S. and {Vestergaard}, M. and {Breeveld}, A. and {Barth}, A.~J. and {Bentz}, M. and {Bottorff}, M. and {Brandt}, W.~N. and {Crawford}, S.~M. and {Dalla Bont{\`a}}, E. and {Emmanoulopoulos}, D. and {Evans}, P. and {Figuera Jaimes}, R. and {Filippenko}, A.~V. and {Ferland}, G. and {Grupe}, D. and {Joner}, M. and {Kennea}, J. and {Korista}, K.~T. and {Krimm}, H.~A. and {Kriss}, G. and {Leonard}, D.~C. and {Mathur}, S. and {Netzer}, H. and {Nousek}, J. and {Page}, K. and {Romero-Colmenero}, E. and {Siegel}, M. and {Starkey}, D.~A. and {Treu}, T. and {Vogler}, H.~A. and {Winkler}, H. and {Zheng}, W.},
        title = "{Swift Monitoring of NGC 4151: Evidence for a Second X-Ray/UV Reprocessing}",
      journal = {\apj},
         year = 2017,
        month = may,
       volume = {840},
       number = {1},
          eid = {41},
        pages = {41},
          doi = {10.3847/1538-4357/aa6890},
archivePrefix = {arXiv},
       eprint = {1703.06901},
 primaryClass = {astro-ph.HE},
       adsurl = {https://ui.adsabs.harvard.edu/abs/2017ApJ...840...41E}
}

@ARTICLE{hernandezsantisteban2020,
       author = {{Hern{\'a}ndez Santisteban}, J.~V. and {Edelson}, R. and {Horne}, K. and {Gelbord}, J.~M. and {Barth}, A.~J. and {Cackett}, E.~M. and {Goad}, M.~R. and {Netzer}, H. and {Starkey}, D. and {Uttley}, P. and {Brandt}, W.~N. and {Korista}, K. and {Lohfink}, A.~M. and {Onken}, C.~A. and {Page}, K.~L. and {Siegel}, M. and {Vestergaard}, M. and {Bisogni}, S. and {Breeveld}, A.~A. and {Cenko}, S.~B. and {Dalla Bont{\`a}}, E. and {Evans}, P.~A. and {Ferland}, G. and {Gonzalez-Buitrago}, D.~H. and {Grupe}, D. and {Joner}, M.~D. and {Kriss}, G. and {LaPorte}, S.~J. and {Mathur}, S. and {Marshall}, F. and {Mehdipour}, M. and {Mudd}, D. and {Peterson}, B.~M. and {Schmidt}, T. and {Vaughan}, S. and {Valenti}, S.},
        title = "{Intensive disc-reverberation mapping of Fairall 9: first year of Swift and LCO monitoring}",
      journal = {\mnras},
         year = 2020,
        month = nov,
       volume = {498},
       number = {4},
        pages = {5399-5416},
          doi = {10.1093/mnras/staa2365},
archivePrefix = {arXiv},
       eprint = {2008.02134},
 primaryClass = {astro-ph.GA},
       adsurl = {https://ui.adsabs.harvard.edu/abs/2020MNRAS.498.5399H}
}

@ARTICLE{donnan2023,
       author = {{Donnan}, Fergus R. and {Hern{\'a}ndez Santisteban}, Juan V. and {Horne}, Keith and {Hu}, Chen and {Du}, Pu and {Li}, Yan-Rong and {Xiao}, Ming and {Ho}, Luis C. and {Aceituno}, Jes{\'u}s and {Wang}, Jian-Min and {Guo}, Wei-Jian and {Yang}, Sen and {Jiang}, Bo-Wei and {Yao}, Zhu-Heng},
        title = "{Testing super-eddington accretion on to a supermassive black hole: reverberation mapping of PG 1119+120}",
      journal = {\mnras},
         year = 2023,
        month = jul,
       volume = {523},
       number = {1},
        pages = {545-567},
          doi = {10.1093/mnras/stad1409},
archivePrefix = {arXiv},
       eprint = {2302.09370},
 primaryClass = {astro-ph.GA},
       adsurl = {https://ui.adsabs.harvard.edu/abs/2023MNRAS.523..545D}
}

@ARTICLE{vincentelli2021,
       author = {{Vincentelli}, F.~M. and {McHardy}, I. and {Cackett}, E.~M. and {Barth}, A.~J. and {Horne}, K. and {Goad}, M. and {Korista}, K. and {Gelbord}, J. and {Brandt}, W. and {Edelson}, R. and {Miller}, J.~A. and {Pahari}, M. and {Peterson}, B.~M. and {Schmidt}, T. and {Baldi}, R.~D. and {Breedt}, E. and {Hern{\'a}ndez Santisteban}, J.~V. and {Romero-Colmenero}, E. and {Ward}, M. and {Williams}, D.~R.~A.},
        title = "{On the multiwavelength variability of Mrk 110: two components acting at different time-scales}",
      journal = {\mnras},
         year = 2021,
        month = jul,
       volume = {504},
       number = {3},
        pages = {4337-4353},
          doi = {10.1093/mnras/stab1033},
archivePrefix = {arXiv},
       eprint = {2104.04530},
 primaryClass = {astro-ph.HE},
       adsurl = {https://ui.adsabs.harvard.edu/abs/2021MNRAS.504.4337V}
}

@ARTICLE{lewin2024,
       author = {{Lewin}, Collin and {Kara}, Erin and {Barth}, Aaron J. and {Cackett}, Edward M. and {De Rosa}, Gisella and {Homayouni}, Yasaman and {Horne}, Keith and {Kriss}, Gerard A. and {Landt}, Hermine and {Gelbord}, Jonathan and {Montano}, John and {Arav}, Nahum and {Bentz}, Misty C. and {Boizelle}, Benjamin D. and {Dalla Bont{\`a}}, Elena and {Brotherton}, Michael S. and {Dehghanian}, Maryam and {Ferland}, Gary J. and {Fian}, Carina and {Goad}, Michael R. and {Hern{\'a}ndez Santisteban}, Juan V. and {Ili{\'c}}, Dragana and {Kaastra}, Jelle and {Kaspi}, Shai and {Korista}, Kirk T. and {Kosec}, Peter and {Kova{\v{c}}evi{\'c}}, Andjelka and {Mehdipour}, Missagh and {Miller}, Jake A. and {Netzer}, Hagai and {Neustadt}, Jack M.~M. and {Panagiotou}, Christos and {Partington}, Ethan R. and {Popovi{\'c}}, Luka {\v{C}}. and {Sanmartim}, David and {Vestergaard}, Marianne and {Ward}, Martin J. and {Zaidouni}, Fatima},
        title = "{AGN STORM 2. VII. A Frequency-resolved Map of the Accretion Disk in Mrk 817: Simultaneous X-Ray Reverberation and UVOIR Disk Reprocessing Time Lags}",
      journal = {\apj},
         year = 2024,
        month = oct,
       volume = {974},
       number = {2},
          eid = {271},
        pages = {271},
          doi = {10.3847/1538-4357/ad6b08},
archivePrefix = {arXiv},
       eprint = {2409.09115},
 primaryClass = {astro-ph.HE},
       adsurl = {https://ui.adsabs.harvard.edu/abs/2024ApJ...974..271L}
}

@ARTICLE{miller2023,
       author = {{Miller}, Jake A. and {Cackett}, Edward M. and {Goad}, Michael R. and {Horne}, Keith and {Barth}, Aaron J. and {Romero-Colmenero}, Encarni and {Fausnaugh}, Michael and {Gelbord}, Jonathan and {Korista}, Kirk T. and {Landt}, Hermine and {Treu}, Tommaso and {Winkler}, Hartmut},
        title = "{Continuum Reverberation Mapping of Mrk 876 over Three Years with Remote Robotic Observatories}",
      journal = {\apj},
         year = 2023,
        month = aug,
       volume = {953},
       number = {2},
          eid = {137},
        pages = {137},
          doi = {10.3847/1538-4357/ace342},
archivePrefix = {arXiv},
       eprint = {2307.02630},
 primaryClass = {astro-ph.GA},
       adsurl = {https://ui.adsabs.harvard.edu/abs/2023ApJ...953..137M}
}

@ARTICLE{cackett2007,
       author = {{Cackett}, Edward M. and {Horne}, Keith and {Winkler}, Hartmut},
        title = "{Testing thermal reprocessing in active galactic nuclei accretion discs}",
      journal = {\mnras},
         year = 2007,
        month = sep,
       volume = {380},
       number = {2},
        pages = {669-682},
          doi = {10.1111/j.1365-2966.2007.12098.x},
archivePrefix = {arXiv},
       eprint = {0706.1464},
 primaryClass = {astro-ph},
       adsurl = {https://ui.adsabs.harvard.edu/abs/2007MNRAS.380..669C}
}

@ARTICLE{korista2001,
       author = {{Korista}, Kirk T. and {Goad}, Michael R.},
        title = "{The Variable Diffuse Continuum Emission of Broad-Line Clouds}",
      journal = {\apj},
         year = 2001,
        month = jun,
       volume = {553},
       number = {2},
        pages = {695-708},
          doi = {10.1086/320964},
archivePrefix = {arXiv},
       eprint = {astro-ph/0101117},
 primaryClass = {astro-ph},
       adsurl = {https://ui.adsabs.harvard.edu/abs/2001ApJ...553..695K}
}

@ARTICLE{korista2019,
       author = {{Korista}, K.~T. and {Goad}, M.~R.},
        title = "{Quantifying the impact of variable BLR diffuse continuum contributions on measured continuum interband delays}",
      journal = {\mnras},
         year = 2019,
        month = nov,
       volume = {489},
       number = {4},
        pages = {5284-5300},
          doi = {10.1093/mnras/stz2330},
archivePrefix = {arXiv},
       eprint = {1908.07757},
 primaryClass = {astro-ph.GA},
       adsurl = {https://ui.adsabs.harvard.edu/abs/2019MNRAS.489.5284K}
}

@ARTICLE{lawther2018,
       author = {{Lawther}, D. and {Goad}, M.~R. and {Korista}, K.~T. and {Ulrich}, O. and {Vestergaard}, M.},
        title = "{Quantifying the diffuse continuum contribution of BLR Clouds to AGN Continuum Inter-band Delays}",
      journal = {\mnras},
         year = 2018,
        month = nov,
       volume = {481},
       number = {1},
        pages = {533-554},
          doi = {10.1093/mnras/sty2242},
archivePrefix = {arXiv},
       eprint = {1808.04798},
 primaryClass = {astro-ph.GA},
       adsurl = {https://ui.adsabs.harvard.edu/abs/2018MNRAS.481..533L}
}

@ARTICLE{cackett2023,
       author = {{Cackett}, Edward M. and {Gelbord}, Jonathan and {Barth}, Aaron J. and {De Rosa}, Gisella and {Edelson}, Rick and {Goad}, Michael R. and {Homayouni}, Yasaman and {Horne}, Keith and {Kara}, Erin A. and {Kriss}, Gerard A. and {Korista}, Kirk T. and {Landt}, Hermine and {Plesha}, Rachel and {Arav}, Nahum and {Bentz}, Misty C. and {Boizelle}, Benjamin D. and {Dalla Bont{\`a}}, Elena and {Dehghanian}, Maryam and {Donnan}, Fergus and {Du}, Pu and {Ferland}, Gary J. and {Fian}, Carina and {Filippenko}, Alexei V. and {Gonz{\'a}lez Buitrago}, Diego H. and {Grier}, Catherine J. and {Hall}, Patrick B. and {Hu}, Chen and {Ili{\'c}}, Dragana and {Kaastra}, Jelle and {Kaspi}, Shai and {Kochanek}, Christopher S. and {Kova{\v{c}}evi{\'c}}, Andjelka B. and {Kynoch}, Daniel and {Li}, Yan-Rong and {McLane}, Jacob N. and {Mehdipour}, Missagh and {Miller}, Jake A. and {Montano}, John and {Netzer}, Hagai and {Panagiotou}, Christos and {Partington}, Ethan and {{\v{C}}. Popovi{\'c}}, Luka and {Proga}, Daniel and {Rogantini}, Daniele and {Sanmartim}, David and {Siebert}, Matthew R. and {Storchi-Bergmann}, Thaisa and {Vestergaard}, Marianne and {Wang}, Jian-Min and {Waters}, Tim and {Zaidouni}, Fatima},
        title = "{AGN STORM 2. IV. Swift X-Ray and Ultraviolet/Optical Monitoring of Mrk 817}",
      journal = {\apj},
         year = 2023,
        month = dec,
       volume = {958},
       number = {2},
          eid = {195},
        pages = {195},
          doi = {10.3847/1538-4357/acfdac},
archivePrefix = {arXiv},
       eprint = {2306.17663},
 primaryClass = {astro-ph.HE},
       adsurl = {https://ui.adsabs.harvard.edu/abs/2023ApJ...958..195C}
}

@ARTICLE{edelson2015,
       author = {{Edelson}, R. and {Gelbord}, J.~M. and {Horne}, K. and {McHardy}, I.~M. and {Peterson}, B.~M. and {Ar{\'e}valo}, P. and {Breeveld}, A.~A. and {De Rosa}, G. and {Evans}, P.~A. and {Goad}, M.~R. and {Kriss}, G.~A. and {Brandt}, W.~N. and {Gehrels}, N. and {Grupe}, D. and {Kennea}, J.~A. and {Kochanek}, C.~S. and {Nousek}, J.~A. and {Papadakis}, I. and {Siegel}, M. and {Starkey}, D. and {Uttley}, P. and {Vaughan}, S. and {Young}, S. and {Barth}, A.~J. and {Bentz}, M.~C. and {Brewer}, B.~J. and {Crenshaw}, D.~M. and {Dalla Bont{\`a}}, E. and {De Lorenzo-C{\'a}ceres}, A. and {Denney}, K.~D. and {Dietrich}, M. and {Ely}, J. and {Fausnaugh}, M.~M. and {Grier}, C.~J. and {Hall}, P.~B. and {Kaastra}, J. and {Kelly}, B.~C. and {Korista}, K.~T. and {Lira}, P. and {Mathur}, S. and {Netzer}, H. and {Pancoast}, A. and {Pei}, L. and {Pogge}, R.~W. and {Schimoia}, J.~S. and {Treu}, T. and {Vestergaard}, M. and {Villforth}, C. and {Yan}, H. and {Zu}, Y.},
        title = "{Space Telescope and Optical Reverberation Mapping Project. II. Swift and HST Reverberation Mapping of the Accretion Disk of NGC 5548}",
      journal = {\apj},
         year = 2015,
        month = jun,
       volume = {806},
       number = {1},
          eid = {129},
        pages = {129},
          doi = {10.1088/0004-637X/806/1/129},
archivePrefix = {arXiv},
       eprint = {1501.05951},
 primaryClass = {astro-ph.GA},
       adsurl = {https://ui.adsabs.harvard.edu/abs/2015ApJ...806..129E}
}

@ARTICLE{abramowicz1988,
       author = {{Abramowicz}, M.~A. and {Czerny}, B. and {Lasota}, J.~P. and {Szuszkiewicz}, E.},
        title = "{Slim Accretion Disks}",
      journal = {\apj},
         year = 1988,
        month = sep,
       volume = {332},
        pages = {646},
          doi = {10.1086/166683},
       adsurl = {https://ui.adsabs.harvard.edu/abs/1988ApJ...332..646A}
}

@ARTICLE{cackett2020,
       author = {{Cackett}, Edward M. and {Gelbord}, Jonathan and {Li}, Yan-Rong and {Horne}, Keith and {Wang}, Jian-Min and {Barth}, Aaron J. and {Bai}, Jin-Ming and {Bian}, Wei-Hao and {Carroll}, Russell W. and {Du}, Pu and {Edelson}, Rick and {Goad}, Michael R. and {Ho}, Luis C. and {Hu}, Chen and {Khatu}, Viraja C. and {Luo}, Bin and {Miller}, Jake and {Yuan}, Ye-Fei},
        title = "{Supermassive Black Holes with High Accretion Rates in Active Galactic Nuclei. XI. Accretion Disk Reverberation Mapping of Mrk 142}",
      journal = {\apj},
         year = 2020,
        month = jun,
       volume = {896},
       number = {1},
          eid = {1},
        pages = {1},
          doi = {10.3847/1538-4357/ab91b5},
archivePrefix = {arXiv},
       eprint = {2005.03685},
 primaryClass = {astro-ph.HE},
       adsurl = {https://ui.adsabs.harvard.edu/abs/2020ApJ...896....1C}
}

@ARTICLE{pahari2020,
       author = {{Pahari}, Mayukh and {McHardy}, I.~M. and {Vincentelli}, Federico and {Cackett}, Edward and {Peterson}, Bradley M. and {Goad}, Mike and {G{\"u}ltekin}, Kayhan and {Horne}, Keith},
        title = "{Evidence for variability time-scale-dependent UV/X-ray delay in Seyfert 1 AGN NGC 7469}",
      journal = {\mnras},
         year = 2020,
        month = may,
       volume = {494},
       number = {3},
        pages = {4057-4068},
          doi = {10.1093/mnras/staa1055},
archivePrefix = {arXiv},
       eprint = {2004.07901},
 primaryClass = {astro-ph.HE},
       adsurl = {https://ui.adsabs.harvard.edu/abs/2020MNRAS.494.4057P}
}

@ARTICLE{wang1999,
       author = {{Wang}, Jian-Min and {Zhou}, You-Yuan},
        title = "{Self-similar Solution of Optically Thick Advection-dominated Flows}",
      journal = {\apj},
         year = 1999,
        month = may,
       volume = {516},
       number = {1},
        pages = {420-424},
          doi = {10.1086/307080},
       adsurl = {https://ui.adsabs.harvard.edu/abs/1999ApJ...516..420W}
}

@ARTICLE{rogantini2022,
       author = {{Rogantini}, D. and {Costantini}, E. and {Gallo}, L.~C. and {Wilkins}, D.~R. and {Brandt}, W.~N. and {Mehdipour}, M.},
        title = "{The multi-epoch X-ray tale of I Zwicky 1 outflows}",
      journal = {\mnras},
         year = 2022,
        month = nov,
       volume = {516},
       number = {4},
        pages = {5171-5186},
          doi = {10.1093/mnras/stac2552},
archivePrefix = {arXiv},
       eprint = {2209.02747},
 primaryClass = {astro-ph.HE},
       adsurl = {https://ui.adsabs.harvard.edu/abs/2022MNRAS.516.5171R}
}

@ARTICLE{juranova2024,
       author = {{Jur{\'a}{\v{n}}ov{\'a}}, A. and {Costantini}, E. and {Kriss}, G.~A. and {Mehdipour}, M. and {Brandt}, W.~N. and {Di Gesu}, L. and {Fabian}, A.~C. and {Gallo}, L. and {Giustini}, M. and {Rogantini}, D. and {Wilkins}, D.~R.},
        title = "{The outflowing ionised gas of I Zw 1 observed by HST COS}",
      journal = {\aap},
         year = 2024,
        month = jun,
       volume = {686},
          eid = {A99},
        pages = {A99},
          doi = {10.1051/0004-6361/202449544},
archivePrefix = {arXiv},
       eprint = {2404.10060},
 primaryClass = {astro-ph.GA},
       adsurl = {https://ui.adsabs.harvard.edu/abs/2024A&A...686A..99J}
}

@ARTICLE{huang2019,
       author = {{Huang}, Ying-Ke and {Hu}, Chen and {Zhao}, Yu-Lin and {Zhang}, Zhi-Xiang and {Lu}, Kai-Xing and {Wang}, Kai and {Zhang}, Yue and {Du}, Pu and {Li}, Yan-Rong and {Bai}, Jin-Ming and {Ho}, Luis C. and {Bian}, Wei-Hao and {Yuan}, Ye-Fei and {Wang}, Jian-Min},
        title = "{Reverberation Mapping of the Narrow-line Seyfert 1 Galaxy I Zwicky 1: Black Hole Mass}",
      journal = {\apj},
         year = 2019,
        month = may,
       volume = {876},
       number = {2},
          eid = {102},
        pages = {102},
          doi = {10.3847/1538-4357/ab16ef},
archivePrefix = {arXiv},
       eprint = {1904.06146},
 primaryClass = {astro-ph.GA},
       adsurl = {https://ui.adsabs.harvard.edu/abs/2019ApJ...876..102H}
}

@ARTICLE{burtscher2013,
       author = {{Burtscher}, L. and {Meisenheimer}, K. and {Tristram}, K.~R.~W. and {Jaffe}, W. and {H{\"o}nig}, S.~F. and {Davies}, R.~I. and {Kishimoto}, M. and {Pott}, J. -U. and {R{\"o}ttgering}, H. and {Schartmann}, M. and {Weigelt}, G. and {Wolf}, S.},
        title = "{A diversity of dusty AGN tori. Data release for the VLTI/MIDI AGN Large Program and first results for 23 galaxies}",
      journal = {\aap},
         year = 2013,
        month = oct,
       volume = {558},
          eid = {A149},
        pages = {A149},
          doi = {10.1051/0004-6361/201321890},
archivePrefix = {arXiv},
       eprint = {1307.2068},
 primaryClass = {astro-ph.CO},
       adsurl = {https://ui.adsabs.harvard.edu/abs/2013A&A...558A.149B}
}

@ARTICLE{Gehrels:2004,
       author = {{Gehrels}, N. and {Chincarini}, G. and {Giommi}, P. and {Mason}, K.~O. and
         {Nousek}, J.~A. and {Wells}, A.~A. and {White}, N.~E. and
         {Barthelmy}, S.~D. and {Burrows}, D.~N. and {Cominsky}, L.~R. and
         {Hurley}, K.~C. and {Marshall}, F.~E. and {M{\'e}sz{\'a}ros}, P. and
         {Roming}, P.~W.~A. and {Angelini}, L. and {Barbier}, L.~M. and
         {Belloni}, T. and {Campana}, S. and {Caraveo}, P.~A. and
         {Chester}, M.~M. and {Citterio}, O. and {Cline}, T.~L. and
         {Cropper}, M.~S. and {Cummings}, J.~R. and {Dean}, A.~J. and
         {Feigelson}, E.~D. and {Fenimore}, E.~E. and {Frail}, D.~A. and
         {Fruchter}, A.~S. and {Garmire}, G.~P. and {Gendreau}, K. and
         {Ghisellini}, G. and {Greiner}, J. and {Hill}, J.~E. and
         {Hunsberger}, S.~D. and {Krimm}, H.~A. and {Kulkarni}, S.~R. and
         {Kumar}, P. and {Lebrun}, F. and {Lloyd-Ronning}, N.~M. and
         {Markwardt}, C.~B. and {Mattson}, B.~J. and {Mushotzky}, R.~F. and
         {Norris}, J.~P. and {Osborne}, J. and {Paczynski}, B. and
         {Palmer}, D.~M. and {Park}, H. -S. and {Parsons}, A.~M. and {Paul}, J. and
         {Rees}, M.~J. and {Reynolds}, C.~S. and {Rhoads}, J.~E. and
         {Sasseen}, T.~P. and {Schaefer}, B.~E. and {Short}, A.~T. and
         {Smale}, A.~P. and {Smith}, I.~A. and {Stella}, L. and
         {Tagliaferri}, G. and {Takahashi}, T. and {Tashiro}, M. and
         {Townsley}, L.~K. and {Tueller}, J. and {Turner}, M.~J.~L. and
         {Vietri}, M. and {Voges}, W. and {Ward}, M.~J. and {Willingale}, R. and
         {Zerbi}, F.~M. and {Zhang}, W.~W.},
        title = "{The Swift Gamma-Ray Burst Mission}",
      journal = {\apj},
         year = "2004",
        month = "Aug",
       volume = {611},
       number = {2},
        pages = {1005-1020},
          doi = {10.1086/422091},
archivePrefix = {arXiv},
       eprint = {astro-ph/0405233},
 primaryClass = {astro-ph},
       adsurl = {https://ui.adsabs.harvard.edu/abs/2004ApJ...611.1005G}
}

@ARTICLE{Roming:2005,
       author = {{Roming}, Peter W.~A. and {Kennedy}, Thomas E. and {Mason}, Keith O. and
         {Nousek}, John A. and {Ahr}, Lindy and {Bingham}, Richard E. and
         {Broos}, Patrick S. and {Carter}, Mary J. and {Hancock}, Barry K. and
         {Huckle}, Howard E. and {Hunsberger}, S.~D. and {Kawakami}, Hajime and
         {Killough}, Ronnie and {Koch}, T. Scott and {McLelland}, Michael K. and
         {Smith}, Kelly and {Smith}, Philip J. and {Soto}, Juan Carlos and
         {Boyd}, Patricia T. and {Breeveld}, Alice A. and {Holland}, Stephen T. and
         {Ivanushkina}, Mariya and {Pryzby}, Michael S. and {Still}, Martin D. and
         {Stock}, Joseph},
        title = "{The Swift Ultra-Violet/Optical Telescope}",
      journal = {\ssr},
         year = 2005,
        month = oct,
       volume = {120},
       number = {3-4},
        pages = {95-142},
          doi = {10.1007/s11214-005-5095-4},
archivePrefix = {arXiv},
       eprint = {astro-ph/0507413},
 primaryClass = {astro-ph},
       adsurl = {https://ui.adsabs.harvard.edu/abs/2005SSRv..120...95R}
}

@article{Astropy-Collaboration:2013aa,
	Author = {{Astropy Collaboration} and {Robitaille}, T.~P. and {Tollerud}, E.~J. and {Greenfield}, P. and {Droettboom}, M. and {Bray}, E. and {Aldcroft}, T. and {Davis}, M. and {Ginsburg}, A. and {Price-Whelan}, A.~M. and {Kerzendorf}, W.~E. and {Conley}, A. and {Crighton}, N. and {Barbary}, K. and {Muna}, D. and {Ferguson}, H. and {Grollier}, F. and {Parikh}, M.~M. and {Nair}, P.~H. and {Unther}, H.~M. and {Deil}, C. and {Woillez}, J. and {Conseil}, S. and {Kramer}, R. and {Turner}, J.~E.~H. and {Singer}, L. and {Fox}, R. and {Weaver}, B.~A. and {Zabalza}, V. and {Edwards}, Z.~I. and {Azalee Bostroem}, K. and {Burke}, D.~J. and {Casey}, A.~R. and {Crawford}, S.~M. and {Dencheva}, N. and {Ely}, J. and {Jenness}, T. and {Labrie}, K. and {Lim}, P.~L. and {Pierfederici}, F. and {Pontzen}, A. and {Ptak}, A. and {Refsdal}, B. and {Servillat}, M. and {Streicher}, O.},
	Journal = {\aap},
	Month = oct,
	Pages = {A33},
	Title = {{Astropy: A community Python package for astronomy}},
	Volume = 558,
	Year = 2013}

@article{Hunter:2007aa,
	Author = {{Hunter, J. D.}},
	Journal = {{Computing In Science \& Engineering}},
	Number = {3},
	Pages = {90--95},
	Title = {{Matplotlib: A 2D graphics environment}},
	Volume = {9},
	Year = {2007}}

@ARTICLE{donnan2021,
       author = {{Donnan}, Fergus R. and {Horne}, Keith and {Hern{\'a}ndez Santisteban}, Juan V.},
        title = "{Bayesian analysis of quasar light curves with a running optimal average: new time delay measurements of COSMOGRAIL gravitationally lensed quasars}",
      journal = {\mnras},
         year = 2021,
        month = dec,
       volume = {508},
       number = {4},
        pages = {5449-5467},
          doi = {10.1093/mnras/stab2832},
archivePrefix = {arXiv},
       eprint = {2107.12318},
 primaryClass = {astro-ph.IM},
       adsurl = {https://ui.adsabs.harvard.edu/abs/2021MNRAS.508.5449D}
}

@ARTICLE{peterson2004,
       author = {{Peterson}, B.~M. and {Ferrarese}, L. and {Gilbert}, K.~M. and {Kaspi}, S. and {Malkan}, M.~A. and {Maoz}, D. and {Merritt}, D. and {Netzer}, H. and {Onken}, C.~A. and {Pogge}, R.~W. and {Vestergaard}, M. and {Wandel}, A.},
        title = "{Central Masses and Broad-Line Region Sizes of Active Galactic Nuclei. II. A Homogeneous Analysis of a Large Reverberation-Mapping Database}",
      journal = {\apj},
         year = 2004,
        month = oct,
       volume = {613},
       number = {2},
        pages = {682-699},
          doi = {10.1086/423269},
archivePrefix = {arXiv},
       eprint = {astro-ph/0407299},
 primaryClass = {astro-ph},
       adsurl = {https://ui.adsabs.harvard.edu/abs/2004ApJ...613..682P}
}

@misc{sun2018,
       author = {{Sun}, Mouyuan and {Grier}, C.~J. and {Peterson}, B.~M.},
        title = "{PyCCF: Python Cross Correlation Function for reverberation mapping studies}",
 howpublished = {Astrophysics Source Code Library, record ascl:1805.032},
         year = 2018,
        month = may,
          eid = {ascl:1805.032},
       adsurl = {https://ui.adsabs.harvard.edu/abs/2018ascl.soft05032S}
}

@ARTICLE{drewes2025,
       author = {{Drewes}, Farin and {Leftley}, James H. and {H{\"o}nig}, Sebastian F. and {Tristram}, Konrad R.~W. and {Kishimoto}, Makoto},
        title = "{I Zw 1 and H0557-385: the dusty tori of two high Eddington AGNs observed in the MATISSE LM bands}",
      journal = {\mnras},
         year = 2025,
        month = feb,
       volume = {537},
       number = {2},
        pages = {1369-1384},
          doi = {10.1093/mnras/staf110},
       adsurl = {https://ui.adsabs.harvard.edu/abs/2025MNRAS.537.1369D}
}

@ARTICLE{gaskell2004,
       author = {{Gaskell}, C. Martin and {Goosmann}, Ren{\'e} W. and {Antonucci}, Robert R.~J. and {Whysong}, David H.},
        title = "{The Nuclear Reddening Curve for Active Galactic Nuclei and the Shape of the Infrared to X-Ray Spectral Energy Distribution}",
      journal = {\apj},
         year = 2004,
        month = nov,
       volume = {616},
       number = {1},
        pages = {147-156},
          doi = {10.1086/423885},
archivePrefix = {arXiv},
       eprint = {astro-ph/0309595},
 primaryClass = {astro-ph},
       adsurl = {https://ui.adsabs.harvard.edu/abs/2004ApJ...616..147G}
}

@ARTICLE{fitzpatrick1999,
       author = {{Fitzpatrick}, Edward L.},
        title = "{Correcting for the Effects of Interstellar Extinction}",
      journal = {\pasp},
         year = 1999,
        month = jan,
       volume = {111},
       number = {755},
        pages = {63-75},
          doi = {10.1086/316293},
archivePrefix = {arXiv},
       eprint = {astro-ph/9809387},
 primaryClass = {astro-ph},
       adsurl = {https://ui.adsabs.harvard.edu/abs/1999PASP..111...63F}
}

@ARTICLE{gordon2003,
       author = {{Gordon}, Karl D. and {Clayton}, Geoffrey C. and {Misselt}, K.~A. and {Landolt}, Arlo U. and {Wolff}, Michael J.},
        title = "{A Quantitative Comparison of the Small Magellanic Cloud, Large Magellanic Cloud, and Milky Way Ultraviolet to Near-Infrared Extinction Curves}",
      journal = {\apj},
         year = 2003,
        month = sep,
       volume = {594},
       number = {1},
        pages = {279-293},
          doi = {10.1086/376774},
archivePrefix = {arXiv},
       eprint = {astro-ph/0305257},
 primaryClass = {astro-ph},
       adsurl = {https://ui.adsabs.harvard.edu/abs/2003ApJ...594..279G}
}

@ARTICLE{chiar2006,
       author = {{Chiar}, J.~E. and {Tielens}, A.~G.~G.~M.},
        title = "{Pixie Dust: The Silicate Features in the Diffuse Interstellar Medium}",
      journal = {\apj},
         year = 2006,
        month = feb,
       volume = {637},
       number = {2},
        pages = {774-785},
          doi = {10.1086/498406},
archivePrefix = {arXiv},
       eprint = {astro-ph/0510156},
 primaryClass = {astro-ph},
       adsurl = {https://ui.adsabs.harvard.edu/abs/2006ApJ...637..774C}
}

@ARTICLE{schlafly2011,
       author = {{Schlafly}, Edward F. and {Finkbeiner}, Douglas P.},
        title = "{Measuring Reddening with Sloan Digital Sky Survey Stellar Spectra and Recalibrating SFD}",
      journal = {\apj},
         year = 2011,
        month = aug,
       volume = {737},
       number = {2},
          eid = {103},
        pages = {103},
          doi = {10.1088/0004-637X/737/2/103},
archivePrefix = {arXiv},
       eprint = {1012.4804},
 primaryClass = {astro-ph.GA},
       adsurl = {https://ui.adsabs.harvard.edu/abs/2011ApJ...737..103S}
}

@ARTICLE{gravity2020,
       author = {{GRAVITY Collaboration} and {Dexter}, J. and {Shangguan}, J. and {H{\"o}nig}, S. and {Kishimoto}, M. and {Lutz}, D. and {Netzer}, H. and {Davies}, R. and {Sturm}, E. and {Pfuhl}, O. and {Amorim}, A. and {Baub{\"o}ck}, M. and {Brandner}, W. and {Cl{\'e}net}, Y. and {de Zeeuw}, P.~T. and {Eckart}, A. and {Eisenhauer}, F. and {F{\"o}rster Schreiber}, N.~M. and {Gao}, F. and {Garcia}, P.~J.~V. and {Genzel}, R. and {Gillessen}, S. and {Gratadour}, D. and {Jim{\'e}nez-Rosales}, A. and {Lacour}, S. and {Millour}, F. and {Ott}, T. and {Paumard}, T. and {Perraut}, K. and {Perrin}, G. and {Peterson}, B.~M. and {Petrucci}, P.~O. and {Prieto}, M.~A. and {Rouan}, D. and {Schartmann}, M. and {Shimizu}, T. and {Sternberg}, A. and {Straub}, O. and {Straubmeier}, C. and {Tacconi}, L.~J. and {Tristram}, K. and {Vermot}, P. and {Waisberg}, I. and {Widmann}, F. and {Woillez}, J.},
        title = "{The resolved size and structure of hot dust in the immediate vicinity of AGN}",
      journal = {\aap},
         year = 2020,
        month = mar,
       volume = {635},
          eid = {A92},
        pages = {A92},
          doi = {10.1051/0004-6361/201936767},
archivePrefix = {arXiv},
       eprint = {1910.00593},
 primaryClass = {astro-ph.GA},
       adsurl = {https://ui.adsabs.harvard.edu/abs/2020A&A...635A..92G}
}

@ARTICLE{lyndenbell1969,
       author = {{Lynden-Bell}, D.},
        title = "{Galactic Nuclei as Collapsed Old Quasars}",
      journal = {\nat},
         year = 1969,
        month = aug,
       volume = {223},
       number = {5207},
        pages = {690-694},
          doi = {10.1038/223690a0},
       adsurl = {https://ui.adsabs.harvard.edu/abs/1969Natur.223..690L}
}

@ARTICLE{cai2018,
       author = {{Cai}, Zhen-Yi and {Wang}, Jun-Xian and {Zhu}, Fei-Fan and {Sun}, Mou-Yuan and {Gu}, Wei-Min and {Cao}, Xin-Wu and {Yuan}, Feng},
        title = "{EUCLIA{\textemdash}Exploring the UV/Optical Continuum Lag in Active Galactic Nuclei. I. A Model without Light Echoing}",
      journal = {\apj},
         year = 2018,
        month = mar,
       volume = {855},
       number = {2},
          eid = {117},
        pages = {117},
          doi = {10.3847/1538-4357/aab091},
archivePrefix = {arXiv},
       eprint = {1711.06266},
 primaryClass = {astro-ph.HE},
       adsurl = {https://ui.adsabs.harvard.edu/abs/2018ApJ...855..117C}
}

@ARTICLE{hagen2024,
       author = {{Hagen}, Scott and {Done}, Chris and {Edelson}, Rick},
        title = "{What drives the variability in AGN? Explaining the UV-Xray disconnect through propagating fluctuations}",
      journal = {\mnras},
         year = 2024,
        month = jun,
       volume = {530},
       number = {4},
        pages = {4850-4867},
          doi = {10.1093/mnras/stae1177},
archivePrefix = {arXiv},
       eprint = {2401.03452},
 primaryClass = {astro-ph.HE},
       adsurl = {https://ui.adsabs.harvard.edu/abs/2024MNRAS.530.4850H}
}

@ARTICLE{trakhtenbrot2017,
       author = {{Trakhtenbrot}, Benny and {Ricci}, Claudio and {Koss}, Michael J. and {Schawinski}, Kevin and {Mushotzky}, Richard and {Ueda}, Yoshihiro and {Veilleux}, Sylvain and {Lamperti}, Isabella and {Oh}, Kyuseok and {Treister}, Ezequiel and {Stern}, Daniel and {Harrison}, Fiona and {Balokovi{\'c}}, Mislav and {Gehrels}, Neil},
        title = "{BAT AGN Spectroscopic Survey (BASS) - VI. The {\ensuremath{\Gamma}}$_{X}$-L/L$_{Edd}$ relation}",
      journal = {\mnras},
         year = 2017,
        month = sep,
       volume = {470},
       number = {1},
        pages = {800-814},
          doi = {10.1093/mnras/stx1117},
archivePrefix = {arXiv},
       eprint = {1705.01550},
 primaryClass = {astro-ph.GA},
       adsurl = {https://ui.adsabs.harvard.edu/abs/2017MNRAS.470..800T}
}

@ARTICLE{du2015,
       author = {{Du}, Pu and {Hu}, Chen and {Lu}, Kai-Xing and {Huang}, Ying-Ke and {Cheng}, Cheng and {Qiu}, Jie and {Li}, Yan-Rong and {Zhang}, Yang-Wei and {Fan}, Xu-Liang and {Bai}, Jin-Ming and {Bian}, Wei-Hao and {Yuan}, Ye-Fei and {Kaspi}, Shai and {Ho}, Luis C. and {Netzer}, Hagai and {Wang}, Jian-Min and {SEAMBH Collaboration}},
        title = "{Supermassive Black Holes with High Accretion Rates in Active Galactic Nuclei. IV. H{\ensuremath{\beta}} Time Lags and Implications for Super-Eddington Accretion}",
      journal = {\apj},
         year = 2015,
        month = jun,
       volume = {806},
       number = {1},
          eid = {22},
        pages = {22},
          doi = {10.1088/0004-637X/806/1/22},
archivePrefix = {arXiv},
       eprint = {1504.01844},
 primaryClass = {astro-ph.GA},
       adsurl = {https://ui.adsabs.harvard.edu/abs/2015ApJ...806...22D}
}

@ARTICLE{uttley2014,
       author = {{Uttley}, P. and {Cackett}, E.~M. and {Fabian}, A.~C. and {Kara}, E. and {Wilkins}, D.~R.},
        title = "{X-ray reverberation around accreting black holes}",
      journal = {\aapr},
         year = 2014,
        month = aug,
       volume = {22},
          eid = {72},
        pages = {72},
          doi = {10.1007/s00159-014-0072-0},
archivePrefix = {arXiv},
       eprint = {1405.6575},
 primaryClass = {astro-ph.HE},
       adsurl = {https://ui.adsabs.harvard.edu/abs/2014A&ARv..22...72U}
}

@misc{mccully2018,
       author = {{McCully}, Curtis and {Crawford}, Steve and {Kovacs}, Gabor and {Tollerud}, Erik and {Betts}, Edward and {Bradley}, Larry and {Craig}, Matt and {Turner}, James and {Streicher}, Ole and {Sipocz}, Brigitta and {Robitaille}, Thomas and {Deil}, Christoph},
        title = "{astropy/astroscrappy: v1.0.5 Zenodo Release}",
         year = 2018,
        month = nov,
          eid = {10.5281/zenodo.1482019},
          doi = {10.5281/zenodo.1482019},
      version = {v1.0.5},
    publisher = {Zenodo},
       adsurl = {https://ui.adsabs.harvard.edu/abs/2018zndo...1482019M}
}

@ARTICLE{bertin1996,
       author = {{Bertin}, E. and {Arnouts}, S.},
        title = "{SExtractor: Software for source extraction.}",
      journal = {\aaps},
         year = 1996,
        month = jun,
       volume = {117},
        pages = {393-404},
          doi = {10.1051/aas:1996164},
       adsurl = {https://ui.adsabs.harvard.edu/abs/1996A&AS..117..393B}
}

@ARTICLE{jansen2001,
       author = {{Jansen}, F. and {Lumb}, D. and {Altieri}, B. and {Clavel}, J. and {Ehle}, M. and {Erd}, C. and {Gabriel}, C. and {Guainazzi}, M. and {Gondoin}, P. and {Much}, R. and {Munoz}, R. and {Santos}, M. and {Schartel}, N. and {Texier}, D. and {Vacanti}, G.},
        title = "{XMM-Newton observatory. I. The spacecraft and operations}",
      journal = {\aap},
         year = 2001,
        month = jan,
       volume = {365},
        pages = {L1-L6},
          doi = {10.1051/0004-6361:20000036},
       adsurl = {https://ui.adsabs.harvard.edu/abs/2001A&A...365L...1J}
}

@ARTICLE{mason2001,
       author = {{Mason}, K.~O. and {Breeveld}, A. and {Much}, R. and {Carter}, M. and {Cordova}, F.~A. and {Cropper}, M.~S. and {Fordham}, J. and {Huckle}, H. and {Ho}, C. and {Kawakami}, H. and {Kennea}, J. and {Kennedy}, T. and {Mittaz}, J. and {Pandel}, D. and {Priedhorsky}, W.~C. and {Sasseen}, T. and {Shirey}, R. and {Smith}, P. and {Vreux}, J. -M.},
        title = "{The XMM-Newton optical/UV monitor telescope}",
      journal = {\aap},
         year = 2001,
        month = jan,
       volume = {365},
        pages = {L36-L44},
          doi = {10.1051/0004-6361:20000044},
archivePrefix = {arXiv},
       eprint = {astro-ph/0011216},
 primaryClass = {astro-ph},
       adsurl = {https://ui.adsabs.harvard.edu/abs/2001A&A...365L..36M}
}

@ARTICLE{rudy2000,
       author = {{Rudy}, Richard J. and {Mazuk}, S. and {Puetter}, R.~C. and {Hamann}, F.},
        title = "{The 1 Micron Fe II Lines of the Seyfert Galaxy I Zw 1}",
      journal = {\apj},
         year = 2000,
        month = aug,
       volume = {539},
       number = {1},
        pages = {166-171},
          doi = {10.1086/309222},
       adsurl = {https://ui.adsabs.harvard.edu/abs/2000ApJ...539..166R}
}

@INPROCEEDINGS{henden2018,
       author = {{Henden}, Arne A. and {Levine}, Stephen and {Terrell}, Dirk and {Welch}, Douglas L. and {Munari}, Ulisse and {Kloppenborg}, Brian K.},
        title = "{APASS Data Release 10}",
    booktitle = {American Astronomical Society Meeting Abstracts \#232},
         year = 2018,
       series = {American Astronomical Society Meeting Abstracts},
       volume = {232},
        month = jun,
          eid = {223.06},
        pages = {223.06},
       adsurl = {https://ui.adsabs.harvard.edu/abs/2018AAS...23222306H}
}

@ARTICLE{ahumada2020,
       author = {{Ahumada}, Romina and {Allende Prieto}, Carlos and {Almeida}, Andr{\'e}s and {Anders}, Friedrich and {Anderson}, Scott F. and {Andrews}, Brett H. and {Anguiano}, Borja and {Arcodia}, Riccardo and {Armengaud}, Eric and {Aubert}, Marie and {Avila}, Santiago and {Avila-Reese}, Vladimir and {Badenes}, Carles and {Balland}, Christophe and {Barger}, Kat and {Barrera-Ballesteros}, Jorge K. and {Basu}, Sarbani and {Bautista}, Julian and {Beaton}, Rachael L. and {Beers}, Timothy C. and {Benavides}, B. Izamar T. and {Bender}, Chad F. and {Bernardi}, Mariangela and {Bershady}, Matthew and {Beutler}, Florian and {Bidin}, Christian Moni and {Bird}, Jonathan and {Bizyaev}, Dmitry and {Blanc}, Guillermo A. and {Blanton}, Michael R. and {Boquien}, M{\'e}d{\'e}ric and {Borissova}, Jura and {Bovy}, Jo and {Brandt}, W.~N. and {Brinkmann}, Jonathan and {Brownstein}, Joel R. and {Bundy}, Kevin and {Bureau}, Martin and {Burgasser}, Adam and {Burtin}, Etienne and {Cano-D{\'\i}az}, Mariana and {Capasso}, Raffaella and {Cappellari}, Michele and {Carrera}, Ricardo and {Chabanier}, Sol{\`e}ne and {Chaplin}, William and {Chapman}, Michael and {Cherinka}, Brian and {Chiappini}, Cristina and {Doohyun Choi}, Peter and {Chojnowski}, S. Drew and {Chung}, Haeun and {Clerc}, Nicolas and {Coffey}, Damien and {Comerford}, Julia M. and {Comparat}, Johan and {da Costa}, Luiz and {Cousinou}, Marie-Claude and {Covey}, Kevin and {Crane}, Jeffrey D. and {Cunha}, Katia and {Ilha}, Gabriele da Silva and {Dai}, Yu Sophia and {Damsted}, Sanna B. and {Darling}, Jeremy and {Davidson}, Jr., James W. and {Davies}, Roger and {Dawson}, Kyle and {De}, Nikhil and {de la Macorra}, Axel and {De Lee}, Nathan and {Queiroz}, Anna B{\'a}rbara de Andrade and {Deconto Machado}, Alice and {de la Torre}, Sylvain and {Dell'Agli}, Flavia and {du Mas des Bourboux}, H{\'e}lion and {Diamond-Stanic}, Aleksandar M. and {Dillon}, Sean and {Donor}, John and {Drory}, Niv and {Duckworth}, Chris and {Dwelly}, Tom and {Ebelke}, Garrett and {Eftekharzadeh}, Sarah and {Davis Eigenbrot}, Arthur and {Elsworth}, Yvonne P. and {Eracleous}, Mike and {Erfanianfar}, Ghazaleh and {Escoffier}, Stephanie and {Fan}, Xiaohui and {Farr}, Emily and {Fern{\'a}ndez-Trincado}, Jos{\'e} G. and {Feuillet}, Diane and {Finoguenov}, Alexis and {Fofie}, Patricia and {Fraser-McKelvie}, Amelia and {Frinchaboy}, Peter M. and {Fromenteau}, Sebastien and {Fu}, Hai and {Galbany}, Llu{\'\i}s and {Garcia}, Rafael A. and {Garc{\'\i}a-Hern{\'a}ndez}, D.~A. and {Garma Oehmichen}, Luis Alberto and {Ge}, Junqiang and {Geimba Maia}, Marcio Antonio and {Geisler}, Doug and {Gelfand}, Joseph and {Goddy}, Julian and {Gonzalez-Perez}, Violeta and {Grabowski}, Kathleen and {Green}, Paul and {Grier}, Catherine J. and {Guo}, Hong and {Guy}, Julien and {Harding}, Paul and {Hasselquist}, Sten and {Hawken}, Adam James and {Hayes}, Christian R. and {Hearty}, Fred and {Hekker}, S. and {Hogg}, David W. and {Holtzman}, Jon A. and {Horta}, Danny and {Hou}, Jiamin and {Hsieh}, Bau-Ching and {Huber}, Daniel and {Hunt}, Jason A.~S. and {Ider Chitham}, J. and {Imig}, Julie and {Jaber}, Mariana and {Jimenez Angel}, Camilo Eduardo and {Johnson}, Jennifer A. and {Jones}, Amy M. and {J{\"o}nsson}, Henrik and {Jullo}, Eric and {Kim}, Yerim and {Kinemuchi}, Karen and {Kirkpatrick}, IV, Charles C. and {Kite}, George W. and {Klaene}, Mark and {Kneib}, Jean-Paul and {Kollmeier}, Juna A. and {Kong}, Hui and {Kounkel}, Marina and {Krishnarao}, Dhanesh and {Lacerna}, Ivan and {Lan}, Ting-Wen and {Lane}, Richard R. and {Law}, David R. and {Le Goff}, Jean-Marc and {Leung}, Henry W. and {Lewis}, Hannah and {Li}, Cheng and {Lian}, Jianhui and {Lin}, Lihwai and {Long}, Dan and {Longa-Pe{\~n}a}, Pen{\'e}lope and {Lundgren}, Britt and {Lyke}, Brad W. and {Mackereth}, J. Ted and {MacLeod}, Chelsea L. and {Majewski}, Steven R. and {Manchado}, Arturo and {Maraston}, Claudia and {Martini}, Paul and {Masseron}, Thomas and {Masters}, Karen L. and {Mathur}, Savita and {McDermid}, Richard M. and {Merloni}, Andrea and {Merrifield}, Michael and {M{\'e}sz{\'a}ros}, Szabolcs and {Miglio}, Andrea and {Minniti}, Dante and {Minsley}, Rebecca and {Miyaji}, Takamitsu and {Mohammad}, Faizan Gohar and {Mosser}, Benoit and {Mueller}, Eva-Maria and {Muna}, Demitri and {Mu{\~n}oz-Guti{\'e}rrez}, Andrea and {Myers}, Adam D. and {Nadathur}, Seshadri and {Nair}, Preethi and {Nandra}, Kirpal and {Correa do Nascimento}, Janaina and {Nevin}, Rebecca Jean and {Newman}, Jeffrey A. and {Nidever}, David L. and {Nitschelm}, Christian and {Noterdaeme}, Pasquier and {O'Connell}, Julia E. and {Olmstead}, Matthew D. and {Oravetz}, Daniel and {Oravetz}, Audrey and {Osorio}, Yeisson and {Pace}, Zachary J. and {Padilla}, Nelson and {Palanque-Delabrouille}, Nathalie and {Palicio}, Pedro A.},
        title = "{The 16th Data Release of the Sloan Digital Sky Surveys: First Release from the APOGEE-2 Southern Survey and Full Release of eBOSS Spectra}",
      journal = {\apjs},
         year = 2020,
        month = jul,
       volume = {249},
       number = {1},
          eid = {3},
        pages = {3},
          doi = {10.3847/1538-4365/ab929e},
archivePrefix = {arXiv},
       eprint = {1912.02905},
 primaryClass = {astro-ph.GA},
       adsurl = {https://ui.adsabs.harvard.edu/abs/2020ApJS..249....3A}
}

@ARTICLE{silva2018,
       author = {{Silva}, C.~V. and {Costantini}, E. and {Giustini}, M. and {Kriss}, G.~A. and {Brandt}, W.~N. and {Gallo}, L.~C. and {Wilkins}, D.~R.},
        title = "{The variability of the warm absorber in I Zwicky 1 as seen by XMM-Newton}",
      journal = {\mnras},
         year = 2018,
        month = oct,
       volume = {480},
       number = {2},
        pages = {2334-2342},
          doi = {10.1093/mnras/sty1938},
archivePrefix = {arXiv},
       eprint = {1807.07294},
 primaryClass = {astro-ph.GA},
       adsurl = {https://ui.adsabs.harvard.edu/abs/2018MNRAS.480.2334S}
}

@ARTICLE{gravity2024,
       author = {{GRAVITY Collaboration} and {Amorim}, A. and {Bourdarot}, G. and {Brandner}, W. and {Cao}, Y. and {Cl{\'e}net}, Y. and {Davies}, R. and {de Zeeuw}, P.~T. and {Dexter}, J. and {Drescher}, A. and {Eckart}, A. and {Eisenhauer}, F. and {Fabricius}, M. and {Feuchtgruber}, H. and {F{\"o}rster Schreiber}, N.~M. and {Garcia}, P.~J.~V. and {Genzel}, R. and {Gillessen}, S. and {Gratadour}, D. and {H{\"o}nig}, S. and {Kishimoto}, M. and {Lacour}, S. and {Lutz}, D. and {Millour}, F. and {Netzer}, H. and {Ott}, T. and {Perraut}, K. and {Perrin}, G. and {Peterson}, B.~M. and {Petrucci}, P.~O. and {Pfuhl}, O. and {Prieto}, A. and {Rabien}, S. and {Rouan}, D. and {Santos}, D.~J.~D. and {Shangguan}, J. and {Shimizu}, T. and {Sternberg}, A. and {Straubmeier}, C. and {Sturm}, E. and {Tacconi}, L.~J. and {Tristram}, K.~R.~W. and {Widmann}, F. and {Woillez}, J.},
        title = "{VLTI/GRAVITY interferometric measurements of the innermost dust structure sizes around active galactic nuclei}",
      journal = {\aap},
         year = 2024,
        month = oct,
       volume = {690},
          eid = {A76},
        pages = {A76},
          doi = {10.1051/0004-6361/202450746},
archivePrefix = {arXiv},
       eprint = {2407.13458},
 primaryClass = {astro-ph.GA},
       adsurl = {https://ui.adsabs.harvard.edu/abs/2024A&A...690A..76G}
}

@ARTICLE{lewin2023,
       author = {{Lewin}, Collin and {Kara}, Erin and {Cackett}, Edward M. and {Wilkins}, Dan and {Panagiotou}, Christos and {Garc{\'\i}a}, Javier A. and {Gelbord}, Jonathan},
        title = "{X-Ray/UVOIR Frequency-resolved Time Lag Analysis of Mrk 335 Reveals Accretion Disk Reprocessing}",
      journal = {\apj},
         year = 2023,
        month = sep,
       volume = {954},
       number = {1},
          eid = {33},
        pages = {33},
          doi = {10.3847/1538-4357/ace77b},
archivePrefix = {arXiv},
       eprint = {2307.11145},
 primaryClass = {astro-ph.HE},
       adsurl = {https://ui.adsabs.harvard.edu/abs/2023ApJ...954...33L}
}

@ARTICLE{thorne2025,
       author = {{Thorne}, James P. and {Landt}, Hermine and {Huang}, Jiamu and {Hern{\'a}ndez Santisteban}, Juan V. and {Horne}, Keith and {Cackett}, Edward M. and {Winkler}, Hartmut and {Sanmartim}, David},
        title = "{Accretion disc reverberation mapping of the quasar 3C 273}",
      journal = {\mnras},
         year = 2025,
        month = mar,
       volume = {537},
       number = {4},
        pages = {3746-3768},
          doi = {10.1093/mnras/staf258},
archivePrefix = {arXiv},
       eprint = {2502.08366},
 primaryClass = {astro-ph.GA},
       adsurl = {https://ui.adsabs.harvard.edu/abs/2025MNRAS.537.3746T}
}

@ARTICLE{bentz2021,
       author = {{Bentz}, Misty C. and {Williams}, Peter R. and {Street}, Rachel and {Onken}, Christopher A. and {Valluri}, Monica and {Treu}, Tommaso},
        title = "{A Detailed View of the Broad-line Region in NGC 3783 from Velocity-resolved Reverberation Mapping}",
      journal = {\apj},
         year = 2021,
        month = oct,
       volume = {920},
       number = {2},
          eid = {112},
        pages = {112},
          doi = {10.3847/1538-4357/ac19af},
archivePrefix = {arXiv},
       eprint = {2108.00482},
 primaryClass = {astro-ph.GA},
       adsurl = {https://ui.adsabs.harvard.edu/abs/2021ApJ...920..112B}
}

@ARTICLE{gravity2018,
       author = {{GRAVITY Collaboration} and {Sturm}, E. and {Dexter}, J. and {Pfuhl}, O. and {Stock}, M.~R. and {Davies}, R.~I. and {Lutz}, D. and {Cl{\'e}net}, Y. and {Eckart}, A. and {Eisenhauer}, F. and {Genzel}, R. and {Gratadour}, D. and {H{\"o}nig}, S.~F. and {Kishimoto}, M. and {Lacour}, S. and {Millour}, F. and {Netzer}, H. and {Perrin}, G. and {Peterson}, B.~M. and {Petrucci}, P.~O. and {Rouan}, D. and {Waisberg}, I. and {Woillez}, J. and {Amorim}, A. and {Brandner}, W. and {F{\"o}rster Schreiber}, N.~M. and {Garcia}, P.~J.~V. and {Gillessen}, S. and {Ott}, T. and {Paumard}, T. and {Perraut}, K. and {Scheithauer}, S. and {Straubmeier}, C. and {Tacconi}, L.~J. and {Widmann}, F.},
        title = "{Spatially resolved rotation of the broad-line region of a quasar at sub-parsec scale}",
      journal = {\nat},
         year = 2018,
        month = nov,
       volume = {563},
       number = {7733},
        pages = {657-660},
          doi = {10.1038/s41586-018-0731-9},
archivePrefix = {arXiv},
       eprint = {1811.11195},
 primaryClass = {astro-ph.GA},
       adsurl = {https://ui.adsabs.harvard.edu/abs/2018Natur.563..657G}
}

@ARTICLE{laor1997,
       author = {{Laor}, Ari and {Jannuzi}, Buell T. and {Green}, Richard F. and {Boroson}, Todd A.},
        title = "{The Ultraviolet Properties of the Narrow-Line Quasar I Zw 1}",
      journal = {\apj},
         year = 1997,
        month = nov,
       volume = {489},
       number = {2},
        pages = {656-671},
          doi = {10.1086/304816},
archivePrefix = {arXiv},
       eprint = {astro-ph/9706264},
 primaryClass = {astro-ph},
       adsurl = {https://ui.adsabs.harvard.edu/abs/1997ApJ...489..656L}
}

@ARTICLE{lobban2022,
       author = {{Lobban}, Andrew and {King}, Andrew},
        title = "{AGN light echoes and the accretion disc self-gravity limit}",
      journal = {\mnras},
         year = 2022,
        month = apr,
       volume = {511},
       number = {2},
        pages = {1992-1998},
          doi = {10.1093/mnras/stac155},
archivePrefix = {arXiv},
       eprint = {2201.08375},
 primaryClass = {astro-ph.GA},
       adsurl = {https://ui.adsabs.harvard.edu/abs/2022MNRAS.511.1992L}
}

@ARTICLE{honig2019,
       author = {{H{\"o}nig}, Sebastian F.},
        title = "{Redefining the Torus: A Unifying View of AGNs in the Infrared and Submillimeter}",
      journal = {\apj},
         year = 2019,
        month = oct,
       volume = {884},
       number = {2},
          eid = {171},
        pages = {171},
          doi = {10.3847/1538-4357/ab4591},
archivePrefix = {arXiv},
       eprint = {1909.08639},
 primaryClass = {astro-ph.GA},
       adsurl = {https://ui.adsabs.harvard.edu/abs/2019ApJ...884..171H}
}

@ARTICLE{baskin2018,
       author = {{Baskin}, Alexei and {Laor}, Ari},
        title = "{Dust inflated accretion disc as the origin of the broad line region in active galactic nuclei}",
      journal = {\mnras},
         year = 2018,
        month = feb,
       volume = {474},
       number = {2},
        pages = {1970-1994},
          doi = {10.1093/mnras/stx2850},
archivePrefix = {arXiv},
       eprint = {1711.00025},
 primaryClass = {astro-ph.GA},
       adsurl = {https://ui.adsabs.harvard.edu/abs/2018MNRAS.474.1970B}
}

@ARTICLE{czerny2011,
       author = {{Czerny}, B. and {Hryniewicz}, K.},
        title = "{The origin of the broad line region in active galactic nuclei}",
      journal = {\aap},
         year = 2011,
        month = jan,
       volume = {525},
          eid = {L8},
        pages = {L8},
          doi = {10.1051/0004-6361/201016025},
archivePrefix = {arXiv},
       eprint = {1010.6201},
 primaryClass = {astro-ph.CO},
       adsurl = {https://ui.adsabs.harvard.edu/abs/2011A&A...525L...8C}
}

@ARTICLE{Morgan2010,
       author = {{Morgan}, Christopher W. and {Kochanek}, C.~S. and {Morgan}, Nicholas D. and {Falco}, Emilio E.},
        title = "{The Quasar Accretion Disk Size-Black Hole Mass Relation}",
      journal = {\apj},
         year = 2010,
        month = apr,
       volume = {712},
       number = {2},
        pages = {1129-1136},
          doi = {10.1088/0004-637X/712/2/1129},
archivePrefix = {arXiv},
       eprint = {1002.4160},
 primaryClass = {astro-ph.CO},
       adsurl = {https://ui.adsabs.harvard.edu/abs/2010ApJ...712.1129M}
}

@ARTICLE{Netzer2022,
       author = {{Netzer}, Hagai},
        title = "{Continuum reverberation mapping and a new lag-luminosity relationship for AGN}",
      journal = {\mnras},
         year = 2022,
        month = jan,
       volume = {509},
       number = {2},
        pages = {2637-2646},
          doi = {10.1093/mnras/stab3133},
archivePrefix = {arXiv},
       eprint = {2110.05512},
 primaryClass = {astro-ph.GA},
       adsurl = {https://ui.adsabs.harvard.edu/abs/2022MNRAS.509.2637N}
}

@ARTICLE{Starkey2016,
       author = {{Starkey}, D.~A. and {Horne}, Keith and {Villforth}, C.},
        title = "{Accretion disc time lag distributions: applying CREAM to simulated AGN light curves}",
      journal = {\mnras},
         year = 2016,
        month = feb,
       volume = {456},
       number = {2},
        pages = {1960-1973},
          doi = {10.1093/mnras/stv2744},
archivePrefix = {arXiv},
       eprint = {1511.06162},
 primaryClass = {astro-ph.GA},
       adsurl = {https://ui.adsabs.harvard.edu/abs/2016MNRAS.456.1960S}
}

@ARTICLE{Brown:2013,
       author = {{Brown}, T.~M. and {Baliber}, N. and {Bianco}, F.~B. and {Bowman}, M. and
         {Burleson}, B. and {Conway}, P. and {Crellin}, M. and
         {Depagne}, {\'E}. and {De Vera}, J. and {Dilday}, B. and
         {Dragomir}, D. and {Dubberley}, M. and {Eastman}, J.~D. and
         {Elphick}, M. and {Falarski}, M. and {Foale}, S. and {Ford}, M. and
         {Fulton}, B.~J. and {Garza}, J. and {Gomez}, E.~L. and {Graham}, M. and
         {Greene}, R. and {Haldeman}, B. and {Hawkins}, E. and {Haworth}, B. and
         {Haynes}, R. and {Hidas}, M. and {Hjelstrom}, A.~E. and
         {Howell}, D.~A. and {Hygelund}, J. and {Lister}, T.~A. and
         {Lobdill}, R. and {Martinez}, J. and {Mullins}, D.~S. and
         {Norbury}, M. and {Parrent}, J. and {Paulson}, R. and {Petry}, D.~L. and
         {Pickles}, A. and {Posner}, V. and {Rosing}, W.~E. and {Ross}, R. and
         {Sand}, D.~J. and {Saunders}, E.~S. and {Shobbrook}, J. and
         {Shporer}, A. and {Street}, R.~A. and {Thomas}, D. and {Tsapras}, Y. and
         {Tufts}, J.~R. and {Valenti}, S. and {Vander Horst}, K. and
         {Walker}, Z. and {White}, G. and {Willis}, M.},
        title = "{Las Cumbres Observatory Global Telescope Network}",
      journal = {\pasp},
         year = "2013",
        month = "Sep",
       volume = {125},
       number = {931},
        pages = {1031},
          doi = {10.1086/673168},
archivePrefix = {arXiv},
       eprint = {1305.2437},
 primaryClass = {astro-ph.IM},
       adsurl = {https://ui.adsabs.harvard.edu/abs/2013PASP..125.1031B}
}

@ARTICLE{Winkler1992,
       author = {{Winkler}, H. and {Glass}, I.~S. and {van Wyk}, F. and {Marang}, F. and {Jones}, J.~H.~S. and {Buckley}, D.~A.~H. and {Sekiguchi}, K.},
        title = "{Variability studies of seyfert galaxies - I. Broad-band optical photometry .}",
      journal = {\mnras},
         year = 1992,
        month = aug,
       volume = {257},
        pages = {659-676},
          doi = {10.1093/mnras/257.4.659},
       adsurl = {https://ui.adsabs.harvard.edu/abs/1992MNRAS.257..659W}
}

@ARTICLE{Winkler1997,
       author = {{Winkler}, H.},
        title = "{The extinction, flux distribution and luminosity of Seyfert 1 nuclei derived from UBV(RI)\_C aperture photometry}",
      journal = {\mnras},
         year = 1997,
        month = dec,
       volume = {292},
       number = {2},
        pages = {273-288},
          doi = {10.1093/mnras/292.2.273},
       adsurl = {https://ui.adsabs.harvard.edu/abs/1997MNRAS.292..273W}
}

@ARTICLE{Collier1999,
       author = {{Collier}, Stefan and {Horne}, Keith and {Wanders}, Ignaz and {Peterson}, Bradley M.},
        title = "{A new direct method for measuring the Hubble constant from reverberating accretion discs in active galaxies}",
      journal = {\mnras},
         year = 1999,
        month = jan,
       volume = {302},
       number = {1},
        pages = {L24-L28},
          doi = {10.1046/j.1365-8711.1999.02250.x},
archivePrefix = {arXiv},
       eprint = {astro-ph/9811278},
 primaryClass = {astro-ph},
       adsurl = {https://ui.adsabs.harvard.edu/abs/1999MNRAS.302L..24C}
}

@ARTICLE{Antonucci2015,
       author = {{Antonucci}, Robert},
        title = "{Active Galactic Nuclei and Quasars: Why Still a Puzzle after 50 years?}",
      journal = {arXiv e-prints},
         year = 2015,
        month = jan,
          eid = {arXiv:1501.02001},
        pages = {arXiv:1501.02001},
archivePrefix = {arXiv},
       eprint = {1501.02001},
 primaryClass = {astro-ph.GA},
       adsurl = {https://ui.adsabs.harvard.edu/abs/2015arXiv150102001A}
}

@ARTICLE{Lawrence2018,
       author = {{Lawrence}, Andy},
        title = "{Quasar viscosity crisis}",
      journal = {Nature Astronomy},
         year = 2018,
        month = feb,
       volume = {2},
        pages = {102-103},
          doi = {10.1038/s41550-017-0372-1},
archivePrefix = {arXiv},
       eprint = {1802.00408},
 primaryClass = {astro-ph.HE},
       adsurl = {https://ui.adsabs.harvard.edu/abs/2018NatAs...2..102L}
}

@ARTICLE{Kammoun2021,
       author = {{Kammoun}, E.~S. and {Dov{\v{c}}iak}, M. and {Papadakis}, I.~E. and {Caballero-Garc{\'\i}a}, M.~D. and {Karas}, V.},
        title = "{UV/Optical Disk Thermal Reverberation in Active Galactic Nuclei: An In-depth Study with an Analytic Prescription for Time-lag Spectra}",
      journal = {\apj},
         year = 2021,
        month = jan,
       volume = {907},
       number = {1},
          eid = {20},
        pages = {20},
          doi = {10.3847/1538-4357/abcb93},
archivePrefix = {arXiv},
       eprint = {2011.08563},
 primaryClass = {astro-ph.HE},
       adsurl = {https://ui.adsabs.harvard.edu/abs/2021ApJ...907...20K}
}

@ARTICLE{cackett2022,
       author = {{Cackett}, Edward M. and {Zoghbi}, Abderahmen and {Ulrich}, Otho},
        title = "{Frequency-resolved Lags in UV/Optical Continuum Reverberation Mapping}",
      journal = {\apj},
         year = 2022,
        month = jan,
       volume = {925},
       number = {1},
          eid = {29},
        pages = {29},
          doi = {10.3847/1538-4357/ac3913},
archivePrefix = {arXiv},
       eprint = {2109.02155},
 primaryClass = {astro-ph.GA},
       adsurl = {https://ui.adsabs.harvard.edu/abs/2022ApJ...925...29C}
}

@ARTICLE{edelson2019,
       author = {{Edelson}, R. and {Gelbord}, J. and {Cackett}, E. and {Peterson}, B.~M. and {Horne}, K. and {Barth}, A.~J. and {Starkey}, D.~A. and {Bentz}, M. and {Brandt}, W.~N. and {Goad}, M. and {Joner}, M. and {Korista}, K. and {Netzer}, H. and {Page}, K. and {Uttley}, P. and {Vaughan}, S. and {Breeveld}, A. and {Cenko}, S.~B. and {Done}, C. and {Evans}, P. and {Fausnaugh}, M. and {Ferland}, G. and {Gonzalez-Buitrago}, D. and {Gropp}, J. and {Grupe}, D. and {Kaastra}, J. and {Kennea}, J. and {Kriss}, G. and {Mathur}, S. and {Mehdipour}, M. and {Mudd}, D. and {Nousek}, J. and {Schmidt}, T. and {Vestergaard}, M. and {Villforth}, C.},
        title = "{The First Swift Intensive AGN Accretion Disk Reverberation Mapping Survey}",
      journal = {\apj},
         year = 2019,
        month = jan,
       volume = {870},
       number = {2},
          eid = {123},
        pages = {123},
          doi = {10.3847/1538-4357/aaf3b4},
archivePrefix = {arXiv},
       eprint = {1811.07956},
 primaryClass = {astro-ph.HE},
       adsurl = {https://ui.adsabs.harvard.edu/abs/2019ApJ...870..123E}
}

@ARTICLE{landt2023,
       author = {{Landt}, Hermine},
        title = "{The outer dusty edge of accretion disks in active galactic nuclei}",
      journal = {Frontiers in Astronomy and Space Sciences},
         year = 2023,
        month = sep,
       volume = {10},
          eid = {1256088},
        pages = {1256088},
          doi = {10.3389/fspas.2023.1256088},
archivePrefix = {arXiv},
       eprint = {2309.15931},
 primaryClass = {astro-ph.GA},
       adsurl = {https://ui.adsabs.harvard.edu/abs/2023FrASS..1056088L}
}

@ARTICLE{lyu2019,
       author = {{Lyu}, Jianwei and {Rieke}, George H. and {Smith}, Paul S.},
        title = "{Mid-IR Variability and Dust Reverberation Mapping of Low-z Quasars. I. Data, Methods, and Basic Results}",
      journal = {\apj},
         year = 2019,
        month = nov,
       volume = {886},
       number = {1},
          eid = {33},
        pages = {33},
          doi = {10.3847/1538-4357/ab481d},
archivePrefix = {arXiv},
       eprint = {1909.11101},
 primaryClass = {astro-ph.GA},
       adsurl = {https://ui.adsabs.harvard.edu/abs/2019ApJ...886...33L}
}

@ARTICLE{clavel1991,
       author = {{Clavel}, J. and {Reichert}, G.~A. and {Alloin}, D. and {Crenshaw}, D.~M. and {Kriss}, G. and {Krolik}, J.~H. and {Malkan}, M.~A. and {Netzer}, H. and {Peterson}, B.~M. and {Wamsteker}, W. and {Altamore}, A. and {Baribaud}, T. and {Barr}, P. and {Beck}, S. and {Binette}, L. and {Bromage}, G.~E. and {Brosch}, N. and {Diaz}, A.~I. and {Filippenko}, A.~V. and {Fricke}, K. and {Gaskell}, C.~M. and {Giommi}, P. and {Glass}, I.~S. and {Gondhalekar}, P. and {Hackney}, R.~L. and {Halpern}, J.~P. and {Hutter}, D.~J. and {Joersaeter}, S. and {Kinney}, A.~L. and {Kollatschny}, W. and {Koratkar}, A. and {Korista}, K.~T. and {Laor}, A. and {Lasota}, J. -P. and {Leibowitz}, E. and {Maoz}, D. and {Martin}, P.~G. and {Mazeh}, T. and {Meurs}, E.~J.~A. and {Nair}, A.~D. and {O'Brien}, P. and {Pelat}, D. and {Perez}, E. and {Perola}, G.~C. and {Ptak}, R.~L. and {Rodriguez-Pascual}, P. and {Rosenblatt}, E.~I. and {Sadun}, A.~C. and {Santos-Lleo}, M. and {Shaw}, R.~A. and {Smith}, P.~S. and {Stirpe}, G.~M. and {Stoner}, R. and {Sun}, W.~H. and {Ulrich}, M. -H. and {van Groningen}, E. and {Zheng}, W.},
        title = "{Steps toward Determination of the Size and Structure of the Broad-Line Region in Active Galactic Nuclei. I. an 8 Month Campaign of Monitoring NGC 5548 with IUE}",
      journal = {\apj},
         year = 1991,
        month = jan,
       volume = {366},
        pages = {64},
          doi = {10.1086/169540},
       adsurl = {https://ui.adsabs.harvard.edu/abs/1991ApJ...366...64C}
}

@ARTICLE{panagiotou2025,
       author = {{Panagiotou}, Christos and {Papadakis}, Iossif and {Kara}, Erin and {Papoutsis}, Marios and {Cackett}, Edward M. and {Dov{\v{c}}iak}, Michal and {Garc{\'\i}a}, Javier A. and {Kammoun}, Elias and {Lewin}, Collin},
        title = "{Frequency-resolved Time Lags due to X-Ray Disk Reprocessing in AGN}",
      journal = {\apj},
         year = 2025,
        month = apr,
       volume = {983},
       number = {2},
          eid = {132},
        pages = {132},
          doi = {10.3847/1538-4357/adbf95},
archivePrefix = {arXiv},
       eprint = {2503.09036},
 primaryClass = {astro-ph.GA},
       adsurl = {https://ui.adsabs.harvard.edu/abs/2025ApJ...983..132P}
}

@ARTICLE{yu2020,
       author = {{Yu}, Z. and {Kochanek}, C.~S. and {Peterson}, B.~M. and {Zu}, Y. and {Brandt}, W.~N. and {Cackett}, E.~M. and {Fausnaugh}, M.~M. and {McHardy}, I.~M.},
        title = "{On reverberation mapping lag uncertainties}",
      journal = {\mnras},
         year = 2020,
        month = feb,
       volume = {491},
       number = {4},
        pages = {6045-6064},
          doi = {10.1093/mnras/stz3464},
archivePrefix = {arXiv},
       eprint = {1909.03072},
 primaryClass = {astro-ph.GA},
       adsurl = {https://ui.adsabs.harvard.edu/abs/2020MNRAS.491.6045Y}
}

@ARTICLE{veron-cetty2004,
       author = {{V{\'e}ron-Cetty}, M. -P. and {Joly}, M. and {V{\'e}ron}, P.},
        title = "{The unusual emission line spectrum of I Zw 1}",
      journal = {\aap},
         year = 2004,
        month = apr,
       volume = {417},
        pages = {515-525},
          doi = {10.1051/0004-6361:20035714},
archivePrefix = {arXiv},
       eprint = {astro-ph/0312654},
 primaryClass = {astro-ph},
       adsurl = {https://ui.adsabs.harvard.edu/abs/2004A&A...417..515V}
}

@ARTICLE{boroson1992,
       author = {{Boroson}, Todd A. and {Green}, Richard F.},
        title = "{The Emission-Line Properties of Low-Redshift Quasi-stellar Objects}",
      journal = {\apjs},
         year = 1992,
        month = may,
       volume = {80},
        pages = {109},
          doi = {10.1086/191661},
       adsurl = {https://ui.adsabs.harvard.edu/abs/1992ApJS...80..109B}
}

@ARTICLE{gaskell2022,
       author = {{Gaskell}, Martin and {Thakur}, Neha and {Tian}, Betsy and {Saravanan}, Anjana},
        title = "{Fe II emission in active galactic nuclei}",
      journal = {Astronomische Nachrichten},
         year = 2022,
        month = jan,
       volume = {343},
       number = {1-2},
          eid = {e210112},
        pages = {e210112},
          doi = {10.1002/asna.20210112},
archivePrefix = {arXiv},
       eprint = {2112.06559},
 primaryClass = {astro-ph.GA},
       adsurl = {https://ui.adsabs.harvard.edu/abs/2022AN....34310112G}
}

@ARTICLE{hu2015,
       author = {{Hu}, Chen and {Du}, Pu and {Lu}, Kai-Xing and {Li}, Yan-Rong and {Wang}, Fang and {Qiu}, Jie and {Bai}, Jin-Ming and {Kaspi}, Shai and {Ho}, Luis C. and {Netzer}, Hagai and {Wang}, Jian-Min and {SEAMBH Collaboration}},
        title = "{Supermassive Black Holes with High Accretion Rates in Active Galactic Nuclei. III. Detection of Fe II Reverberation in Nine Narrow-line Seyfert 1 Galaxies}",
      journal = {\apj},
         year = 2015,
        month = may,
       volume = {804},
       number = {2},
          eid = {138},
        pages = {138},
          doi = {10.1088/0004-637X/804/2/138},
archivePrefix = {arXiv},
       eprint = {1503.03611},
 primaryClass = {astro-ph.GA},
       adsurl = {https://ui.adsabs.harvard.edu/abs/2015ApJ...804..138H}
}

@ARTICLE{zhang2019,
       author = {{Zhang}, Zhi-Xiang and {Du}, Pu and {Smith}, Paul S. and {Zhao}, Yulin and {Hu}, Chen and {Xiao}, Ming and {Li}, Yan-Rong and {Huang}, Ying-Ke and {Wang}, Kai and {Bai}, Jin-Ming and {Ho}, Luis C. and {Wang}, Jian-Min},
        title = "{Kinematics of the Broad-line Region of 3C 273 from a 10 yr Reverberation Mapping Campaign}",
      journal = {\apj},
         year = 2019,
        month = may,
       volume = {876},
       number = {1},
          eid = {49},
        pages = {49},
          doi = {10.3847/1538-4357/ab1099},
archivePrefix = {arXiv},
       eprint = {1811.03812},
 primaryClass = {astro-ph.GA},
       adsurl = {https://ui.adsabs.harvard.edu/abs/2019ApJ...876...49Z}
}

@ARTICLE{barth2013,
       author = {{Barth}, Aaron J. and {Pancoast}, Anna and {Bennert}, Vardha N. and {Brewer}, Brendon J. and {Canalizo}, Gabriela and {Filippenko}, Alexei V. and {Gates}, Elinor L. and {Greene}, Jenny E. and {Li}, Weidong and {Malkan}, Matthew A. and {Sand}, David J. and {Stern}, Daniel and {Treu}, Tommaso and {Woo}, Jong-Hak and {Assef}, Roberto J. and {Bae}, Hyun-Jin and {Buehler}, Tabitha and {Cenko}, S. Bradley and {Clubb}, Kelsey I. and {Cooper}, Michael C. and {Diamond-Stanic}, Aleksandar M. and {H{\"o}nig}, Sebastian F. and {Joner}, Michael D. and {Laney}, C. David and {Lazarova}, Mariana S. and {Nierenberg}, A.~M. and {Silverman}, Jeffrey M. and {Tollerud}, Erik J. and {Walsh}, Jonelle L.},
        title = "{The Lick AGN Monitoring Project 2011: Fe II Reverberation from the Outer Broad-line Region}",
      journal = {\apj},
         year = 2013,
        month = jun,
       volume = {769},
       number = {2},
          eid = {128},
        pages = {128},
          doi = {10.1088/0004-637X/769/2/128},
archivePrefix = {arXiv},
       eprint = {1304.4643},
 primaryClass = {astro-ph.CO},
       adsurl = {https://ui.adsabs.harvard.edu/abs/2013ApJ...769..128B}
}

@ARTICLE{vestergaard2001,
       author = {{Vestergaard}, M. and {Wilkes}, B.~J.},
        title = "{An Empirical Ultraviolet Template for Iron Emission in Quasars as Derived from I Zwicky 1}",
      journal = {\apjs},
         year = 2001,
        month = may,
       volume = {134},
       number = {1},
        pages = {1-33},
          doi = {10.1086/320357},
archivePrefix = {arXiv},
       eprint = {astro-ph/0104320},
 primaryClass = {astro-ph},
       adsurl = {https://ui.adsabs.harvard.edu/abs/2001ApJS..134....1V}
}

@ARTICLE{Horne2021,
       author = {{Horne}, Keith and {De Rosa}, G. and {Peterson}, B.~M. and {Barth}, A.~J. and {Ely}, J. and {Fausnaugh}, M.~M. and {Kriss}, G.~A. and {Pei}, L. and {Bentz}, M.~C. and {Cackett}, E.~M. and {Edelson}, R. and {Eracleous}, M. and {Goad}, M.~R. and {Grier}, C.~J. and {Kaastra}, J. and {Kochanek}, C.~S. and {Krongold}, Y. and {Mathur}, S. and {Netzer}, H. and {Proga}, D. and {Tejos}, N. and {Vestergaard}, M. and {Villforth}, C. and {Adams}, S.~M. and {Anderson}, M.~D. and {Ar{\'e}valo}, P. and {Beatty}, T.~G. and {Bennert}, V.~N. and {Bigley}, A. and {Bisogni}, S. and {Borman}, G.~A. and {Boroson}, T.~A. and {Bottorff}, M.~C. and {Brandt}, W.~N. and {Breeveld}, A.~A. and {Brotherton}, M. and {Brown}, J.~E. and {Brown}, J.~S. and {Canalizo}, G. and {Carini}, M.~T. and {Clubb}, K.~I. and {Comerford}, J.~M. and {Corsini}, E.~M. and {Crenshaw}, D.~M. and {Croft}, S. and {Croxall}, K.~V. and {Dalla Bont{\`a}}, E. and {Deason}, A.~J. and {Dehghanian}, M. and {De Lorenzo-C{\'a}ceres}, A. and {Denney}, K.~D. and {Dietrich}, M. and {Done}, C. and {Efimova}, N.~V. and {Evans}, P.~A. and {Ferland}, G.~J. and {Filippenko}, A.~V. and {Flatland}, K. and {Fox}, O.~D. and {Gardner}, E. and {Gates}, E.~L. and {Gehrels}, N. and {Geier}, S. and {Gelbord}, J.~M. and {Gonzalez}, L. and {Gorjian}, V. and {Greene}, J.~E. and {Grupe}, D. and {Gupta}, A. and {Hall}, P.~B. and {Henderson}, C.~B. and {Hicks}, S. and {Holmbeck}, E. and {Holoien}, T.~W. -S. and {Hutchison}, T. and {Im}, M. and {Jensen}, J.~J. and {Johnson}, C.~A. and {Joner}, M.~D. and {Jones}, J. and {Kaspi}, S. and {Kelly}, P.~L. and {Kennea}, J.~A. and {Kim}, M. and {Kim}, S. and {Kim}, S.~C. and {King}, A. and {Klimanov}, S.~A. and {Korista}, K.~T. and {Lau}, M.~W. and {Lee}, J.~C. and {Leonard}, D.~C. and {Li}, Miao and {Lira}, P. and {Lochhaas}, C. and {Ma}, Zhiyuan and {MacInnis}, F. and {Malkan}, M.~A. and {Manne-Nicholas}, E.~R. and {Mauerhan}, J.~C. and {McGurk}, R. and {McHardy}, I.~M. and {Montuori}, C. and {Morelli}, L. and {Mosquera}, A. and {Mudd}, D. and {M{\"u}ller-S{\'a}nchez}, F. and {Nazarov}, S.~V. and {Norris}, R.~P. and {Nousek}, J.~A. and {Nguyen}, M.~L. and {Ochner}, P. and {Okhmat}, D.~N. and {Pancoast}, A. and {Papadakis}, I. and {Parks}, J.~R. and {Penny}, M.~T. and {Pizzella}, A. and {Pogge}, R.~W. and {Poleski}, R. and {Pott}, J. -U. and {Rafter}, S.~E. and {Rix}, H. -W. and {Runnoe}, J. and {Saylor}, D.~A. and {Schimoia}, J.~S. and {Schn{\"u}lle}, K. and {Scott}, B. and {Sergeev}, S.~G. and {Shappee}, B.~J. and {Shivvers}, I. and {Siegel}, M. and {Simonian}, G.~V. and {Siviero}, A. and {Skielboe}, A. and {Somers}, G. and {Spencer}, M. and {Starkey}, D. and {Stevens}, D.~J. and {Sung}, H. -I. and {Tayar}, J. and {Treu}, T. and {Turner}, C.~S. and {Uttley}, P. and {Van Saders}, J. and {Vican}, L. and {Villanueva}, S., Jr. and {Weiss}, Y. and {Woo}, J. -H. and {Yan}, H. and {Young}, S. and {Yuk}, H. and {Zheng}, W. and {Zhu}, W. and {Zu}, Y.},
        title = "{Space Telescope and Optical Reverberation Mapping Project. IX. Velocity-Delay Maps for Broad Emission Lines in NGC 5548}",
      journal = {\apj},
         year = 2021,
        month = feb,
       volume = {907},
       number = {2},
          eid = {76},
        pages = {76},
          doi = {10.3847/1538-4357/abce60},
archivePrefix = {arXiv},
       eprint = {2003.01448},
 primaryClass = {astro-ph.GA},
       adsurl = {https://ui.adsabs.harvard.edu/abs/2021ApJ...907...76H}
}

@ARTICLE{fei2023,
       author = {{Fei}, Qinyue and {Wang}, Ran and {Molina}, Juan and {Shangguan}, Jinyi and {Ho}, Luis C. and {Bauer}, Franz E. and {Treister}, Ezequiel},
        title = "{Dynamics of Molecular Gas in the Central Region of the Quasar I Zwicky 1}",
      journal = {\apj},
         year = 2023,
        month = mar,
       volume = {946},
       number = {1},
          eid = {45},
        pages = {45},
          doi = {10.3847/1538-4357/acbb05},
archivePrefix = {arXiv},
       eprint = {2302.04003},
 primaryClass = {astro-ph.GA},
       adsurl = {https://ui.adsabs.harvard.edu/abs/2023ApJ...946...45F}
}

@ARTICLE{Laha2021,
       author = {{Laha}, Sibasish and {Reynolds}, Christopher S. and {Reeves}, James and {Kriss}, Gerard and {Guainazzi}, Matteo and {Smith}, Randall and {Veilleux}, Sylvain and {Proga}, Daniel},
        title = "{Ionized outflows from active galactic nuclei as the essential elements of feedback}",
      journal = {Nature Astronomy},
         year = 2021,
        month = jan,
       volume = {5},
        pages = {13-24},
          doi = {10.1038/s41550-020-01255-2},
archivePrefix = {arXiv},
       eprint = {2012.06945},
 primaryClass = {astro-ph.GA},
       adsurl = {https://ui.adsabs.harvard.edu/abs/2021NatAs...5...13L}
}

@ARTICLE{Haardt1991,
       author = {{Haardt}, F. and {Maraschi}, L.},
        title = "{A Two-Phase Model for the X-Ray Emission from Seyfert Galaxies}",
      journal = {\apjl},
         year = 1991,
        month = oct,
       volume = {380},
        pages = {L51},
          doi = {10.1086/186171},
       adsurl = {https://ui.adsabs.harvard.edu/abs/1991ApJ...380L..51H}
}

@ARTICLE{Gravity2022,
       author = {{GRAVITY+ Collaboration} and {Abuter}, R. and {Alarcon}, P. and {Allouche}, F. and {Amorim}, A. and {Bailet}, C. and {Bedigan}, H. and {Berdeu}, A. and {Berger}, J. -P. and {Berio}, P. and {Bigioli}, A. and {Blaho}, R. and {Boebion}, O. and {Bolzer}, M. -L. and {Bonnet}, H. and {Bourdarot}, G. and {Bourget}, P. and {Brandner}, W. and {Cardenas}, C. and {Conzelmann}, R. and {Comin}, M. and {Cl{\'e}net}, Y. and {Courtney-Barrer}, B. and {Dallilar}, Y. and {Davies}, R. and {Defr{\`e}re}, D. and {Delboulb{\'e}}, A. and {Delplancke-Str{\"o}bele}, F. and {Dembet}, R. and {de Zeeuw}, T. and {Drescher}, A. and {Eckart}, A. and {{\'E}douard}, C. and {Eisenhauer}, F. and {Fabricius}, M. and {Feuchtgruber}, H. and {Finger}, G. and {F{\"o}rster Schreiber}, N.~M. and {Fuenteseca}, E. and {Garcia}, E. and {Garcia}, P. and {Gao}, F. and {Gendron}, E. and {Genzel}, R. and {Gil}, J.~P. and {Gillessen}, S. and {Gomes}, T. and {Gont{\'e}}, F. and {Gouvret}, C. and {Guajardo}, P. and {Guidolin}, I. and {Guieu}, S. and {Guzmann}, R. and {Hackenberg}, W. and {Haddad}, N. and {Hartl}, M. and {Haubois}, X. and {Hau{\ss}mann}, F. and {Hei{\ss}el}, G. and {Henning}, T. and {Hippler}, S. and {H{\"o}nig}, S. and {Horrobin}, M. and {Hubin}, N. and {Jacqmart}, E. and {Jocou}, L. and {Kaufer}, A. and {Kervella}, P. and {Kirchbauer}, J. -P. and {Kolb}, J. and {Korhonen}, H. and {Kreidberg}, L. and {Krempl}, P. and {Lacour}, S. and {Lagarde}, S. and {Lai}, O. and {Lapeyr{\`e}re}, V. and {Laugier}, R. and {Le Bouquin}, J. -B. and {Leftley}, J. and {L{\'e}na}, P. and {Lewis}, S. and {Lutz}, D. and {Magnard}, Y. and {Mang}, F. and {Marcotto}, A. and {Maurel}, D. and {M{\'e}rand}, A. and {Millour}, F. and {More}, N. and {Nowacki}, H. and {Nowak}, M. and {Oberti}, S. and {Olivares}, F. and {Ott}, T. and {Pallanca}, L. and {Paumard}, T. and {Perraut}, K. and {Perrin}, G. and {Petrov}, R. and {Pfuhl}, O. and {Pourr{\'e}}, N. and {Rabien}, S. and {Rau}, C. and {Riquelme}, M. and {Robbe-Dubois}, S. and {Rochat}, S. and {Salman}, M. and {Scherbarth}, M. and {Sch{\"o}ller}, M. and {Schubert}, J. and {Schuhler}, N. and {Shangguan}, J. and {Shchekaturov}, P. and {Shimizu}, T. and {Scheithauer}, S. and {Sevin}, A. and {Soenke}, C. and {Soulez}, F. and {Spang}, A. and {Stadler}, E. and {Straubmeier}, C. and {Sturm}, E. and {Sykes}, C. and {Tacconi}, L. and {Tischer}, H. and {Tristram}, K. and {Vincent}, F. and {von Fellenberg}, S. and {Uysal}, S. and {Widmann}, F. and {Wieprecht}, E. and {Wiezorrek}, E. and {Woillez}, J. and {Yaz{\i}c{\i}}, {\c{S}}. and {Zins}, G.},
        title = "{The GRAVITY+ Project: Towards All-sky, Faint-Science, High-Contrast Near-Infrared Interferometry at the VLTI}",
      journal = {The Messenger},
         year = 2022,
        month = dec,
       volume = {189},
        pages = {17-22},
          doi = {10.18727/0722-6691/5285},
archivePrefix = {arXiv},
       eprint = {2301.08071},
 primaryClass = {astro-ph.IM},
       adsurl = {https://ui.adsabs.harvard.edu/abs/2022Msngr.189...17A}
}

@ARTICLE{vieliute2025,
       author = {{Vieliute}, Roberta and {Hern{\'a}ndez Santisteban}, Juan V. and {Horne}, Keith and {Cornfield}, Hannah},
        title = "{PyTICS: an iterative method for photometric light-curve intercalibration using comparison stars}",
      journal = {RAS Techniques and Instruments},
         year = 2025,
        month = jan,
       volume = {4},
          eid = {rzaf021},
        pages = {rzaf021},
          doi = {10.1093/rasti/rzaf021},
archivePrefix = {arXiv},
       eprint = {2505.23328},
 primaryClass = {astro-ph.IM},
       adsurl = {https://ui.adsabs.harvard.edu/abs/2025RASTI...4...21V}
}

@ARTICLE{wang1999b,
       author = {{Wang}, Jian-Min and {Szuszkiewicz}, Ewa and {Lu}, Fang-Jun and {Zhou}, You-Yuan},
        title = "{Emergent Spectra from Slim Accretion Disks in Active Galactic Nuclei}",
      journal = {\apj},
         year = 1999,
        month = sep,
       volume = {522},
       number = {2},
        pages = {839-845},
          doi = {10.1086/307686},
       adsurl = {https://ui.adsabs.harvard.edu/abs/1999ApJ...522..839W}
}

@ARTICLE{gordon2006,
       author = {{Richards}, Gordon T. and {Lacy}, Mark and {Storrie-Lombardi}, Lisa J. and {Hall}, Patrick B. and {Gallagher}, S.~C. and {Hines}, Dean C. and {Fan}, Xiaohui and {Papovich}, Casey and {Vanden Berk}, Daniel E. and {Trammell}, George B. and {Schneider}, Donald P. and {Vestergaard}, Marianne and {York}, Donald G. and {Jester}, Sebastian and {Anderson}, Scott F. and {Budav{\'a}ri}, Tam{\'a}s and {Szalay}, Alexander S.},
        title = "{Spectral Energy Distributions and Multiwavelength Selection of Type 1 Quasars}",
      journal = {\apjs},
         year = 2006,
        month = oct,
       volume = {166},
       number = {2},
        pages = {470-497},
          doi = {10.1086/506525},
archivePrefix = {arXiv},
       eprint = {astro-ph/0601558},
 primaryClass = {astro-ph},
       adsurl = {https://ui.adsabs.harvard.edu/abs/2006ApJS..166..470R}
}

@ARTICLE{kubota2019,
       author = {{Kubota}, Aya and {Done}, Chris},
        title = "{Modelling the spectral energy distribution of super-Eddington quasars}",
      journal = {\mnras},
         year = 2019,
        month = oct,
       volume = {489},
       number = {1},
        pages = {524-533},
          doi = {10.1093/mnras/stz2140},
archivePrefix = {arXiv},
       eprint = {1905.02920},
 primaryClass = {astro-ph.GA},
       adsurl = {https://ui.adsabs.harvard.edu/abs/2019MNRAS.489..524K}
}

@ARTICLE{trammell2007,
       author = {{Trammell}, George B. and {Vanden Berk}, Daniel E. and {Schneider}, Donald P. and {Richards}, Gordon T. and {Hall}, Patrick B. and {Anderson}, Scott F. and {Brinkmann}, J.},
        title = "{The UV Properties of SDSS-Selected Quasars}",
      journal = {\aj},
         year = 2007,
        month = apr,
       volume = {133},
       number = {4},
        pages = {1780-1794},
          doi = {10.1086/511817},
archivePrefix = {arXiv},
       eprint = {astro-ph/0611549},
 primaryClass = {astro-ph},
       adsurl = {https://ui.adsabs.harvard.edu/abs/2007AJ....133.1780T}
}

@ARTICLE{davis2007,
       author = {{Davis}, Shane W. and {Woo}, Jong-Hak and {Blaes}, Omer M.},
        title = "{The UV Continuum of Quasars: Models and SDSS Spectral Slopes}",
      journal = {\apj},
         year = 2007,
        month = oct,
       volume = {668},
       number = {2},
        pages = {682-698},
          doi = {10.1086/521393},
archivePrefix = {arXiv},
       eprint = {0707.1456},
 primaryClass = {astro-ph},
       adsurl = {https://ui.adsabs.harvard.edu/abs/2007ApJ...668..682D}
}

@ARTICLE{vestergaard2005,
       author = {{Vestergaard}, Marianne and {Peterson}, Bradley M.},
        title = "{Variability of Fe II Emission Features in the Seyfert 1 Galaxy NGC 5548}",
      journal = {\apj},
         year = 2005,
        month = jun,
       volume = {625},
       number = {2},
        pages = {688-698},
          doi = {10.1086/429791},
archivePrefix = {arXiv},
       eprint = {astro-ph/0502433},
 primaryClass = {astro-ph},
       adsurl = {https://ui.adsabs.harvard.edu/abs/2005ApJ...625..688V}
}



\appendix

\section{I Zw 1 Spectrum}
In parallel with our LCO 1\:m photometric monitoring, we monitored variations in the optical spectrum of I~Zw~1 using the LCO 2\:m robotic telescopes.
The Faulkes Telescope North (FTN) at Haleakala in Hawaii, and the Faulkes Telescope South (FTS) at Siding Springs Observatory in Australia are equipped with nearly-identical Floyds spectrographs\footnote{\url{https://lco.global/observatory/instruments/floyds/}}.
We obtained $80$ spectra taken in pairs in $40$ epochs, throughout Years 2--4 ($\sim15$ per year). The spectra are cross-dispersed with first-order red spectra (5400-10000~\AA) and second-order blue spectra (3200-5700~\AA) projected simultaneously on the CCD. The spectral resolution is $R\equiv\lambda/\Delta\lambda=400$ on the blue end and $R=700$ on the red end of each order. We used 600~s exposures, and employed the $6\arcsec$ slit, oriented at the parallactic angle, to minimize wavelength-dependent slit losses due to changes in seeing and airmass. 
Wavelength calibration and flat field lamp exposures were taken on each visit. 

The spectra were extracted using the {\sc AGNFLOYDS} pipeline\footnote{\url{ https://github.com/svalenti/FLOYDS pipeline}}. To establish the absolute flux calibration, divide out telluric absorption features, and reduce fringing artifacts (although still evident at wavelengths larger than 7000\:\AA), we used the spectrum of the closest spectrophotometric standard star observed within 5 days. 

\begin{figure*}
    \centering
    \includegraphics[width=0.9\textwidth]{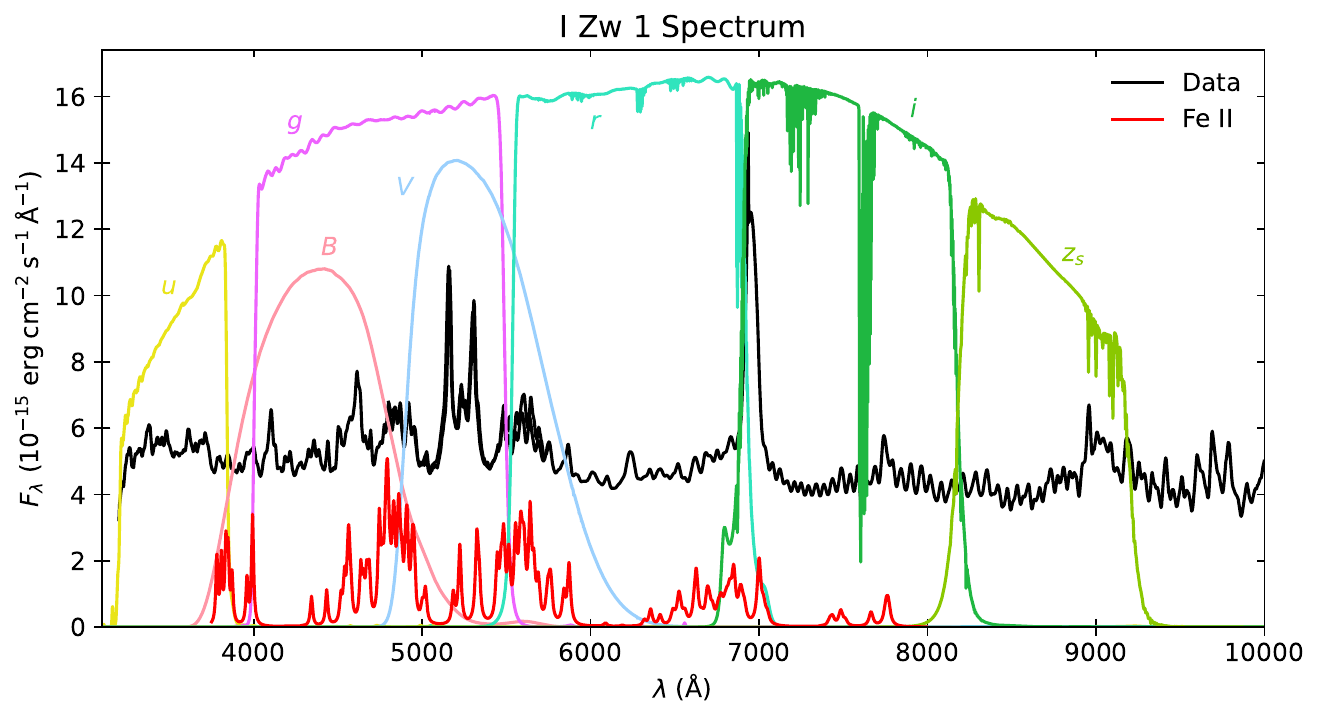}
    \caption{The average spectrum of I~Zw~1 over the LCO campaign in the observed frame (black lines), overlaid with the LCO filter transmission curves. The \ion{Fe}{ii} template by \citet{veron-cetty2004}, which was created based on I~Zw~1, is plotted with the red lines but does not indicate the actual amount of \ion{Fe}{ii} in I~Zw~1.}
    \label{fig:spec}
\end{figure*}

We analysed the time-resolved spectroscopy with {\sc prepspec}\footnote{\url{http://star-www.st-andrews.ac.uk/\~kdh1/lib/prepspec/prepspec.tar.gz}} to quantify variability in several broad emission lines. A similar {\sc prepspec} analysis of NGC~5548 is described in full detail in \citet{Horne2021}. Briefly, {\sc prepspec} performs a maximum likelihood fit to the Floyds spectra at all epochs, adopting a simple (ABC) model for the spectral variations:
\begin{equation}
 F(\lambda,t) = A(\lambda) + B(\lambda,t)
 + C(\lambda,t) \ .
\end{equation}
This fit estimates the mean spectrum $A(\lambda)$, the broad emission-line variations $B(\lambda,t)$, and the continuum variations $C(\lambda,t)$. The full results of this analysis will be published in a following paper. Here, we just show the average spectrum across this campaign in Fig.~\ref{fig:spec} to highlight the contribution of hydrogen, helium and \ion{Fe}{ii} lines to the broadband filters. We note that the oscillation pattern dominant in the continuum at $>7000$~\AA\ are caused by fringing interference artefacts in the detector. We also include the \ion{Fe}{ii} template from \citet{veron-cetty2004} to show the location of these prominent emission lines.

\section{Additional Time Series Analysis Results}

\begin{figure}
    \centering
    \includegraphics[width=0.48\textwidth]{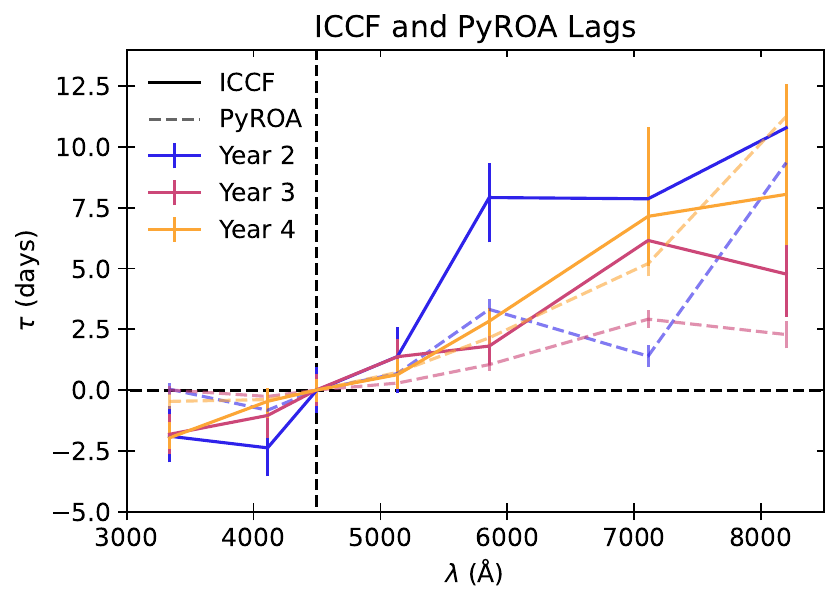}
    \caption{The lag spectrum for Years 2--4 as calculated using ICCF and PyROA for each year separately, with reference to the $g$-band in the AGN rest frame. The ICCF lags are the same as plotted in Fig. \ref{fig:iccf_lags} and presented in Table \ref{tab:lags}. The ICCF lags are denoted by the solid line and the PyROA lags by the dashed line.}
    \label{fig:iccf_pyroa}
\end{figure}

\begin{figure*}
    \centering
    \includegraphics[width=0.9\textwidth]{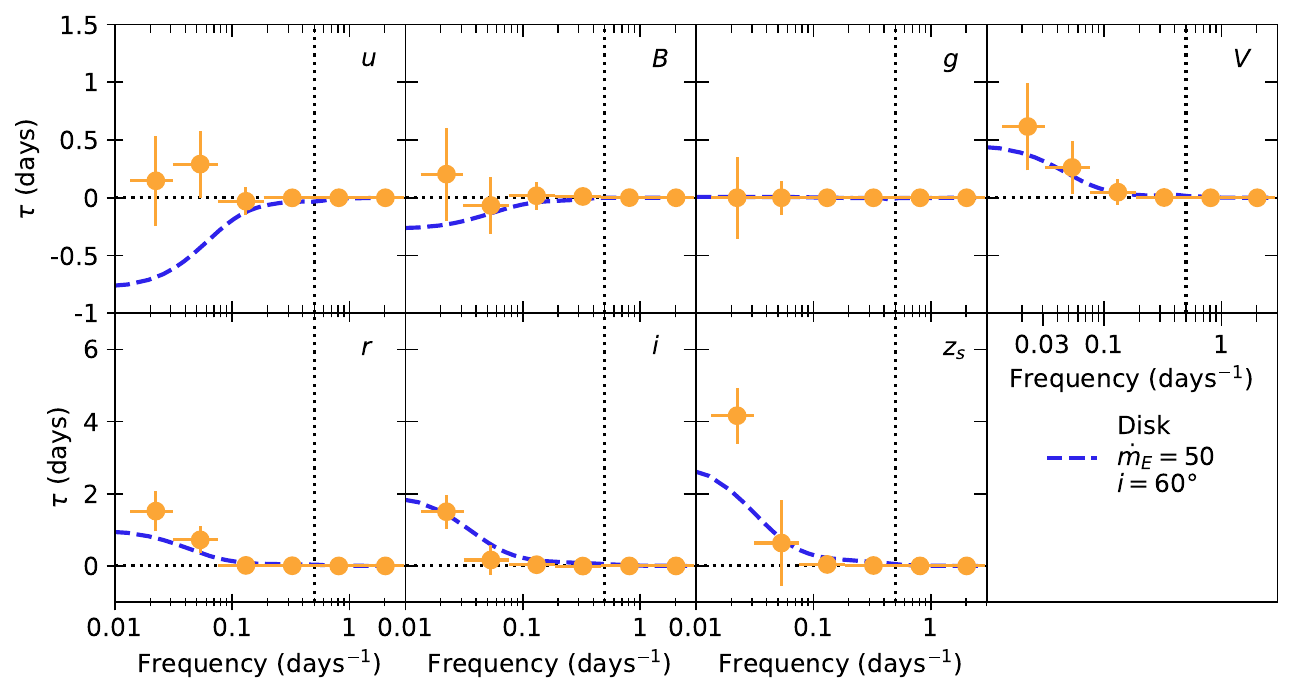}
    \caption{The lag-frequency spectra as plotted in Fig.~\ref{fig:frl} combined with the prediction for a thin accretion disk with an inclination of 60° and $\dot{m}_E = 50$ (blue dashed line).}
    \label{fig:frl_mdot50}
\end{figure*}

This section includes additional results of our time series analysis in Section~\ref{sec:tsa}. In Table~\ref{tab:rmax}, we tabulate the peak correlation coefficients from the interpolated cross correlation analysis in Section~\ref{ccf} for Years 2, 3, and 4. To compare the ICCF and PyROA lags on a per year basis we have plotted these in Fig. \ref{fig:iccf_pyroa}. Here, we have calculated the PyROA lags for each year separately but otherwise identical to the method described in Section \ref{pyroa}. In Table~\ref{tab:var_lags}, we present the lags derived from PyROA modelling at different variability timescales as described in Section~\ref{pyroa}. For the PyROA fits we vary the width $\Delta$ of the Gaussian memory function with $\Delta = 3$, 5, 10, and 20\:days. The lags are plotted in Fig.~\ref{fig:var_ts_lags}. Fig.~\ref{fig:frl_mdot50} shows the same frequency-resolved lags as Fig.~\ref{fig:frl} but overlaid with the model of a thin disk at an inclination of 60° with an Eddington ratio of 50. The frequency-resolved lags are described in Section~\ref{frl}, as well as the thin disk model and transfer function \citep{Collier1999,cackett2007,Starkey2016}.

\begin{table*}
    \centering
    \begin{tabularx}{0.75\textwidth}{ccYYY}
    && Year 2 & Year 3 & Year 4 \\
    Band & $\lambda_{\mathrm{eff}}$ ($\textnormal{\AA}$) & $r_\mathrm{peak}$ &$r_\mathrm{peak}$&$r_\mathrm{peak}$\vspace{1ex}\\ \hline \hline \\[-1ex]
    $u$&$3540$&$0.95$&$0.96$&$0.89$\vspace{1ex}\\
    $B$&$4361$&$0.94$&$0.96$&$0.90$\vspace{1ex}\\
    $g$&$4770$&$1.00$&$1.00$&$1.00$\vspace{1ex}\\
    $V$&$5448$&$0.97$&$0.97$&$0.92$\vspace{1ex}\\
    $r$&$6215$&$0.93$&$0.96$&$0.85$\vspace{1ex}\\
    $i$&$7545$&$0.92$&$0.93$&$0.79$\vspace{1ex}\\
    $z_s$&$8700$&$0.85$&$0.84$&$0.58$\vspace{1ex}\\
    \hline
    \end{tabularx}
    \caption{The peak correlation coefficient $r_\mathrm{peak}$ calculated in the ICCF analysis with reference to the $g$-band in Section~\ref{ccf}. }
    \label{tab:rmax}
\end{table*}

\begin{table*}
    \centering
    \begin{tabularx}{0.8\textwidth}{ccYYYY}
    && $\Delta=3$ & $\Delta = 5$ & $\Delta=10$ & $\Delta=20$ \\
    Band & $\lambda_{\mathrm{eff}}$ ($\textnormal{\AA}$) & $\tau$ (days) & $\tau$ (days)&$\tau$ (days)&$\tau$ (days)\vspace{1ex}\\ \hline \hline \\[-1ex]
    $u$&$3540$&$-0.28\pm0.16$&$-0.64_{-0.20}^{+0.21}$&$-1.23\pm0.30$&$-1.78_{-0.56}^{+0.57}$\vspace{1ex}\\
    $B$&$4361$&$-0.52\pm0.17$&$-0.66_{-0.21}^{+0.22}$&$-0.89\pm0.32$&$-1.62\pm0.60$\vspace{1ex}\\
    $g$&$4770$&$0.00\pm0.13$&$0.00\pm0.17$&$0.00\pm0.27$&$0.00\pm0.55$\vspace{1ex}\\
    $V$&$5448$&$0.52\pm0.14$&$0.59\pm0.19$&$0.79\pm0.28$&$1.50\pm0.53$\vspace{1ex}\\
    $r$&$6215$&$2.18\pm0.20$&$2.35\pm0.24$&$2.31\pm0.33$&$3.11\pm0.56$\vspace{1ex}\\
    $i$&$7545$&$3.50\pm0.24$&$4.23_{-0.30}^{+0.29}$&$5.38\pm0.39$&$8.42\pm0.62$\vspace{1ex}\\
    $z_s$&$8700$&$4.78_{-0.48}^{+0.65}$&$5.73_{-0.51}^{+0.52}$&$6.54_{-0.63}^{+0.66}$&$9.47\pm0.86$\vspace{1ex}\\
    \hline
    \end{tabularx}
    \caption{Lags in days between each light curve and the reference light curve in the $g$-band, calculated with PyROA for different values of the light curve stiffness fitting parameter $\Delta$. The lags are for PyROA fits with $\Delta=3$, 5, 10, and 20 days as described in Section~\ref{pyroa}. The lags are plotted in Fig.~\ref{fig:var_ts_lags}.}
    \label{tab:var_lags}
\end{table*}

\section{LCO SED}

Here we tabulate the LCO SED of the AGN; the faint and bright contours, the RMS, and the host galaxy contribution in Table~\ref{tab:sed}, which were calculated using flux-flux analysis as described in Section~\ref{sed}. Years 2, 3, and 4 of the LCO campaign are included. We present both the observed flux and flux corrected for Galactic extinction using $E(B-V) = 0.057$ \citep{schlafly2011,fitzpatrick1999}. The SED corrected for Galactic and internal extinction is plotted in Figs. \ref{fig:sed_all} and \ref{fig:sed_fits}.

\begin{table*}
    \centering
    \begin{tabularx}{\textwidth}{cc|YYYY|YYYY}
    &&\multicolumn{4}{c|}{observed}&\multicolumn{4}{c}{dereddened}\\
    & $\lambda_{\mathrm{eff}}$ ($\textnormal{\AA}$) & $F_\mathrm{bright}$&$F_\mathrm{faint}$ & $F_\mathrm{RMS}$& $F_\mathrm{gal}$& $F_\mathrm{bright}$&$F_\mathrm{faint}$ & $F_\mathrm{RMS}$& $F_\mathrm{gal}$\vspace{1ex}\\ \hline \hline \\[-1ex]
    $u$ & 3540 & $3.451\pm0.003$ & $2.061\pm0.002$ & $0.419\pm0.001$ & $0.006\pm 0.007$ & $4.453\pm0.003$& $2.659\pm0.002$ & $0.541\pm0.001$ & $0.008\pm0.009$\\
    $B$ & 4361 & $4.798\pm0.013$ & $2.865\pm0.010$ & $0.583\pm0.002$ & $0.268\pm0.009$ & $5.955\pm0.016$ & $3.552\pm0.012$ & $0.722\pm0.002$ & $0.332\pm0.011$\\
    $g$ & 4770 & $5.352\pm0.011$ & $3.196\pm0.009$ & $0.650\pm0.001$ & $0.624\pm0.010$ & $6.502\pm0.014$ & $3.883\pm0.011$ & $0.790\pm0.002$ & $0.758\pm0.013$ \\
    $V$ & 5448 & $6.792\pm0.015$ & $4.056\pm0.012$ & $0.825\pm0.002$ & $1.139\pm0.013$ & $7.981\pm0.017$ & $4.766\pm0.014$ & $0.969\pm0.002$ & $1.338\pm0.015$\\
    $r$ & 6215 & $8.078\pm0.022$ & $4.824\pm0.016$ & $0.981\pm0.003$ & $1.467\pm0.016$ & $9.238\pm0.025$ & $5.517\pm0.019$ & $1.123\pm0.003$ & $1.677\pm0.018$\\
    $i$ & 7545 & $9.871\pm0.032$ & $5.895\pm0.023$ & $1.199\pm0.004$ & $3.350\pm0.019$ & $10.90\pm0.04$ & $6.510\pm0.025$ & $1.324\pm0.004$ & $3.699\pm0.021$\\
    $z_s$ & 8700 & $10.30\pm0.05$ & $6.152\pm0.035$ & $1.251\pm0.006$ & $3.890\pm0.020$ & $11.14\pm0.06$ & $6.651\pm0.038$ & $1.353\pm0.007$ & $4.206\pm0.021$\vspace{1ex}\\
    \hline
    \end{tabularx}
    \caption{The LCO I~Zw~1 SED calculated using the flux-flux analysis, as described in Section~\ref{sed}. All fluxes are in mJy. The `observed' flux is flux as observed, without any extinction corrections. The `dereddened' flux is corrected for line-of-sight Galactic extinction only, with $E(B-V)=0.057$ \citep{schlafly2011,fitzpatrick1999}. $F_\mathrm{bright}$ is the brightest AGN flux over the years 2, 3, and 4 of the campaign, $F_\mathrm{faint}$ is the faintest flux, and $F_\mathrm{RMS}$ is the AGN RMS flux. $F_\mathrm{gal}$ is the host galaxy flux.}
    \label{tab:sed}
\end{table*}


\bsp	
\label{lastpage}
\end{document}